\documentclass{mn2e}
\usepackage{amssymb,mnextra,times}
\input{epsf}
\def \etal {et al.\ }

\nobrackets     

\begin{document}

\title[Black Hole Growth]{Black Hole Growth in Hierarchical Galaxy Formation}
\author[R. K. Malbon et al.] {
\parbox[h]{\textwidth}{Rowena K. Malbon$^1$, C. M. Baugh$^1$,
C. S. Frenk$^1$, C. G. Lacey$^1$}
\vspace*{6pt} \\ 
$^1$Institute for Computational Cosmology, Department of Physics, 
University of Durham, \\
Science Laboratories, South 
Road, Durham DH1 3LE, United Kingdom \\
}

\maketitle
\begin{abstract}
We incorporate a model for black hole growth during galaxy mergers 
into the semi-analytical galaxy formation model based on $\Lambda \rm CDM$ proposed by 
Baugh \etal. Our black hole model has one free parameter, which we 
set by matching the observed zeropoint of the local correlation between 
black hole mass and bulge luminosity. We present predictions for the 
evolution with redshift of the relationships between black 
hole mass and bulge properties. Our simulations reproduce the evolution of the optical luminosity 
function of quasars. We study the demographics of the black hole population 
and address the issue of how black holes acquire their mass. We find 
that the direct accretion of cold gas during starbursts is an important growth 
mechanism for lower mass black holes and at high redshift. 
On the other hand, the re-assembly of pre-existing black hole mass
into larger units via merging 
dominates the growth of more massive black holes at low redshift. 
This prediction could be tested by future gravitational wave experiments. 
As redshift decreases, progressively less massive black holes have the
highest fractional growth rates, in line with recent claims of
``downsizing'' in quasar activity.  
\end{abstract}

\section{Introduction}

In the local Universe, luminous, dusty,
merger-driven starburst activity has long been
suspected to have quasar activity associated with it (\cite{sm96}).
Some authors find that the most powerful Seyfert II active galactic nuclei are usually found in
galaxies which have had a starburst in the past 1-2 Gyr and use this
observation to argue that the brightest quasars are associated with galaxy
mergers, (\cite{kauffmann03}), whilst others claim that the brightest quasars
are hosted in elliptical galaxies which are indistinguishable from the general
elliptical population (\cite{dunlop03}). At high redshift, sources
detected in the submillimeter are thought to be starbursts (\cite{chapman04}), many
are associated with galaxy mergers (\cite{swinbank04}) and many show
evidence of active nuclei when probed deeply in the X-rays, although
it appears that the AGN makes a much smaller contribution to the
powerful submm flux than the starburst (\cite{alexander03}).

Black holes (BH) display strong correlations with the properties of
their host galaxy, particularly those of the galactic bulge
(\cite{kr95,magorrian98,novak06}). Black hole mass is observed
to scale with the bulge's B-band luminosity (\cite{magorrian98,kg01}),
K-band luminosity (\cite{mh03,hr04}), stellar mass (\cite{mh03,hr04})
and velocity dispersion (\cite{fm00,gebhardt00}). We refer to these collectively as the `$M_{\rm BH}-\rm bulge$' relations. It has long been
theorized that galactic bulges form through galaxy mergers
(\cite{tt72}), so it is natural to speculate that these events also
drive the strong correlation between the properties of the bulge and
the mass of the black hole.

There is strong evidence for a link between galactic star formation and
accretion onto central black holes. The evolution with redshift of the global star 
formation rate and the luminosity density of optical quasars are strongly 
correlated (\cite{bt98}). At low redshift, the ratio of the global star 
formation rate to the global black hole accretion rate for bulge 
dominated galaxies, $SFR / \dot M_{\rm BH}$ is $\sim 1000$, which is 
remarkably similar to the ratio of $M_{\rm BH}/M_{\rm bulge}$ 
(\cite{heckman04}). However, it is still an open question whether  
black hole growth is correlated with all star formation equivalently,
or whether its strongest relationship is with star formation in bursts. 

The physical conditions in mergers and starbursts are amenable to fuelling 
the accretion of material onto a central supermassive black hole. 
Numerical simulations of galaxy mergers have shown that the asymmetrical 
gravitational potential present during the merger is responsible for driving 
gas to the centres of the merging galaxies and of their remnant in both 
major mergers (\cite{mh94major}) and minor mergers (\cite{mh94minor}). 
The enhanced supply of gas to the centre of the galaxy leads to rapid star 
formation and is also available to fuel an AGN (\cite{ns88,dsh05,sdh05}). 
Furthermore, the formation of a dense stellar system with a steep 
$R^{1/4}$-law potential well during a gas-rich merger may help to funnel gas 
to the AGN at the very centre. 
Starbursts appear to be required in the high redshift Universe 
to explain observations of various galaxy populations 
(\cite{somerville01,baugh05}). The increased prevalence of starbursts at 
early epochs may be responsible for the accelerated growth of the most 
massive black holes towards high redshift (e.g. Granato et al. 2004,
2006).

Theoretical calculations of the growth of black holes in the Cold Dark
Matter (CDM) cosmology in which structures grow through gravitational
instability have tended to fall into one of three classes: (i)
calculations based on the rate at which dark matter haloes are
assembled, either without any treatment of galaxy formation
(e.g. \cite{er88,hr93,hl98,pm99,wl03,haiman04,kbd04,yoo04,mds05}) or
with very simple estimates of the supply of gas accreted onto the BH
(\cite{vol03,islam03,bromley04,libeskind06}); (ii) numerical
simulations of galaxy mergers, which use a mixture of smooth particle
hydrodynamics and simple recipes to follow the fuelling of a supermassive
black hole
(\cite{cattaneo05sph}; Di Matteo \etal \shortcite{dsh05}; \cite{hopkins05quasar}; Springel \etal \shortcite{sdh05}; \cite{robertson06,hopkins06unified});
(iii) semi-analytical modelling of the formation of galaxies and black
holes
(\cite{kh00,cattaneo01,enoki03,menci04,granato04,monacofontanot05,cattaneo05}).
Recently, the semi-analytical approach has been extended to produce
models in which the evolution of galaxies and black holes are coupled,
with energy released by accretion onto the black hole either
truncating ongoing star formation or suppressing the rate at which gas
can cool in more massive haloes
(\cite{granato04,monacofontanot05,bower06,croton06}).
 
In this paper, we incorporate a model for the growth of black holes
into the Durham semi-analytical galaxy formation code {\tt GALFORM}
(\cite{cole00,benson03}). Our prescription for growing black holes is
tied to galaxy mergers and is similar to the first implementation of
black hole growth in semi-analytical models by Kauffmann \& Haehnelt
(\shortcite{kh00}). Our starting point is the galaxy formation model
introduced by Baugh et~al. (\shortcite{baugh05}). This was the first
model to match the observed properties of galaxies in both the low and high
redshift Universe, following the whole of the galaxy population and
incorporating a self-consistent calculation of the reprocessing of
starlight by dust. In particular, the model reproduces the number
counts of Lyman break galaxies and sub-mm sources, which are both
dominated by starbursts. The success of the model is primarily due to
an increased level of star formation in bursts at high redshift 
compared with previous models, and the adoption of a flat initial mass
function (IMF) for stars produced in starbursts. The same model
also accounts for the metal content of the hot gas in clusters and stars in
ellipticals (\cite{nagashima05a,nagashima05b} and the numbers of
Lyman-alpha emitters (Le~Delliou et al. 2005, 2006). 
Since the Baugh et~al. 
model has been tested extensively, and, in particular, in view 
of the success of this model in reproducing aspects of the 
galaxy population which are associated with starbursts (and hence 
bulge formation), we 
have chosen to focus on the predictions for black hole growth 
and quasar activity.

The use of a semi-analytical model allows us to follow a much wider population 
of objects than is accessible by direct numerical simulation. This means  
that we can follow the demographics of the black hole population and explore 
how black holes acquire their mass. The latter is of great importance in view 
of the recent observational evidence suggesting that the most massive 
black holes acquired the bulk of their mass at early epochs and 
that it is the lower mass black holes which are being built up most rapidly 
today. This phenomenon has been termed ``downsizing'' (\cite{cowie03,steffen03,ueda03,barger05,hasinger05}). At first sight, downsizing appears to imply 
that the growth of black hole mass is ``anti-hierarchical'' and thus
incompatible with the CDM cosmological framework 
(\cite{marconi04,merloni04,shankar04}). We will examine here whether or not
such downsizing is really a problem for hierarchical models of galaxy formation. 

The paper is organized as follows. We provide a description of the 
model in \S\ref{sec:method}. The model contains one free parameter, which 
we set by matching the $z=0$ $M_{\rm BH}-\rm bulge$ relations in
\S\ref{sec:obsmatch}, where 
we also show that our model is consistent with the evolution of the
quasar luminosity function. In \S\ref{sec:results}, we study the
growth histories of black holes, separating the contributions from black hole mergers 
and direct gas accretion, and show how the relative importance of these channels 
varies with black hole mass. We predict the evolution of the 
$M_{\rm BH}-\rm bulge$ relations and compare this to data in
\S\ref{sec:magz}. We demonstrate  
that we are able to produce downsizing in the AGN population in 
\S\ref{sec:downsizing}. We summarize our main results, discuss their 
context and outline future improvements to the model in  \S\ref{sec:discuss}.

\section{Method}
\label{sec:method}

In this section we first give a brief overview of our galaxy formation
model (\S2.1,\S2.2), before explaining how the model has been extended
to follow the formation of black holes (\S2.3). We discuss the
sensitivity of our model predictions to the mass resolution of the dark
matter merger trees in \S2.4. We end this section with a brief
description of how a quasar luminosity is assigned to an accreting black
hole, and present some illustrative results for the quasar luminosity
function at selected redshifts (\S2.5).

\subsection{The semi-analytical galaxy formation model}
\label{sec:GF}

Our starting point is the model for galaxy formation in the CDM
cosmology described by Baugh \etal (\shortcite{baugh05}).  As we have
already pointed out in the Introduction, in addition to
giving a reasonable match to the properties of galaxies in the local
Universe, this model also reproduces the counts of sub-millimetre sources and the
luminosity function of Lyman-break galaxies at high redshift. In both cases, the model
associates these high redshift objects with galaxies which are
undergoing merger-driven starbursts.  The success of the Baugh \etal
(\shortcite{baugh05}) model in reproducing observations linked with
vigorous starbursts and the formation of spheroids is important for
the current analysis. Here we will follow the proposal of Kauffmann \&
Haehnelt (2000) and assume that black hole growth is driven by galaxy
mergers.  For a more exhaustive description of the physics and
methodology behind the semi-analytical model, we refer the reader to
Cole \etal (\shortcite{cole00}) and Benson \etal
(2003b). A gentler introduction to hierarchical galaxy
formation may be found in Baugh (\shortcite{baugh06}).

We will review the aspects of the model which control the outcome of
galaxy mergers in the next subsection, and will limit ourselves here to more
general aspects of the cosmological and galaxy formation models. We assume a standard
$\Lambda$CDM cosmology, with a flat geometry, a matter density
$\Omega_{0}=0.3$, a baryon density $\Omega_{\rm b}=0.04$, a Hubble
constant of $H_{0}=70~{\rm kms}^{-1}{\rm Mpc}^{-1}$ and a fluctuation
amplitude specified by $\sigma_{8}=0.9$. The break in the local galaxy
luminosity function is reproduced by invoking a superwind which drives
cold gas out of galaxies (\cite{benson03,nagashima05a}); an
alternative physical mechanism to produce this break is AGN feedback
in quasi-hydrostatically cooling haloes (\cite{bower06,croton06}). Gas cooling is
prevented below $z = 6$ in low circular velocity haloes ($v_{\rm c}=60~{\rm kms}^{-1}$), to mimic the impact of the presence of a photoionizing
background on the intergalactic medium (\cite{benson02}).  Baugh \etal
(\shortcite{baugh05}) 
adopt a timescale for quiescent star formation in galactic discs which
is independent of the dynamical time, which results in gas rich
mergers at high redshift (see their fig. 1). They assume that stars which
form in bursts are produced with a top-heavy initial mass function; this
choice has no impact on the predictions presented in this paper beyond
the high fraction of cold gas forming stars that is recycled into the IGM.

The parameters of the Baugh \etal (\shortcite{baugh05}) galaxy formation model are held
fixed in this paper; we do not adjust these parameters in any way when
generating predictions for black holes and quasars. This is a clear
strength of our approach and choice of galaxy formation model. Thus,
our results are to be viewed as genuine predictions of the model. The
properties of the quasars and active nuclei in our model can easily be
related to the properties of their host galaxies; such comparisons are
deferred to future papers.

\subsection{Galaxy mergers}
\label{sec:Gmergers}

Mergers between galaxies play an important role in building up the
mass and determining the morphology of galaxies. When dark matter
haloes merge in our model, the galaxies they contain are ranked in
mass. The most massive one is designated as the `central' or `primary'
galaxy in the new dark halo and the remaining galaxies become its
satellites.  The satellites lose any hot gas reservoir that they may
have had prior to the merger and any subsequent accretion of cooling
gas is funnelled into the central galaxy. The orbits of the satellite
galaxies decay through dynamical friction.  If the timescale for a
satellite to sink to the centre of the halo is shorter than the
lifetime of the halo, then the satellite is merged with the central
galaxy at the appropriate time (see e.g. \cite{cole00}).

The result of a galaxy merger is determined by two principal
quantities: (i) the ratio of the mass of
the accreted satellite to the mass of the primary, $f_{\rm merge} =
M_{\rm smaller}/M_{\rm larger}$, and (ii) the fraction of the mass of
the primary disc which is cold gas, $f_{\rm gas} =
M_{\rm cold,primary}/M_{\rm disc,primary}$, where $M_{\rm disc}=M_{\rm stars}+M_{\rm cold}$. If the mass ratio exceeds a threshold $f_{\rm ellip}$, the merger is termed
`violent' or `major'. In this case, all stars present are rearranged
into a spheroid, with a radius determined by arguments based on the
conservation of energy and the virial theorem (see
\cite{cole00}; Almeida, Baugh \& Lacey 2007).  In addition, any cold gas
in the merging galaxies is assumed to undergo a star formation burst
and the stars thus produced are added to the new spheroid. In this paper, 
we use the parameters set by Baugh et~al., who defined major mergers by 
the threshold $f_{\rm ellip}=0.3$. 

In cases where $f_{\rm merge} < f_{\rm ellip}$, the
merger is termed `minor'. In this case, the stars in the accreted 
satellite are added to the spheroid of the primary, leaving intact 
any stellar disc present in the primary. In minor mergers, the fate of the
gas in the merging galaxies depends upon the gas fraction in the primary
disc and on the value of $f_{\rm merge}$. If the primary disc is gas 
rich (if $f_{\rm gas} > f_{\rm gas,burst}$, where, following Baugh et~al., 
we take $f_{\rm gas,burst} = 0.75$), and if $f_{\rm merge} > f_{\rm burst}$ 
(where Baugh et~al. adopted $f_{\rm burst}=0.05$), then we assume that
the perturbation introduced by the merging satellite is sufficient to
drive all the cold gas, from both the primary and the satellite, into the
spheroid, where it takes part in the burst. Otherwise, if in a minor
merger the gas fraction in the primary disc is small, no burst
occurs. Alternatively, if the secondary galaxy is very much less massive 
than the primary (i.e. if $f_{\rm merge} < f_{\rm burst}$) then the
primary disc remains unchanged, the accreted
stars are added to the spheroid and there is no burst, irrespective of
$f_{\rm gas}$. The refinements of the Cole \etal
(\shortcite{cole00}) model relating to minor mergers were described by Baugh \etal
(\shortcite{baugh05}), who also set the 
values of $f_{\rm burst}$, $f_{\rm ellip}$ and $f_{\rm gas,burst}$. 

During a starburst, we assume that all of the cold gas available at
the start, $M_{\rm cold}$, is processed by the burst and suffers
one of three fates: (i) It is reheated by supernova feedback and
returned to the hot ISM. (ii) It is ejected from the dark matter halo
by a superwind. (iii) It forms long-lived stellar remnants. The mass of
gas which forms long-lived stellar remnants in the burst, $\Delta \rm
M _{\rm stars}$, depends on the feedback prescription
used, and is calculated as follows:

\begin{equation} 
\Delta M_{\rm stars} = \frac{(1-e^{-{efold}}) (1-R_{\rm
    burst})}{(1-R_{\rm
    burst})+\beta_{\rm burst}+f_{\rm sw, burst}}M_{\rm cold}
\end{equation} 

For completeness, we now define the parameters in this equation (for 
further details see Cole et~al. 2000, Granato et~al. 2000 and Baugh et~al. 2005): 

$\bullet$ $\beta_{\rm burst}  = (V_{\rm circ, bulge}/200 \rm km \,
s^{-1})^{-2},$ where $V_{\rm circ, bulge}$ is the effective circular velocity of
the bulge. This quantity gives the rate at which cold gas is reheated by 
supernova feedback in units of the star formation rate. 
This reheated gas is returned to the hot gas reservoir and is allowed to recool once 
a new halo forms (i.e. when the halo has doubled in mass). 

$\bullet$ $f_{\rm sw}$ gives the rate of ejection of cold gas by superwinds, in units 
of the star formation rate. This gas is ejected from the dark halo and is not allowed to 
recool. This parameter has the following dependence on the effective circular velocity of the bulge: 
\begin{equation}
f_{\rm sw} = f_{\rm sw0}~~~~~~~~~~~~~~~~~~~~~~~~~~~~~~~~~~~~~\rm{for}~~~~~ V_{\rm circ} < V_{\rm sw}
\end{equation}
\begin{equation}
f_{\rm sw} = f_{\rm sw0} \times (V_{\rm sw}/V_{\rm circ})^2 ~~~~~~~~~~~\rm{for}~~~~~ V_{\rm circ} > V_{\rm sw}
\end{equation}
The superwind feedback model was introduced by Benson \etal (2003b). Baugh \etal (2005) set $f_{\rm sw0}=2$ 
and $V_{\rm sw}=200 \rm km \, s^{-1}$.

$\bullet$ $R_{\rm burst}$ is the fraction of the mass turned into stars which we assume
is instantaneously recycled into high mass supernovae and returned to the cold phase of the ISM. 
For the flat IMF used in bursts, $R_{\rm burst}=0.41$. 

$\bullet$ $efold$ is the number of e-foldings over which star formation (assumed to have an 
exponentially declining rate) is allowed to take place in a burst. We
follow Baugh \etal (2005), taking $efold=3$.  

\begin{figure*}
\begin{picture}(0,240)
\put(-250,0){\epsfxsize=8.5 truecm \epsfbox{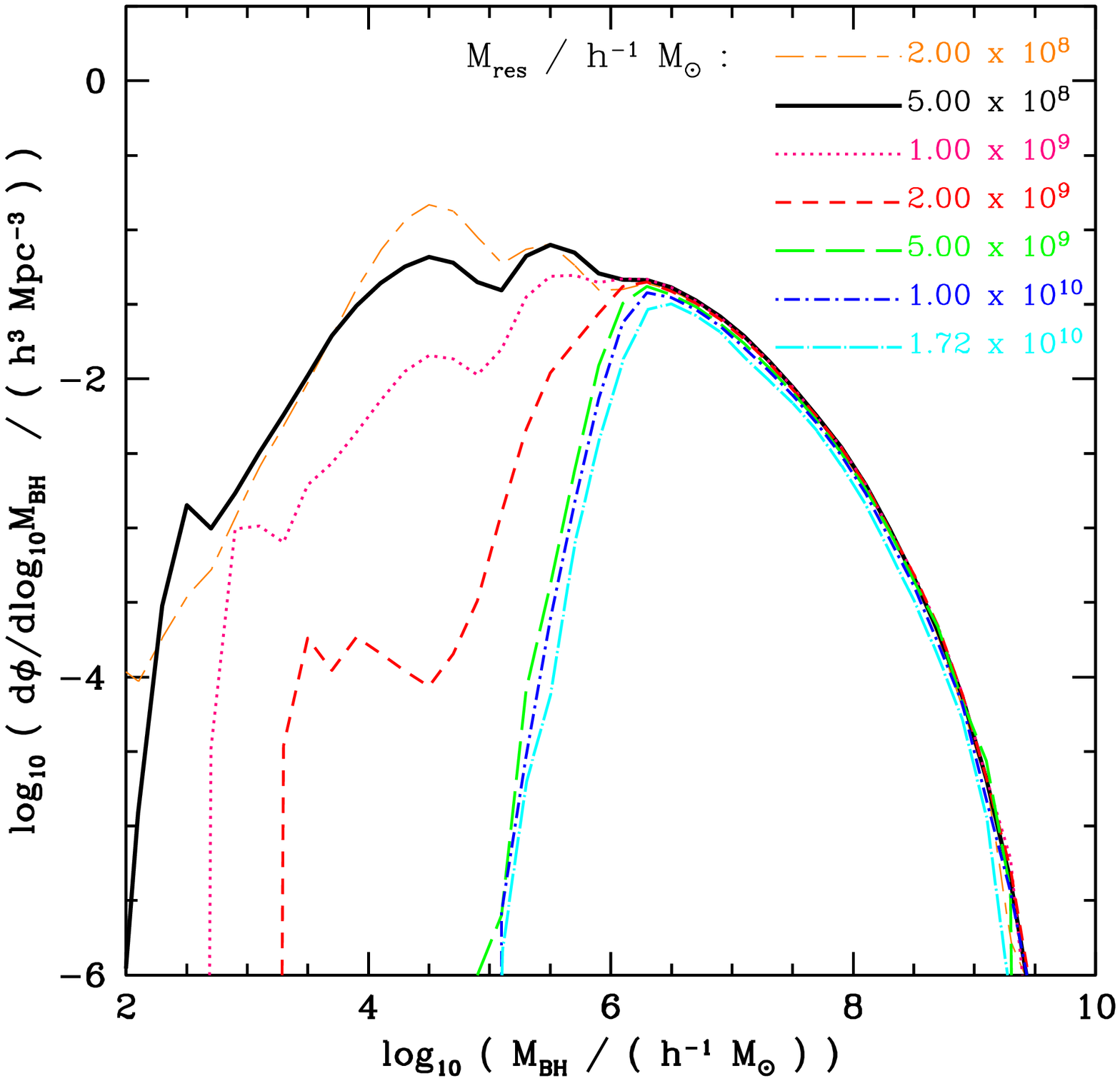}}
\put(0,0){\epsfxsize=8.5 truecm \epsfbox{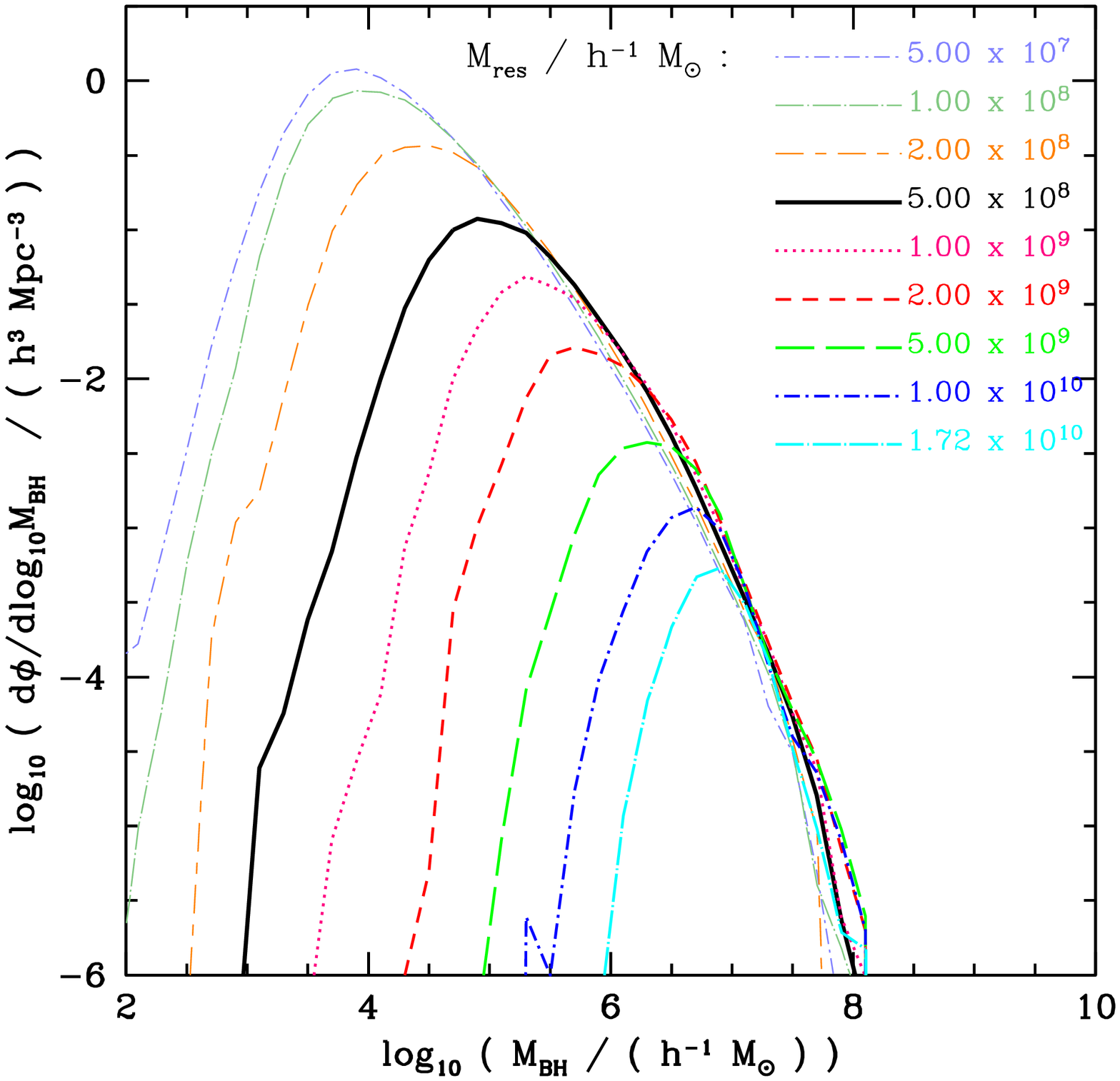}}
\end{picture}
\caption{The mass function of black holes, computed using different 
resolutions for the merger trees of dark matter haloes, as indicated 
by the legend in each panel. Our fiducial resolution is shown by the 
thick solid line. The left hand panel shows the results for $z = 0$ 
and the right hand panel shows $z = 6$.}
\label{fig:bhmf_res_z0}
\end{figure*}

\subsection{The growth of black holes in galaxy centres}
\label{sec:BHmodel}

The observed correlation between the inferred mass of galactic central black holes
and the properties of their host spheroids suggests a common origin for
these two classes of object (e.g. \cite{magorrian98,fm00,gebhardt00}). We adopt
a model of black hole growth similar to that
implemented for the first time in a fully-fledged semi-analytical galaxy
formation by Kauffmann \& Haehnelt (\shortcite{kh00}).

We assume that any contribution to the black hole mass from processes
other than galaxy mergers (e.g. the end products of population III
stars, primordial black holes or accretion onto black holes from
galactic discs or from the hot gas within a halo) is negligible 
compared with the change in black hole mass which occurs during 
merger-driven starbursts or through black hole mergers following 
a galaxy merger.  
Hence, the first galaxies to form in our model, when gas first cools 
into galactic disks, do not contain a significant black hole mass. 
The first important growth of black hole mass is assumed 
to occur during the first merger triggered starburst. Our reason for 
this choice is that the physics
of black hole seeding is very uncertain, and many mechanisms of black
hole seeding have been proposed, with widely
varying associated seed masses
(see \cite{vol06review} and references therein). The
largest mass of seed black holes suggested by models in the
current literature is $\sim
10^5-10^6 M_{\odot}$ (e.g. \cite{bl03,vr05,ln06}). These models
typically apply only to
metal-free and highly biased regions at high redshift and probably
therefore 
only to the seeding of the the most massive black holes (they are
motivated by the difficulty in producing $\sim 10^9
h^{-1} M_{\odot}$ black holes at $z \sim 6$ from smaller seeds, since
at high redshifts the age of the Universe is not long compared to
the Salpeter time). Other models
for seed black holes predict seeds which are less massive than $10^5 h^{-1} M_{\odot}$,
which is our black hole mass resolution limit. Therefore, it is
reasonable for us to neglect the role of seed black holes in our calculations.  In
practice, we assume that if a pre-existing seed is indeed required for
supermassive black hole formation, then the mass is small enough that
it only makes a negligible contribution to the mass of the final black
hole. Futhermore, theoretical considerations
(e.g. \cite{begelman78,begelman02,king02}) and observations
(e.g. \cite{ck04}) suggest that super-Eddington accretion of mass is
possible, and we assume that this occurs during the early stages of
black hole growth, so that the mass of any seed does not affect the
final mass by Eddington limiting of mass accretion.

The mass of black holes is assumed to grow during galaxy mergers via
two channels, accretion of gas during merger-driven starbursts and mergers with
other black holes. (Note that in the recent model by Bower \etal
\shortcite{bower06}, additional modes of black hole growth are
considered: accretion during starbursts triggered by disc
instabilities, and accretion of cooling gas from quasi-hydrostatically
cooling haloes.) As discussed in \S2.2, we allow starbursts, and thus
accretion, in both major and minor mergers. (In contrast, \cite{kh00} 
only allowed starbursts and black hole accretion during major galaxy 
mergers.) The two channels for black hole growth are as follows:
 
Firstly, in starbursts triggered by galaxy mergers, we assume that a
fraction, $F_{\rm BH}$, of the gas mass which is turned into stars is
accreted onto the black hole:

\begin{equation} 
\Delta M_{\rm BH} = F_{\rm BH} \times \Delta M_{\rm stars},  
\end{equation} 
where $\Delta M_{\rm BH}$ is the mass added to the black hole and 
$\Delta M_{\rm stars}$ is the mass of stars produced in the burst
\emph{after} taking into account feedback processes that
may expel gas from the galaxy and the recycling of mass from stars 
(Eq. 1; see \cite{cole00} and \cite{granato00}). 
Typically, we use $F_{\rm BH} <
0.03$ (this is explained in \S\ref{sec:bhdata}), and so for simplicity we ignore the
depletion of the cold gas reservoir by black hole growth when we calculate star
formation. We assume that the growth of black hole mass is not limited
to the Eddington accretion rate appropriate to our chosen radiative
efficiency.

Secondly, if the merging galaxies already host black holes, then we
assume that these black holes merge when the host galaxies
merge. In reality, black holes do not merge instantaneously, but gas-dynamical
processes are likely to speed-up black hole coalescence in gas-rich
mergers (\cite{an02}) and circumstantial observational evidence exists
to suggest that most binary black holes do merge efficiently, even in
gas-poor mergers (\cite{mm05}). Since we only consider binary galaxy
mergers with instantanous central black hole merging, all BH-BH
mergers in our model are binary, and we ignore slingshot ejection of black holes from the galactic
centre (\cite{sva74}). We also ignore the recoil of
the merger products of unequal mass black holes due to the anisotropic
emission of gravitational waves, which may lead to the resultant black
hole being ejected from the galaxy nucleus
(\cite{fitchett83,vol03,libeskind06}). Recent
calculations suggest that most recoil velocities are likely to be in
the range $10-100\,{\rm km\,s}^{-1}$ (\cite{fhh04}), and thus
unimportant except in very low mass galaxies.

Note that we neglect any loss of mass arising from the radiation of
gravitational waves during the merger of two black holes.  Such
radiation could result in the mass of the merger product being less
than the sum of the masses of the black holes from which it formed
(\cite{yt02}).  This effect is very uncertain, but is maximal for
equal mass black holes, and even then it is likely to be small --
approximately 3\% or less of the initial mass energy for equal-mass
non-spinning or Kerr black holes (Baker et al. 2002, 2004). Since most
BH--BH mergers in the Universe have unequal mass ratios, the
cumulative mass loss by gravitational radiation is unlikely to be more
than the figure of 20\% predicted using the most extreme models for
gravitational wave loss in individual BH--BH
mergers (\cite{mh04}). Therefore, we assume that the final black hole
mass is the sum of the mass accreted plus the mass of the two
progenitors.

\subsection{Resolution tests}
\label{sec:res}

The black hole mass down to which our predictions for the properties of black
holes can be trusted depends upon two factors: the accuracy of our
prescriptions for handling the physical ingredients of our galaxy
formation model and the mass resolution of the dark matter halo merger
trees. The semi-analytical galaxy formation model gives a reasonable
match to the field galaxy luminosity function, including its faint end
(\cite{benson03}). Further tests of the modelling of the
phenomena operating in low mass systems are deferred to future
work. This leaves the mass resolution of the halo merger trees as a
numerical parameter that directly influences the properties of
low-mass black holes.

In this paper, we use dark matter halo merger trees generated using
the Monte Carlo scheme described by Cole \etal
(\shortcite{cole00}). Merger trees extracted from N-body simulations
are, in some respects, more accurate
(e.g. \cite{kang05,nagashima05nbody}). However, a
major limitation of the trees extracted from simulations is their
finite mass resolution. Unpublished work by one of us (CGL) and work in preparation by Helly et al. show that the
merger trees in the Durham semi-analytic model agree well with the
merger trees in N-body simulations. Monte-Carlo generated trees can
have far superior mass resolution, because the whole of the computer
memory is devoted to one tree at a time, rather than to a large
ensemble of haloes within a cosmological volume. This also means we
are able to extend our merger trees to high redshifts (we use $z_{\rm
  start} = 20$). Also, Monte-Carlo trees typically have superior time
resolution to those taken from N-body simulations. On the other hand,
Monte Carlo trees tend to become less accurate as the time interval
over which the trees are grown is increased (\cite{somerville00}).

Putting this caveat aside, we have performed extensive tests of the
impact of the choice of resolution of the dark matter merger trees on
our predictions for the mass function of black holes. The results of
this convergence study are presented in Fig.~\ref{fig:bhmf_res_z0} for
$z = 0$ and $z = 6$. With improved mass resolution in the merger tree,
we are able to trace more of the gas which cools in low mass haloes
before reionization. This is the reason for the odd-looking `bumps' at
low BH masses in the $z=0$ panel. Our fiducial choice of halo mass resolution is $5
\times 10^8h^{-1} M_{\odot}$. This is an order of magnitude better
than the resolution used in our standard galaxy formation
calculations, and thirty times better than the resolution of the best
N-body merger trees currently available within a cosmological volume
(the Millennium simulation of \cite{springel05} which can resolve
haloes of mass $1.72 \times 10^{10} h^{-1} M_{\odot}$). With our
fiducial halo mass resolution, our predictions for the mass function
of black holes have converged for masses of $10^5 h^{-1} M_{\odot}$
and above. 

\section{Defining the model: comparison with observational data}
\label{sec:obsmatch}

\begin{figure*}
\begin{picture}(0,520)
\put(-250,280){\epsfxsize=8.5 truecm \epsfbox{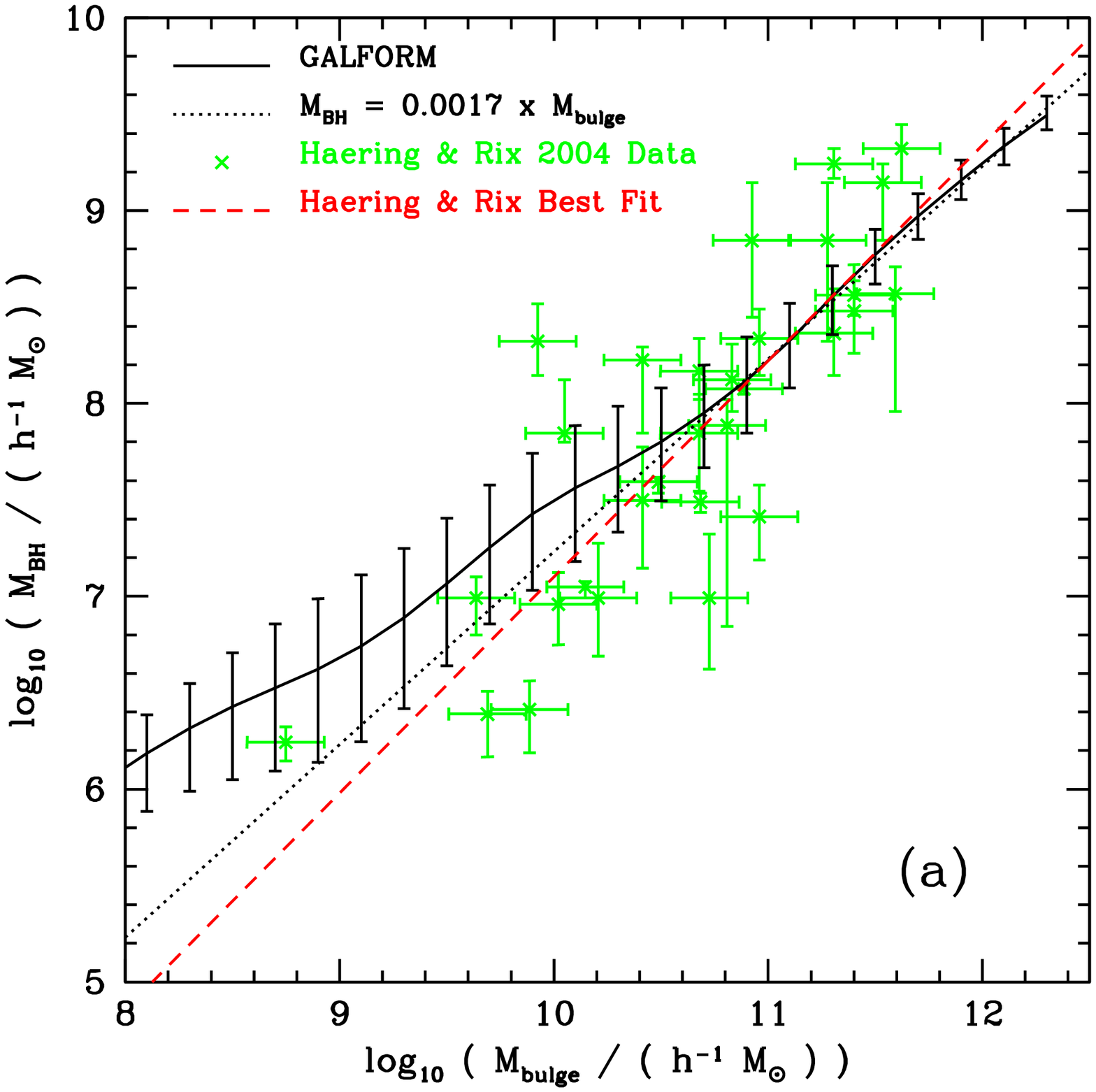}}
\put(0,280){\epsfxsize=8.5 truecm \epsfbox{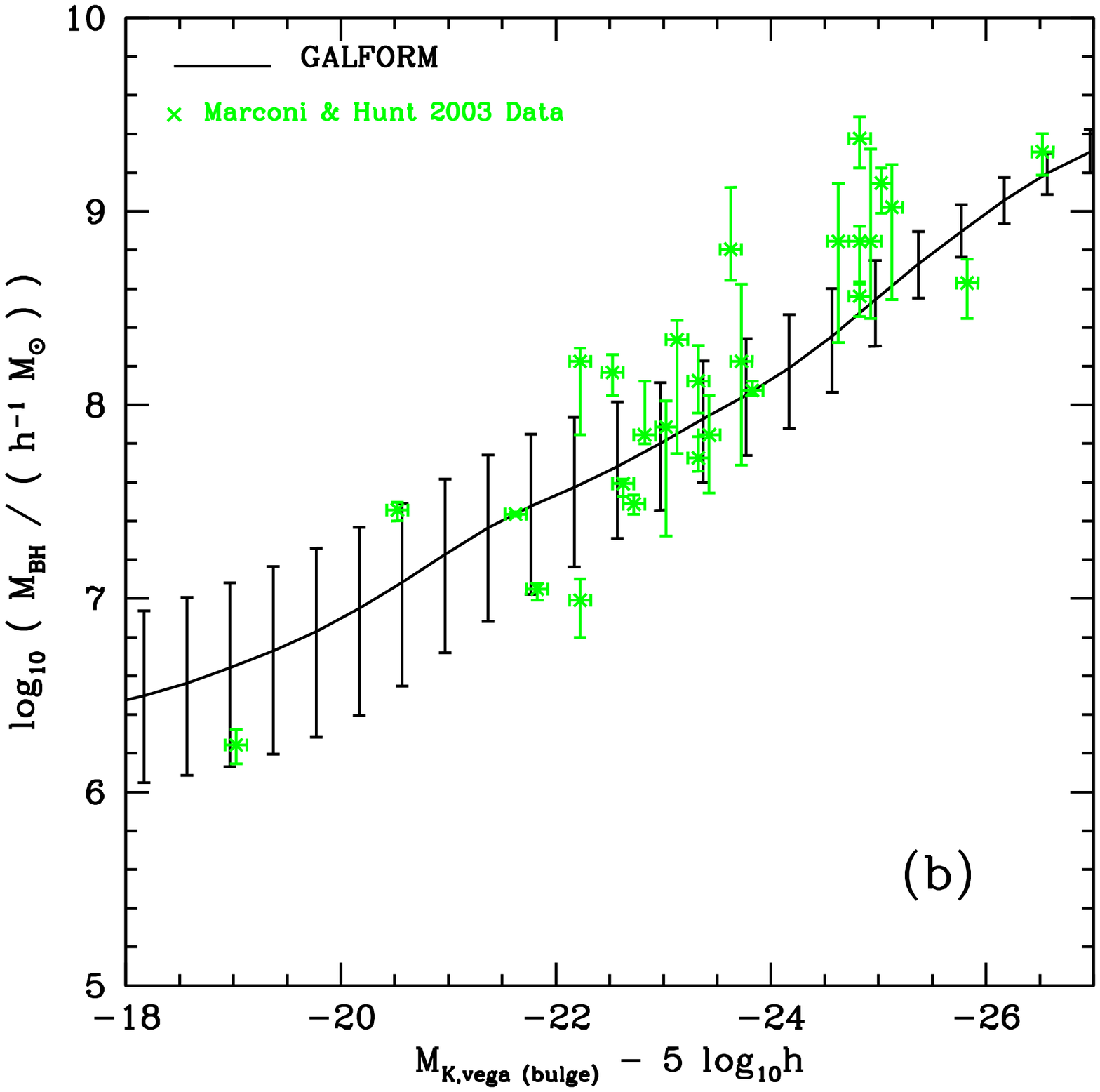}}
\put(-250,20){\epsfxsize=8.5 truecm \epsfbox{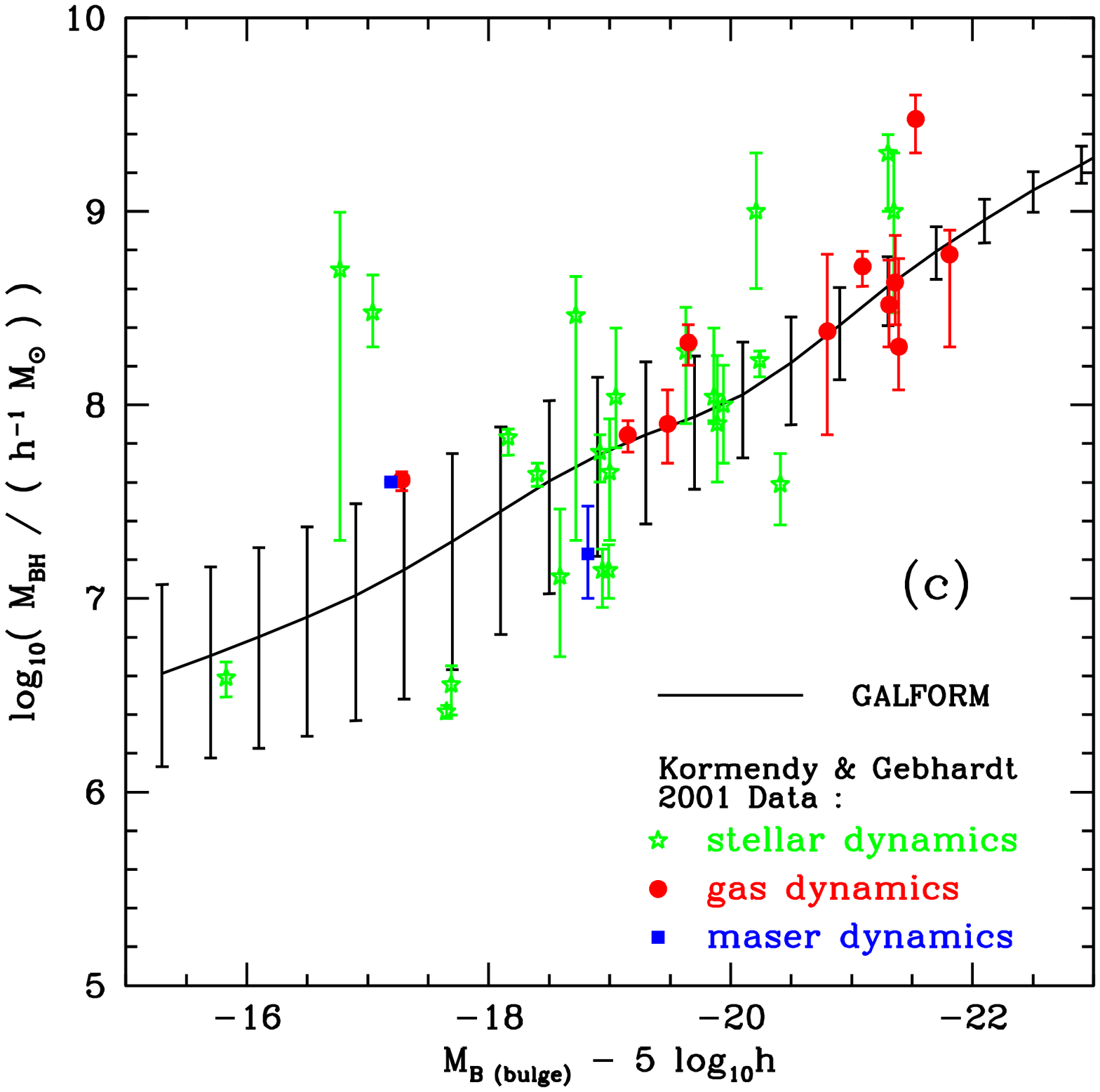}}
\put(0,20){\epsfxsize=8.5 truecm \epsfbox{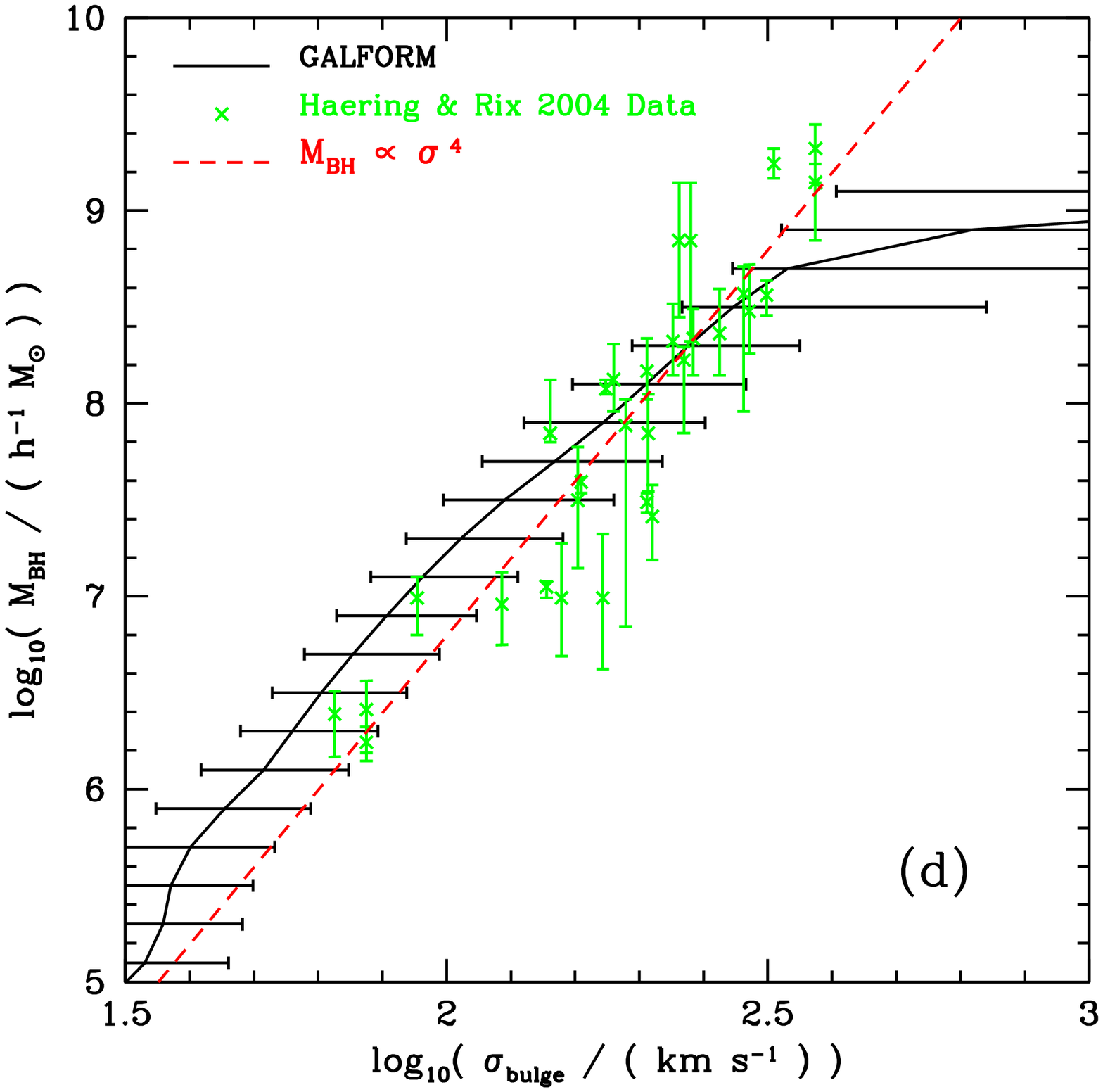}}
\end{picture}
\caption{The relation between black hole mass, $M_{\rm BH}$, and 
a selection of properties of the spheroid of the host galaxy. Each panel 
shows the correlation with a different bulge property: (a) the stellar 
mass of the bulge; (b) the bulge rest-frame K-band magnitude; (c) the 
rest-frame B-band bulge magnitude; (d) the velocity dispersion of the bulge.
The model predictions are shown by the line with errorbars: the line shows 
the median and the errorbars the 10--90 percentile spread of the 
distribution. The observational measurements are shown by symbols, with 
sources indicated in each panel. 
}
\label{fig:z0magor}
\end{figure*}

\begin{figure*}
\begin{picture}(320,320)
\put(-90,-10) {\epsfxsize=17.0 truecm \epsfbox{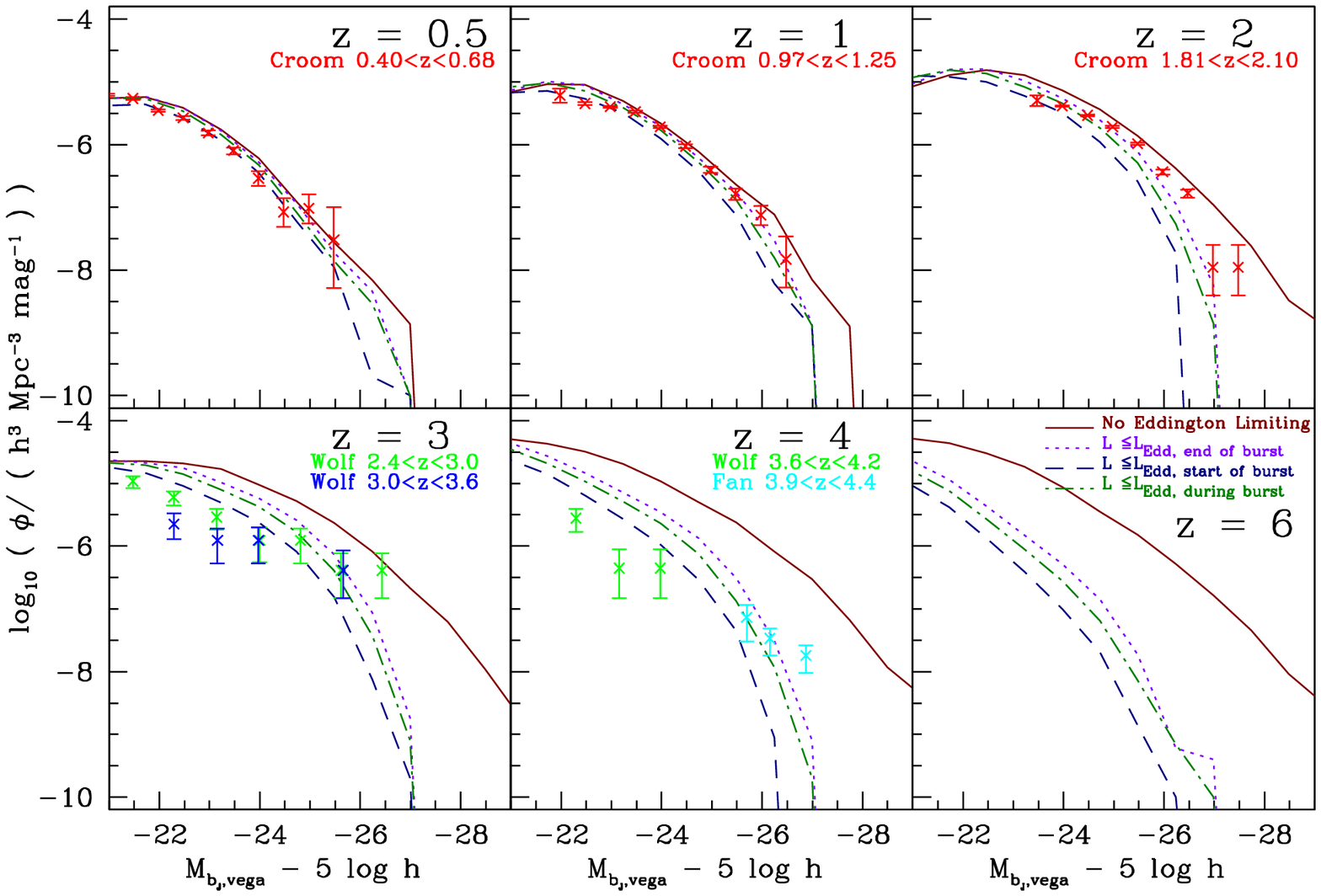}}
\end{picture}
\caption{The quasar luminosity function at selected redshifts, as 
indicated in each panel. The model predictions are shown by lines 
and the data by symbols, with the source indicated in each 
panel. The different line styles correspond to different 
assumptions for how the quasar luminosity depends upon the Eddington 
luminosity of the black hole, as indicated by the legend in the 
bottom right panel. The data are taken from the following papers:
Croom = Croom \etal 2004; Fan = Fan \etal 2001; Wolf = Wolf \etal 2003. }
\label{fig:qlf}
\end{figure*}

We first fix the value of the main parameter in our black hole model, $F_{\rm
BH}$, which determines the mass accreted onto the black hole during a starburst
(see Section~\ref{sec:BHmodel}). In \S3.1, we set $F_{\rm BH}$ by requiring that
the model should reproduce the local observed relationship between black hole mass
($M_{\rm BH}$) and the stellar mass of the bulge ($M_{\rm bulge}$) in which it
resides. We also show the model predictions for how black hole mass scales with
other properties of the bulge. Any viable model of black hole growth should also
be consistent with the observed quasar population. In \S3.2, we briefly describe
how a quasar luminosity can be assigned to accreting black holes, and present
some illustrative results for the quasar luminosity function at selected
redshifts.

\subsection{Setting the model parameter: predictions for the present day 
bulge--black hole relation}
\label{sec:bhdata}

The main parameter of our black hole model is the fraction, $F_{\rm BH}$, of the
mass of stars formed in a starburst which is accreted onto the central black
hole (after taking into account gas ejected from the galaxy by feedback
processes and the recycling of mass in supernova explosions and stellar
winds, as described in \S \ref{sec:GF}). We fit the value of $F_{\rm BH}$ by comparing the model predictions to
the observed correlation between the mass of galactic central black holes and the
stellar mass of the bulge component, {$M_{\rm BH}-M_{\rm bulge}$}, as inferred
by H{\"a}ring \& Rix (\shortcite{hr04}). H{\"a}ring \& Rix make a dynamical
estimate of the stellar mass of the bulge. They compile from the literature black hole mass
estimates made using a variety of techniques (stellar, gas or maser dynamics). A
review of these techniques and their uncertainties can be found in
Kormendy \& Gebhardt (2001).

We find that a value of $F_{\rm BH} = 0.022$ is required for the model to match
the zeropoint of the observed {$M_{\rm BH}-M_{\rm bulge}$} relationship
(Fig.~\ref{fig:z0magor}a). It is important to remember that the normalization of
this relationship is set by the choice of $F_{\rm BH}$. However, the slope and
scatter are genuine model predictions, and as Fig.~\ref{fig:z0magor} shows, these
predictions are in good agreement with the observations.

Na{\"i}vely one might argue that, since we have assumed that a fixed fraction of
the mass of stars formed in a burst is added to the mass of the black hole, it
is hardly surprising that a tight {$M_{\rm BH}-M_{\rm bulge}$} relationship
results. 

We find in the Baugh et~al. model that bursts actually play a fairly 
minor role in the formation of
bulge stars. The dominant channel responsible for building up the mass of
present day spheroids is the re-assembly of pre-existing stellar
fragments during
mergers, not the burst accompanying the most recent major merger
experienced by the galaxy (\cite{baugh96}). We find that only 15\% of
the stellar mass in bulges at redshift zero was formed in bursts. The other 85\% of the
stars in bulges was originally formed quiescently, in discs, and no
black hole accretion is associated with the formation of these stars. Thus, the slope and
scatter of the $M_{\rm
BH}-M_{\rm bulge}$ relation are non-trivial predictions of the model. Essentially, the $M_{\rm BH}-M_{\rm bulge}$ relation results from the evolution in
bulge star formation (and in particular the fraction of bulge stars
which were formed quiescently) in our galaxy formation model. This
topic is discussed extensively by Croton (\shortcite{croton0512}). The
scatter is due to the variation in the fraction of the stellar mass of a bulge
which was formed quiescently in discs, before being rearranged into the
spheroid. The slope originates from how this fraction varies with
stellar mass -- as bulge mass increases, the fraction of stars which
formed in starbursts decreases, so that black hole accretion is
associated with a lower fraction of the stars in the bulge.

Further support for both the galaxy formation model and our new model for the
growth of black holes comes from examining the other relationships between black hole mass
and observable properties of the galactic spheroid, as shown in Fig.~\ref{fig:z0magor}~(b)-(d).

In Fig.~\ref{fig:z0magor}\thinspace (b), we compare our model predictions for
black hole mass as a function of the K-band magnitude of the bulge with the
measurements by Marconi \& Hunt (\shortcite{mh03}). Again, the match is
very good. K-band magnitudes correlate well with stellar mass. In the model,
the K-band magnitude depends upon the star formation and merger history of the
galaxy, taking into account all of the progenitors of the galaxy, its dust
content and linear size. Observationally, this property is completely independent of the bulge
stellar mass estimates based on the velocity profile fitting method used by
H{\"a}ring \& Rix (\shortcite{hr04}).

It is notable that the scatter in both the $M_{\rm BH}-M_{\rm bulge}$ and
$M_{\rm BH}-M_K$ relations decreases significantly as the bulge magnitude gets
brighter. A number of factors may contribute to this result. As shown in
\S\ref{sec:results}, less massive black holes vary far more in their formation
histories than do larger black holes. Therefore, for bulges hosting less massive
black holes, there is more scatter in the time available for stars to form in
progenitor discs before starbursts and black hole accretion occur. Stars in
larger ellipticals and bulges tend to be formed earlier. Once stellar
populations exceed a certain age, scatter in their ages have only a small impact
on colour and luminosity.

In Fig.~\ref{fig:z0magor}\thinspace (c) we plot black hole mass against the
B-band magnitude of the bulge and compare the model with a compilation of data
by Kormendy \& Gebhardt (\shortcite{kg01}). The scatter in this relationship is
the greatest of all four variations on the $M_{\rm BH}-\rm bulge$ relations
shown in Fig.~\ref{fig:z0magor}. This is due to the sensitivity of the B-band
magnitude to the details of the recent star formation history of the bulge which
can vary considerably between galaxies with similar mass black holes.

Finally, in Fig.~\ref{fig:z0magor}\thinspace (d), we compare the model
prediction for the $M_{\rm BH}-\sigma_{\rm bulge}$ relation to data from
H{\"a}ring \& Rix (\shortcite{hr04}). We calculate the velocity dispersion
directly from the circular velocity of the bulge, assuming {$\sigma_{\rm bulge}
= 1.1 \times V_{\rm circ, bulge}/\sqrt{3}$} (see \cite{almeida06} for 
an explanation of the pre-factor). The full details of the
calculation of $V_{\rm circ,bulge}$ are given in Cole \etal
(\shortcite{cole00}). We obtain a reasonable match to the data, reproducing the tightness of the
relationship, except at the largest velocity dispersions. For less massive black holes, 
our model gives $M_{\rm BH} \propto \sigma^{4}_{\rm bulge}$, which compares well 
with the Tremaine \etal (\shortcite{tremaine02}) estimate of the slope of $4.02 \pm 0.32$. 
However, for black holes more massive than $M_{\rm BH} = 10^{7.5} h^{-1} M_{\odot}$, the
slope is shallower than observed, closer to $M_{\rm BH} \propto \sigma^{3}_{\rm
bulge}$. Direct accretion of cooling gas from a hot reservoir may help to bring
the slope of the $M_{\rm BH}-\sigma_{\rm bulge}$ relation closer to that observed
(\cite{bower06}).

The slope of the $M_{\rm BH}-\sigma_{\rm bulge}$ relation should perhaps 
be regarded as one of the less robust predictions of the model, because 
of the complexity of calculating $\sigma_{\rm bulge}$. This quantity depends 
upon the accuracy of the calculation of the radius of the spheroid. 
Cole et~al. (2000) introduced a prescription for computing the size of 
merger remnants, by applying the virial theorem and the conservation of 
energy to the progenitor galaxies and the remnant. The resulting size of the 
spheroid is adjusted to take into account the self-gravity of the disc and 
bulge and the reaction of the dark matter to the presence of condensed 
baryons. This step is carried out using an adiabatic contraction model. 
The assumptions behind this approach are likely to become less valid as 
the mass of the spheroid increases in relation to the mass of the halo. 
Almeida et~al. (2007) tested this prescription against the properties of 
spheroids in the SDSS. Whilst the 
agreement between the observed and predicted Faber-Jackson relation  
(velocity dispersion - luminosity) is encouraging, the predicted 
slope is somewhat steeper than is observed and the brightest galaxies 
in the model will perhaps have too large a velocity dispersion. 

The level of agreement with observations that we find between different bulge
properties and black hole mass is encouraging and suggests that, overall, our
model of galaxy and black hole formation is on a firm footing.

\subsection{Predictions for the luminosity function of quasars}
\label{sec:qlf}
In order to establish further the credentials of our model, we present
some illustrative predictions for the evolution of the quasar luminosity
function. Further assumptions and model parameters are required to assign
a luminosity to the quasar phase which occurs when the black hole accretes
gas during a galaxy merger. In this section, we give a brief outline of
our model for calculating the luminosity of the quasar and present some
results for the quasar luminosity function at different redshifts. These
predictions are included here for completeness and to allow comparison
with previous work (e.g. \cite{kh00}).  We will explore the
form and evolution of the quasar luminosity function in more detail in a
future paper.

There are two basic parameters in our model for quasar luminosity:  the
lifetime of the quasar, $t_{\rm Q}$, and the fraction of the accreted
mass-energy that is turned into the bolometric luminosity of the quasar,
$\epsilon_{\rm Q}$. We assume that the gas available for accretion onto
the black hole in a galaxy merger is accreted at a constant rate,
$\dot M(t)$, over the quasar lifetime:  
\begin{equation} 
\dot M(t) = \Delta M_{\rm BH} / t_{\rm Q} \qquad 
\rm{for}\qquad t < t_{\rm Q}. 
\end{equation} 
(Recall that $\Delta M_{\rm BH}$ is defined by Eq. 1.) 
We note in passing that if we had instead assumed an exponentially 
decaying mass accretion rate, with a timescale given by 
$t_{\rm Q}=0.5 t_{\rm bulge}$, giving a mass accretion rate 
of (i.e. $\dot M(t) = \Delta M \rm exp (-t/ (0.5 t_{\rm bulge} ))~$),  
this would lead to very similar results to those we obtain 
for a constant mass accretion rate.  

The quasar lifetime, {$t_{\rm Q}$}, is assumed to be directly
proportional to the dynamical time of the bulge, $t_{\rm bulge}$.  In
the simplest case, without imposing any further conditions on the
luminosity of the quasar, this assumption results in a top-hat light
curve: \begin{equation} L_{\rm Q}(t) = \epsilon_{\rm Q} \dot M(t)\,
c^2\qquad \rm{for}\qquad t < t_{\rm Q}. \end{equation}

When computing the luminosity of quasars, the Eddington
limit may play an important role. A quasar is said to be radiating at its
Eddington limit when the pressure of the radiation emitted following
accretion onto the black hole balances the gravitational force exerted by
the black hole on new material that is being accreted. The Eddington limit
depends upon the mass of the black hole. Physical mechanisms have been
proposed which permit mass to be accreted at rates which exceed the
Eddington limit (see, for example, \cite{begelman78}). Here, we show the
impact of the Eddington limit on the luminosity of quasars. We
consider four different cases:

Case (1) No Eddington limit is applied to the bolometric luminosity of
      the quasar.

Case (2) The bolometric luminosity is limited by the Eddington luminosity
  corresponding to the black hole mass at the \emph{end} of the
  accretion episode:\begin{equation}L_{\rm Q}(t) = \rm max(\,\epsilon_{\rm Q} \dot
  M(t)\,c^2 ,\,L_{\rm Edd}(M_{\rm final})\:)\end{equation} 

Case (3) The bolometric luminosity is limited by the Eddington luminosity
  corresponding to the black hole mass at the \emph{start} of the
  accretion episode:
\begin{equation}L_{\rm Q}(t) = \rm max(\,\epsilon_{\rm Q} \dot
M(t)\,c^2 ,\,L_{\rm Edd}(M_{\rm 
start})\:)\end{equation}

Case (4) The bolometric luminosity is limited by the Eddington luminosity
  corresponding to the black hole mass calculated \emph{during} the
accretion episode: \begin{equation}L_{\rm Q}(t) =
\rm max(\,\epsilon_{\rm Q} \dot M(t)\,c^2 ,\,L_{\rm
Edd}(M(t)\,)\:)\end{equation}

Case (4) is the most realistic estimate of the luminosity after applying
Eddington limiting. However, there is some uncertainty in the
evolution of the Eddington limit \emph{during} the burst, as 
we do not know in detail how the mass of the black hole changes 
from its initial value to the final value. 
Case (2) corresponds to the maximum possible Eddington limit during 
the accretion episode, being set by the final black hole mass. 
Case (3) is the minimum possible Eddington limit during the accretion
episode, corresponding to the initial black hole mass. 
We find little variation in the quasar luminosity function between 
these three cases, suggesting that the precise growth of the black 
hole over the accretion episode is unimportant. 

We assume that all visible quasars have identical, flat spectra over
the range of wavelengths of interest, and that a fraction, {$f_{\rm
b_J}$}, of the bolometric luminosity is emitted in the B and {$\rm
b_J$}-bands.  We adopt $f_{\rm b_J}=1/15$, which agrees well with 
Elvis \etal (1994). Taking into account the {$\rm b_J$}-band filter 
profile, we can calculate a magnitude for a quasar from its bolometric 
luminosity: 
\begin{equation} M_{\rm b_J,vega} = 13.2 -2.5 \times \log_{10}(f_{\rm b_J}
  \times L_{\rm bol} / {10^{40}}{\rm erg\:s^{-1}) }.
\end{equation}

Finally, we assume that only a fraction $f_{\rm visible}$ of quasars are
detected in optical surveys; the remainder are obscured in the optical. 
We first set $f_{\rm visible} = 0.25$, i.e. only a quarter of quasars are 
visible in optical surveys. This is roughly in line with the results of 
X-ray surveys, which typically find that 20-30\% of quasars are unobscured 
in soft X-rays (and we presume that this is the same fraction visible
in the optical visible in optical) 
(\cite{ueda03,lafranca05,shinozaki06,gilli07}), although this 
fraction is still somewhat uncertain, and furthermore is likely to 
vary with intrinsic quasar luminosity. 
We then set the parameters $t_Q$ and $\epsilon_Q$ in order to produce a
reasonable match to the Croom \etal (\shortcite{croom04}) measurement of
the $\rm b_J$-band luminosity function over the redshift interval $0.5
\leq z \leq 2$, as shown in Fig.~\ref{fig:qlf}.  At higher redshifts, we
show a comparison between our predicted luminosity functions and the
1450\thinspace\AA~rest frame quasar luminosity functions from the SDSS survey
(\cite{fan01}) and the combo-17 survey (\cite{wolf03}), using the
corrections given in the respective papers to convert to the B-band,
and applying a further minor correction to the $b_{\rm J}$ band. 

To achieve the best fit, we require that the quasar lifetime, $t_{\rm Q}$, be 
related to the bulge dynamical time, $t_{\rm bulge}$, by {$t_Q = 1.5\, 
t_{\rm bulge}$} and that the fraction of accreted mass-energy produced 
as bolometric luminosity be {$\epsilon_Q = 0.06$}.
A typical bulge has a dynamical time of $t_{\rm bulge}$ of
$2 \times 10^7 \rm yr$ at $z=1$, $8 \times 10^6 \rm yr$ at $z=3$ and $2.5 
\times 10^6 \rm yr$ at $z=6$, although within each redshift bin, the
distribution in $t_{\rm bulge}$ is very broad. 
As noted by Kauffmann \& Haehnelt
(\shortcite{kh00}), this redshift evolution in $t_{\rm bulge}$ helps
to reproduce the evolution in the quasar luminosity function. Our
timescales agree fairly well with the Martini \& Weinberg
(\shortcite{martiniweinberg01}) estimate of $t_{\rm Q}=4 \times
10^7 \rm yr$ at $z=2$, which is a typical value. We also note that
our adopted radiative efficiency of 0.06 is consistent with standard
disc accretion, which is likely to be required for optically bright
emission.


Our simple model does a reasonable job of reproducing the observed
quasar luminosity function at $z \le 2$, but over-predicts the
luminosity function at higher redshifts.  Our basic prediction for the
quasar luminosity function (shown by the solid lines in
Fig~\ref{fig:qlf}) shows strong evolution with redshift which cannot
be described as pure luminosity evolution. If the Eddington limit is
taken into account, then the form of the model predictions changes,
particularly at bright luminosities, where the abundance of objects is
strongly suppressed, with the result that the predictions match the
data better. The suppression affects more objects at higher
redshifts -- the gas supply then is greater for any given mass of
black hole, and the dynamical timescales are shorter, leading to
higher rates of supply for any given mass of available gas. The
predicted luminosity functions are relatively insensitive to the
precise details of how the Eddington limit is allowed to influence the
quasar luminosity.

\section{The growth of black holes through accretion and mergers}
\label{sec:results}

We present the bulk of our results in this section. The section is quite 
long, so we list the contents here to help the reader navigate through 
the various topics. Firstly, in \S\ref{sec:tree}, 
we give illustrative examples of how black holes acquire mass, 
tracing the mass assembly history of two black holes. 
We then explore the demographics of the black hole population: in 
\S\ref{sec:bhmf}, we present the predictions for the evolution of the 
mass function of black holes and in \S\ref{sec:BHMFmhalo}, we investigate 
how the black holes are distributed between dark haloes of different mass. 
The next few sections deal with how black holes build up their mass. In 
\S\ref{sec:BHprog}, we show the distribution of progenitor masses of 
black holes, and, in \S\ref{sec:growth}, we address the issue of
whether the accretion of gas or mergers is the main mechanism for
accumulating black hole mass. We present results for the formation
redshift of black holes in \S\ref{sec:zform} and for their merger
rates in \S\ref{sec:merrate}. Finally, in \S\ref{sec:univ}, we
compare the amount of baryons locked up in black holes with other
phases, such as cold gas and stars.

\subsection{Illustrations of black hole growth}
\label{sec:tree}

Before concentrating on statistical descriptions, it is instructive to
show some illustrations of how individual black holes grow in our
simulations. These examples serve to provide a qualitative picture of the
model, and to make clear certain definitions and results on black hole
formation histories that will be of use later on. Note that, although
space limitations restrict us to only two examples, there is, in fact,
a rich diversity in black hole formation histories in the model.

\begin{figure*}
\begin{picture}(0,600)
\put(-265,20){\epsfxsize=20.0 truecm \epsfbox{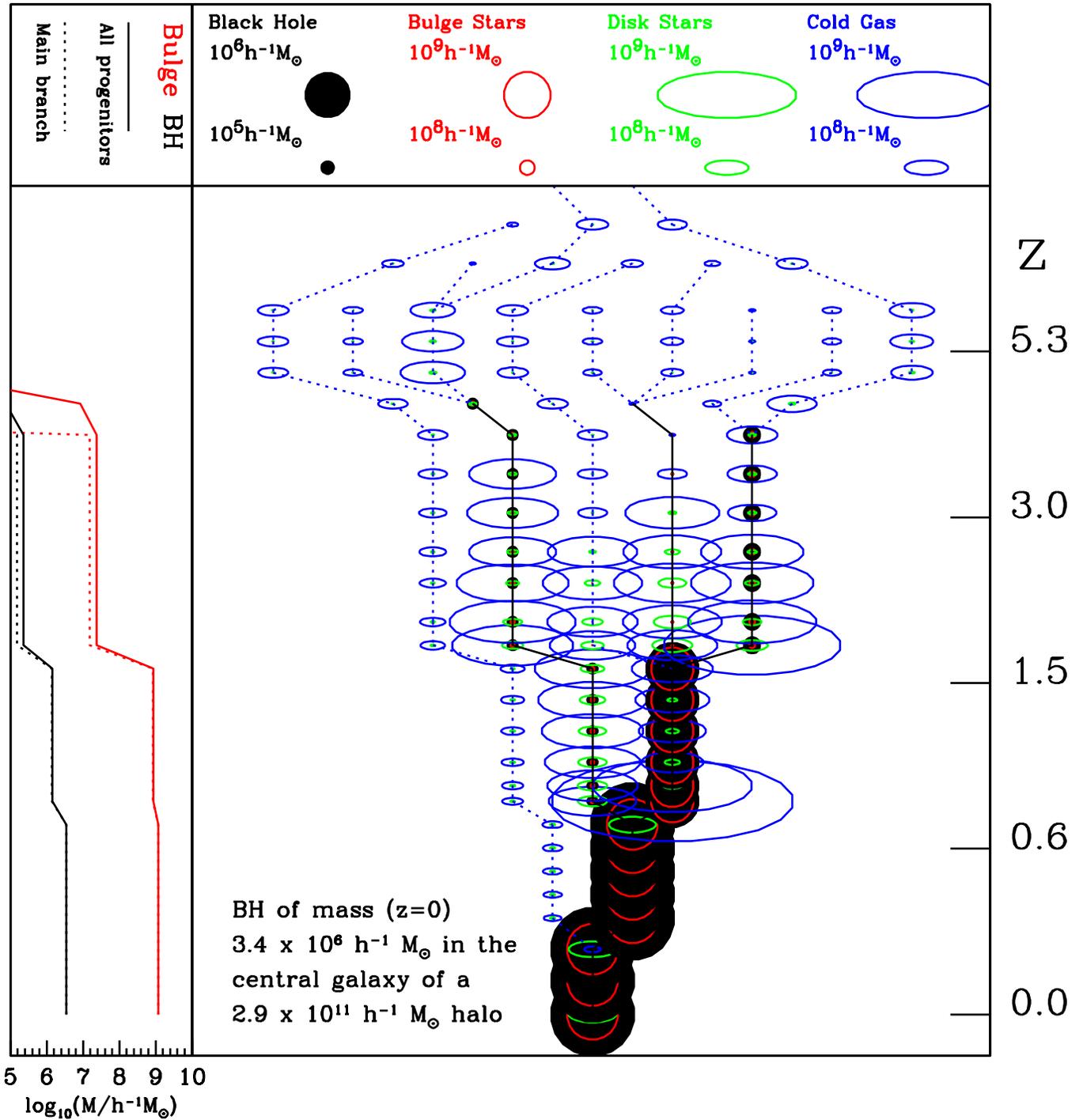}}
\end{picture}
\caption{An example of a mass assembly tree of a black hole and its host
galaxy. Progenitor galaxies without black holes are connected by dotted
lines. The trees show the relative amounts of cold gas, disc stars,
bulge stars and black hole mass, as indicated by the key. The area of
the symbols is proportional to mass. The left-hand side panel shows
the assembly of black hole mass and bulge stars, adding all
progenitors (solid line) and tracing back the main branch, which is
usually the most massive progenitor (dotted line).  The left--right
positioning in the plot is purely schematic and has no relevance to
the spatial positions of galaxies. The final galaxy is the central
galaxy of a halo of mass $2.9 \times 10^{11} h^{-1} M_{\odot}$.}
\label{fig:trees}
\end{figure*}

\begin{figure*}
\begin{picture}(0,600)
\put(-265,20){\epsfxsize=20.0 truecm \epsfbox{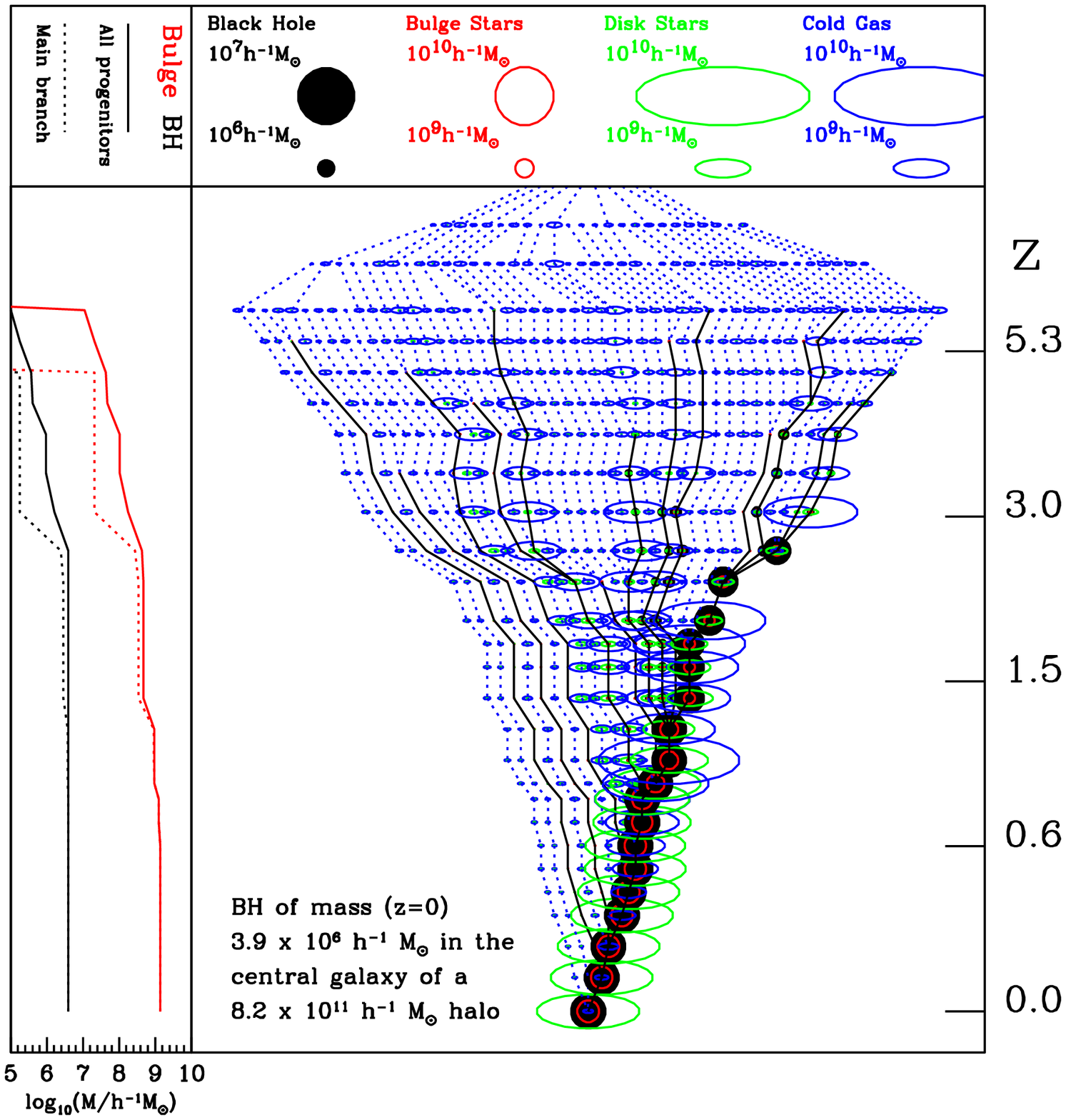}}
\end{picture}
\caption{A second example of a mass assembly tree of a BH and its
  host. The final galaxy is the central
galaxy of a halo of mass $8.2 \times 10^{11} h^{-1} M_{\odot}$.}
\label{fig:treesa}
\end{figure*}

The mass assembly history of two black holes is shown in
Fig.~\ref{fig:trees} and Fig.~\ref{fig:treesa}. Fig.~\ref{fig:trees} shows the central
galaxy in a halo of mass $2.9 \times 10^{11} h^{-1} M_{\odot}$ and
Fig~\ref{fig:treesa} shows the central galaxy in a halo of mass
$8.2 \times 10^{11} h^{-1} M_{\odot}$. The main part of each panel
follows the mass assembly tree. Various components are plotted, as
indicated by the key at the top of each plot: black hole mass, bulge
stars, disc stars and cold gas. The area of the symbols is
proportional to the mass in a given component, with reference
areas/masses provided at the top of each plot. Galaxies containing
black holes are linked by solid lines, while galaxies not containing
black holes are linked to their descendents with dotted lines. The
redshifts plotted are the output redshifts of the simulation. The
left--right positioning in the plot is schematic, and has no relevance
to spatial positions of galaxies within the dark matter halo; the
`main branch' (i.e. the most massive progenitor at each merger) is
always the right-most branch of the merger tree.

In some cases, the black hole in the `main branch', 
which we denote the `main progenitor', may not actually be 
{\it the} most massive of {\it all} the progenitor black holes 
at a given epoch, particularly at higher redshifts. We have 
chosen to avoid jumping from one branch of the black hole merger tree 
to another when following the `main progenitor'  backwards in time. 
Instead, we start from the present day black hole, find its most massive 
progenitor, and then build up a continuous branch by tracking 
the most massive progenitor at each of $\sim 25-30$ output redshifts. 

In the side panel of each figure, we plot the cumulative masses of the black 
hole and bulge stars as a function of time. The 
solid line shows the total mass in these components, adding together 
all of the progenitors. The dotted line shows the mass in the branch 
tracing back the most massive progenitor of the present day black hole (the
`main branch'). 

In general, we find a very wide variety of black hole formation
histories, and we have chosen the ones we plotted to be
illustrative. The formation trees of the most massive black holes tend to
be too large and complicated to plot effectively. Meanwhile, there is
a high abundance of black hole trees with just one burst in their
history, which were not very interesting to plot. All merger trees
are included, however, in the quantitative results we present later.

Inspection of the mass assembly trees, particularly the one for the 
more massive galaxy, reveals that there can be many branches to the 
black hole merger trees at high redshifts. However, most of the black 
hole mass is contained in one or two main branches, as shown by 
the closeness of the solid and dotted lines in the side panels. 
In the Baugh \etal (\shortcite{baugh05}) model, the quiescent star 
formation timescale is independent of the dynamical time. This results 
in discs which are gas rich at early epochs (blue circles), with significant 
quantities of stars only forming at relatively recent epochs (green circles).  
At later times, it is also apparent that the ratio of the stellar mass 
of the bulge to the mass of the black hole is increasing. 
We will present predictions for the evolution of the $M_{\rm BH}-\rm
bulge$ relations in a later section. 


\subsection{Black hole mass function}
\label{sec:bhmf}

\begin{figure}
\epsfxsize=8.5 truecm \epsfbox{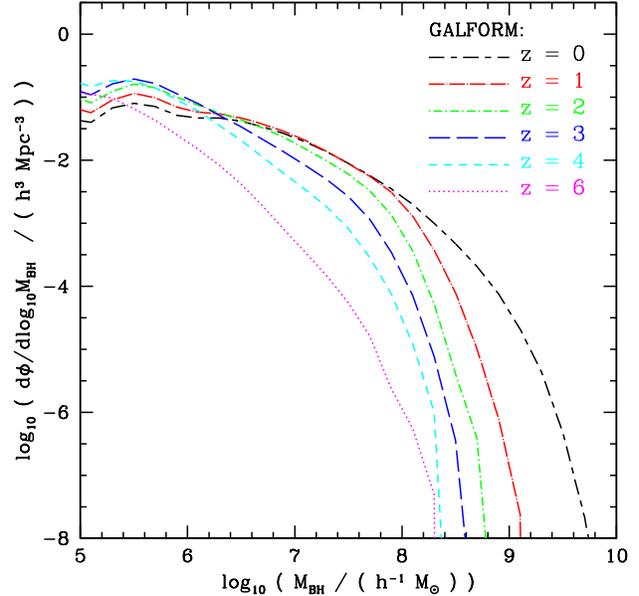}
\caption{The evolution of the black hole mass function with redshift, 
as indicated by the key.}
\label{fig:bhmf}
\end{figure}

The black hole mass function at various redshifts is shown in Fig.
\ref{fig:bhmf}. The high mass end advances  
to higher masses at lower redshifts. This is unsurprising in a
hierarchical galaxy formation model, and reflects the corresponding
evolution of the dark matter halo mass function.  The predicted
evolution in the black hole mass function is quite strong. This is in
contrast with observational claims that the abundance of large black holes does not
vary with redshift (e.g. \cite{md04}). Such studies, however,
typically include only optically selected quasars, and so can only
probe accreting black holes. We examine the relationship of accreting
black holes to the general black hole population in a later section on
downsizing (\S\ref{sec:downsizing}) and consider the implications of
the rare, massive black holes at high redshift inferred from the
observations of Fan \etal (\shortcite{fan01}) in the discussion
(\S\ref{sec:discuss}).

At redshift zero, the break in the black hole mass function occurs
around $10^8 h^{-1}M_{\odot}$. This corresponds to the scale at which
there is a transition between accretion-dominated growth and
merger-dominated growth (as we demonstrate specifically in
\S\ref{sec:growth}). In larger galaxies hosting more massive black
holes, the cold gas has been substantially depleted, so the black hole
mass can only increase significantly through mergers. Gas depletion and suppression
of further cooling by feedback processes is also the likely mechanism
by which a break in the luminosity function of galaxies is produced
(\cite{benson03,bower06,croton06}).

From the observed $M_{\rm BH}-M_{\rm K,bulge}$ relation, $M_{\rm BH} =
10^8 h^{-1}M_{\odot}$ corresponds to $M_{\rm K,vega (bulge)} - 5
\thinspace{\rm log}_{10}h \sim -23.5$. This is very close to the 
break in the K-band luminosity function, $M^{\ast}_{\thinspace K} -
5\thinspace{\rm log}_{10}h = -23.44 \pm 0.03$
(\cite{cole01}). Similarly, from the observed $M_{\rm BH}-M_{\rm
bulge}$ relation, $M_{\rm BH} = 10^8 h^{-1}M_{\odot}$ corresponds to
$M_{\rm bulge} \sim 3 \times 10^{10} M_{\odot}$. This is the stellar
mass at which Kauffmann \etal (\shortcite{kauffmann03}) find a
transition in galaxy properties. Although black hole mass is related
to bulge properties only, the identification of the knee in the black
hole mass function with a transition in the global properties of
galaxies is reasonable since galaxies brighter and more massive than
the transition mass tend to be bulge-dominated. The conclusion is that
galaxies (particularly galactic bulges) and black holes grow together
(as demonstrated graphically in the side panels of
Fig. \ref{fig:trees} and Fig. \ref{fig:treesa}).

As time advances, the black hole mass function becomes progressively
flatter at the low-mass end. To a large extent, this is a generic
feature of a hierarchical mass assembly model in which small objects
merge into larger objects (at least when this effect is not exceeded by the production of new low mass objects). A further contribution to the change in
slope comes from less massive black holes accreting larger amounts of
gas as a fraction of their mass than larger ones (i.e. downsizing,
\S\ref{sec:downsizing}). The combination of these effects is greater
than the effect of the formation of new, lower mass black holes; most
black holes are seeded at high redshift, as discussed in
\S\ref{sec:zform}.

\subsection{Black hole demography: the conditional mass function}
\label{sec:BHMFmhalo}

\begin{figure}
\epsfxsize=8.5 truecm \epsfbox{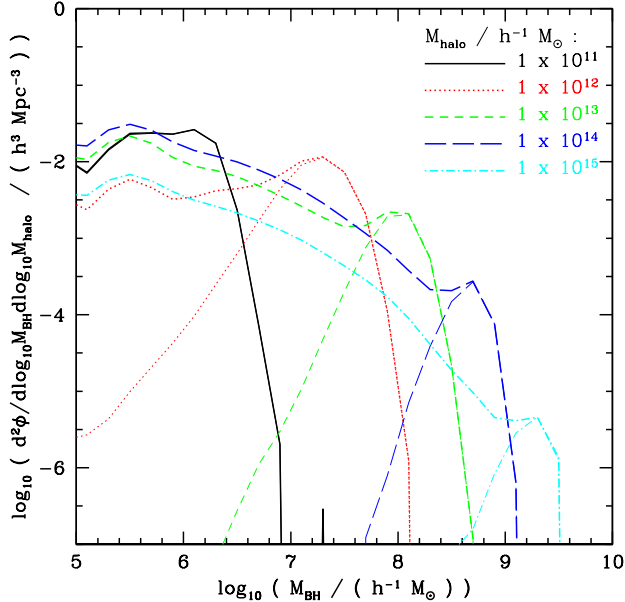}
\caption{The contribution to the black hole mass function from black 
  holes in dark matter haloes of various masses, as indicated by the key. 
  We show the case for black holes contained within any galaxy (thick 
  lines) and in central galaxies only (thin lines). 
  The black hole mass function for 
  each halo mass is normalized to be the mass function per dex in dark
  matter halo mass.}
\label{fig:halobhmf}
\end{figure}

Black holes of a given mass form in haloes with a broad range of
masses. The contribution to the black hole mass function from
different ranges of dark matter halo mass are shown in
Fig. \ref{fig:halobhmf}.  At the high mass end of each of these
conditional mass functions, there is a peak and a cut-off. The peak
corresponds to the mass of the black hole in the central galaxy, which increases as the mass of the central galaxy bulge which is strongly correlated with the mass of its host halo. This phenomenon is not
restricted to black holes. Eke \etal (\shortcite{eke04}) find
tentative evidence for a similar bump in the galaxy luminosity
function of groups and clusters which they attribute to central
galaxies, although this remains controversial (\cite{yang05}).

In galaxy formation models, a bump is sometimes present in 
luminosity functions where only galaxies in a limited range of halo
masses are selected -- this is because of the contribution of central galaxies
(\cite{benson03photoi3}). This reflects the different physical
processes relevant to central and satellite galaxies:
in the model, satellite-satellite mergers are not allowed, while all cooling gas is
funnelled into the central galaxies. These simple assumptions, common
in semi-analytic models, have been validated in gasdynamic simulations
(\cite{zz05}). Models with intense star formation in bursts, such as
the Baugh \etal (\shortcite{baugh05}) model used here, smear out the
bumps somewhat (\cite{eke04}), since the bursts introduce additional
scatter in the properties of galaxies that form in haloes of a given
mass.

\subsection{Mass function of progenitor black holes}
\label{sec:BHprog}

\begin{figure*}
\begin{picture}(0,270)
\put(-250,30){\epsfxsize=8.5 truecm \epsfbox{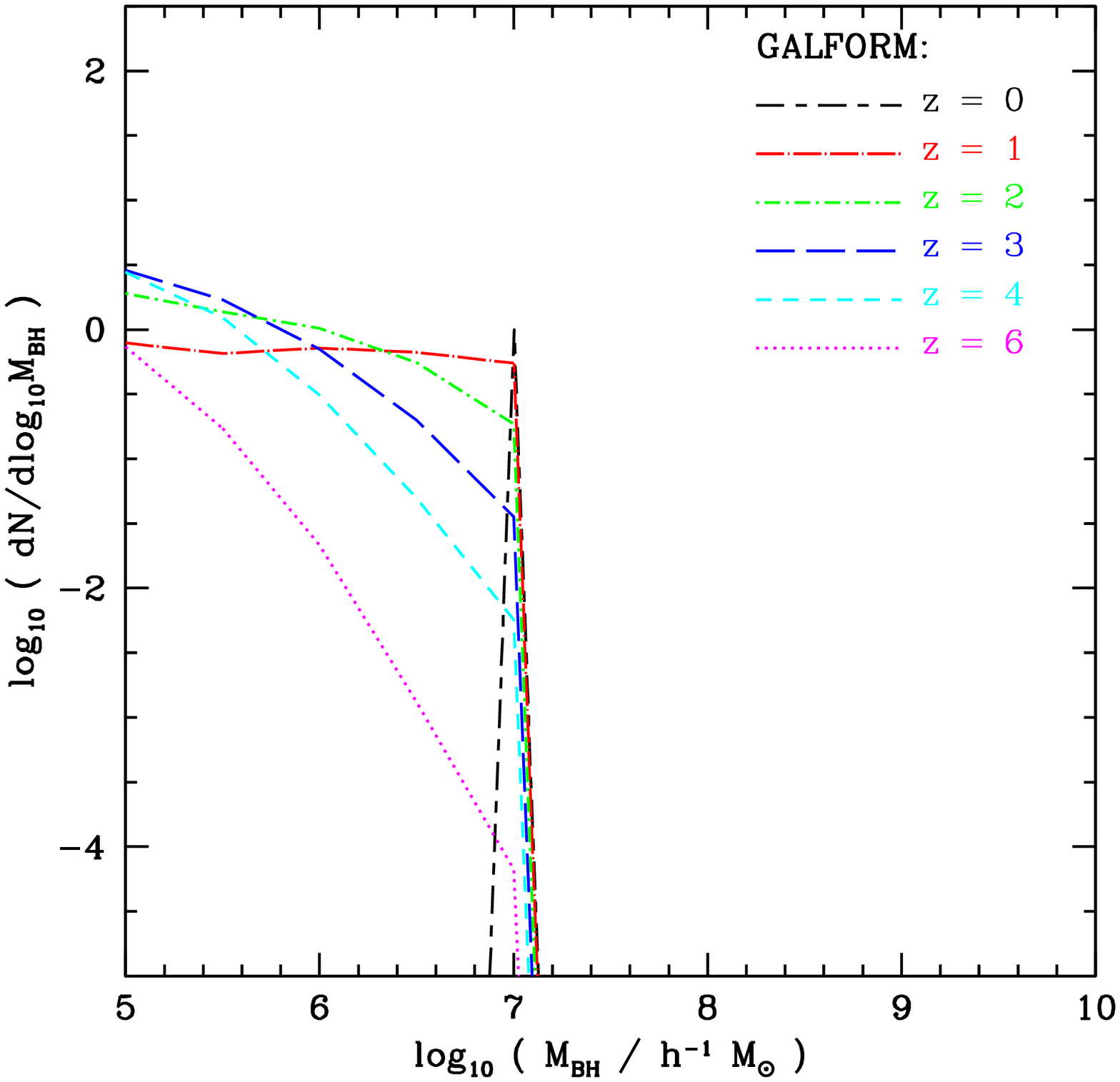}}
\put(10,30){\epsfxsize=8.5 truecm \epsfbox{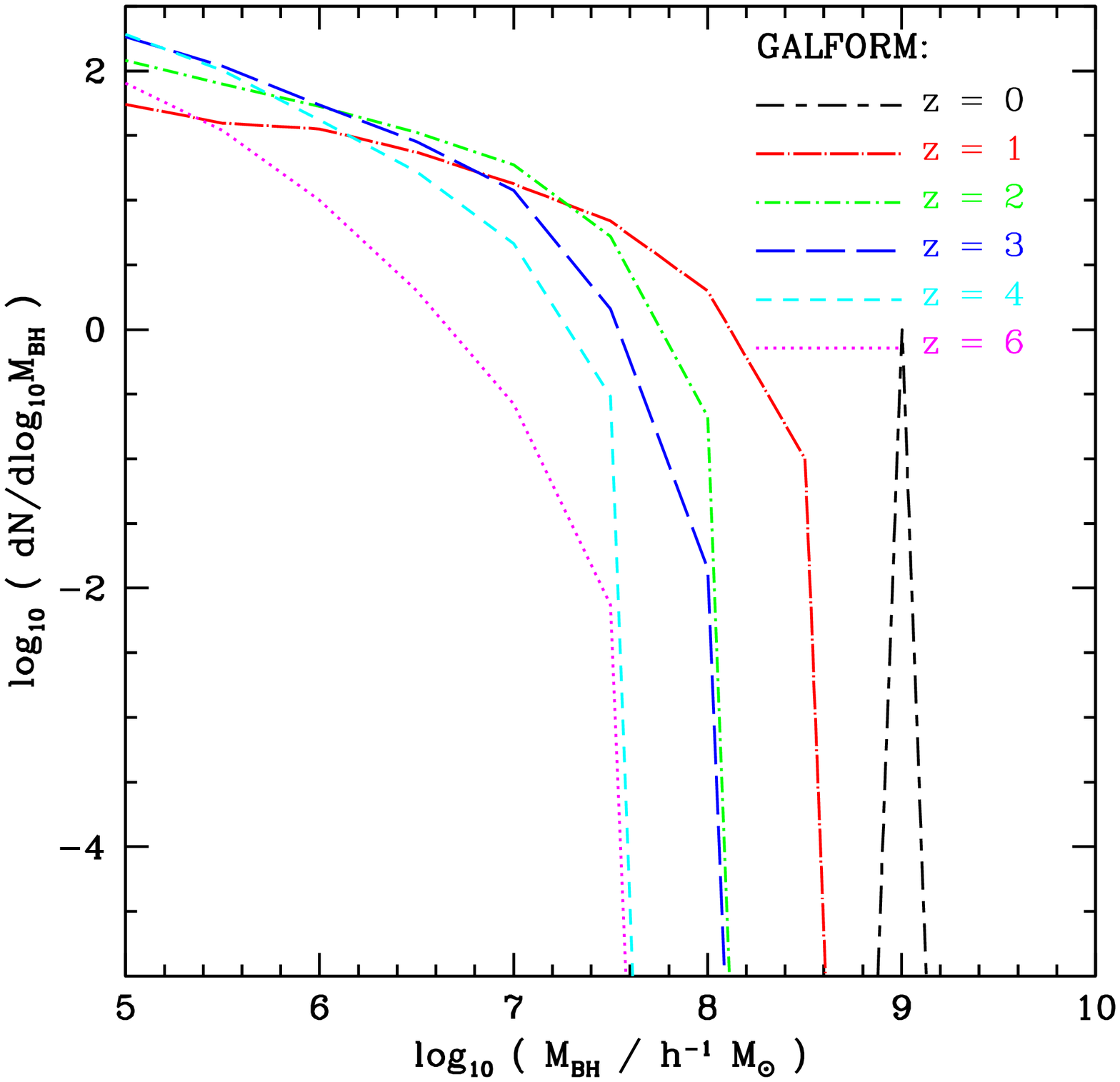}}
\end{picture}
\caption{The mass functions of progenitor black holes for present 
day black holes with mass in the interval $10^7 - 10^{7.5} h^{-1} 
M_{\odot}$ (left panel) and $10^9- 10^{9.5} h^{-1} M_{\odot}$ (right 
panel). The distribution of progenitor masses is plotted at different 
redshifts, as indicated by the key. The mass functions are generated 
by considering a large sample of black holes at $z=0$, and the 
normalization is chosen so that each progenitor mass function 
is the mass function \emph{per black hole} at $z=0$.
\label{fig:mfprogen}
}
\end{figure*}

We now consider the distribution of the masses of black hole progenitors 
at different epochs for present day black holes. 
Fig. \ref{fig:mfprogen} shows the distribution of progenitor masses 
at different redshifts, for two ranges of black hole mass measured 
at $z=0$. The left-hand panel shows the progenitors of $z=0$ black 
holes with masses in the range $10^7 - 10^{7.5} h^{-1}M_{\odot}$ 
and the right-hand panel shows the progenitors of $z=$ black holes 
with masses in the range $10^9 - 10^{9.5} h^{-1} M_{\odot}$.
The different line types in the plot show the progenitor mass 
distributions at different redshifts, as indicated by the key. 
The distributions plotted are averaged over large numbers of black 
holes with the appropriate present day mass. In both panels, 
the $z=0$ distribution is naturally peaked around the present day  
mass of the black hole. 

%
The evolution of these progenitor mass functions
from $z = 6$ to $z = 1$ looks remarkably similar to that of the
universal black hole mass function, albeit truncated at the final $z =
0$ black hole mass, and with an overall normalization which increases
with increasing final black hole mass. 

The similarity of the form and evolution of the progenitor mass
functions with those of the universal mass function is
remarkable. Only at the lowest progenitor redshift plotted ($z = 1$)
for black holes with present day masses in the range $10^7 - 10^{7.5}
h^{-1} M_{\odot}$ do we see a significant deviation from the form of
the overall mass function. The progenitor mass function in this case
is rather flat, with fewer low mass black holes and more high mass
black holes close in mass to the final black hole. The amplitude of
the progenitor mass functions is substantially larger for the black
holes with present day mass of $10^9 - 10^{9.5}h^{-1} M_{\odot}$:
larger black holes have a significantly larger number of progenitor
black holes. This fits in well with our later result
(\S\ref{sec:growth}) that less massive black holes grow primarily by
accretion onto a single main branch, whereas black holes larger than
$5
\times 10^7 h^{-1} M_{\odot}$ grow primarily by mergers of
pre-existing black holes.

\subsection{Black hole growth by mergers and accretion}
\label{sec:growth}

\begin{figure*}
\begin{picture}(0,270)
\put(-250,30){\epsfxsize=8.5 truecm \epsfbox{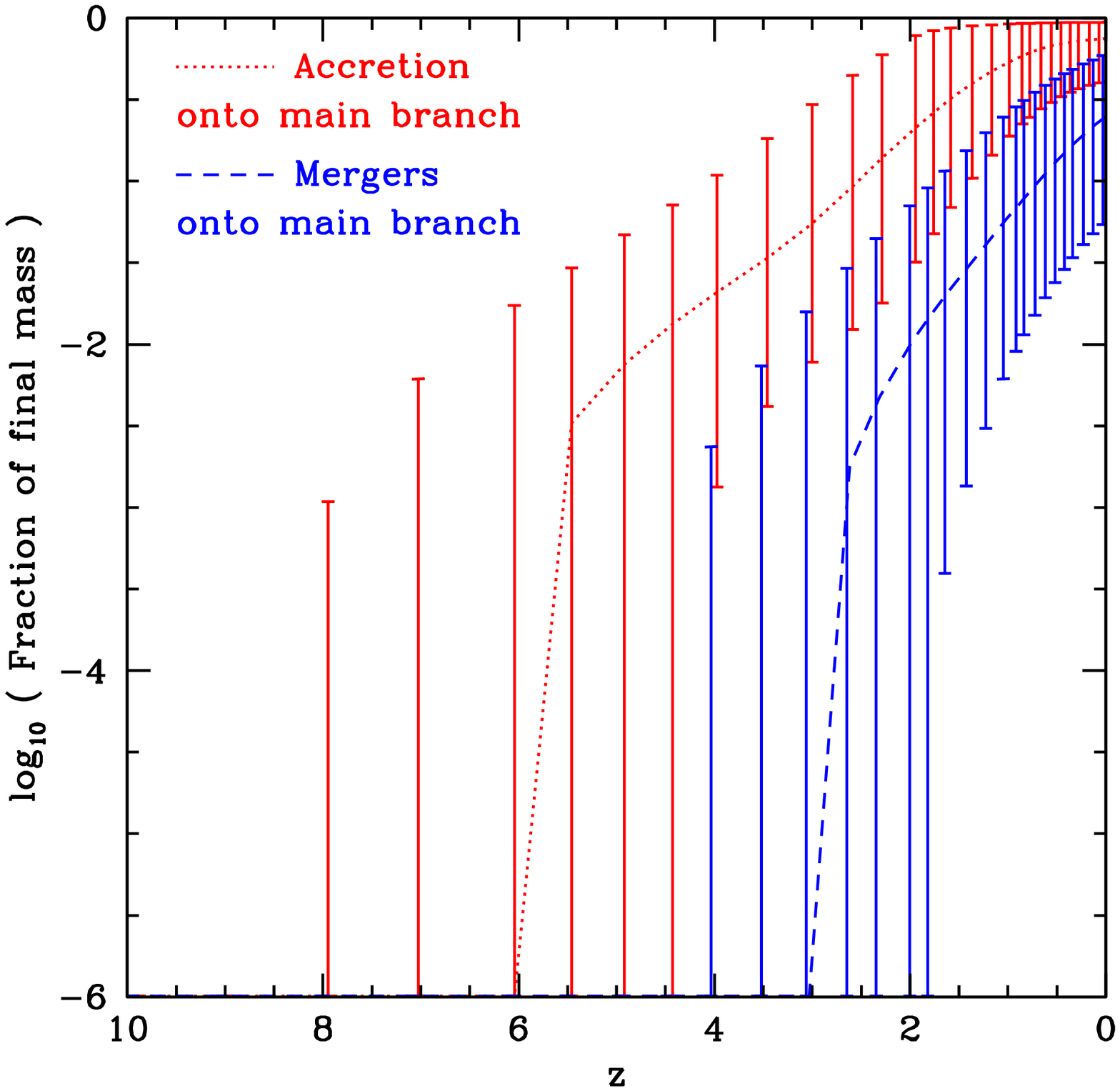}}
\put(10,30){\epsfxsize=8.5 truecm \epsfbox{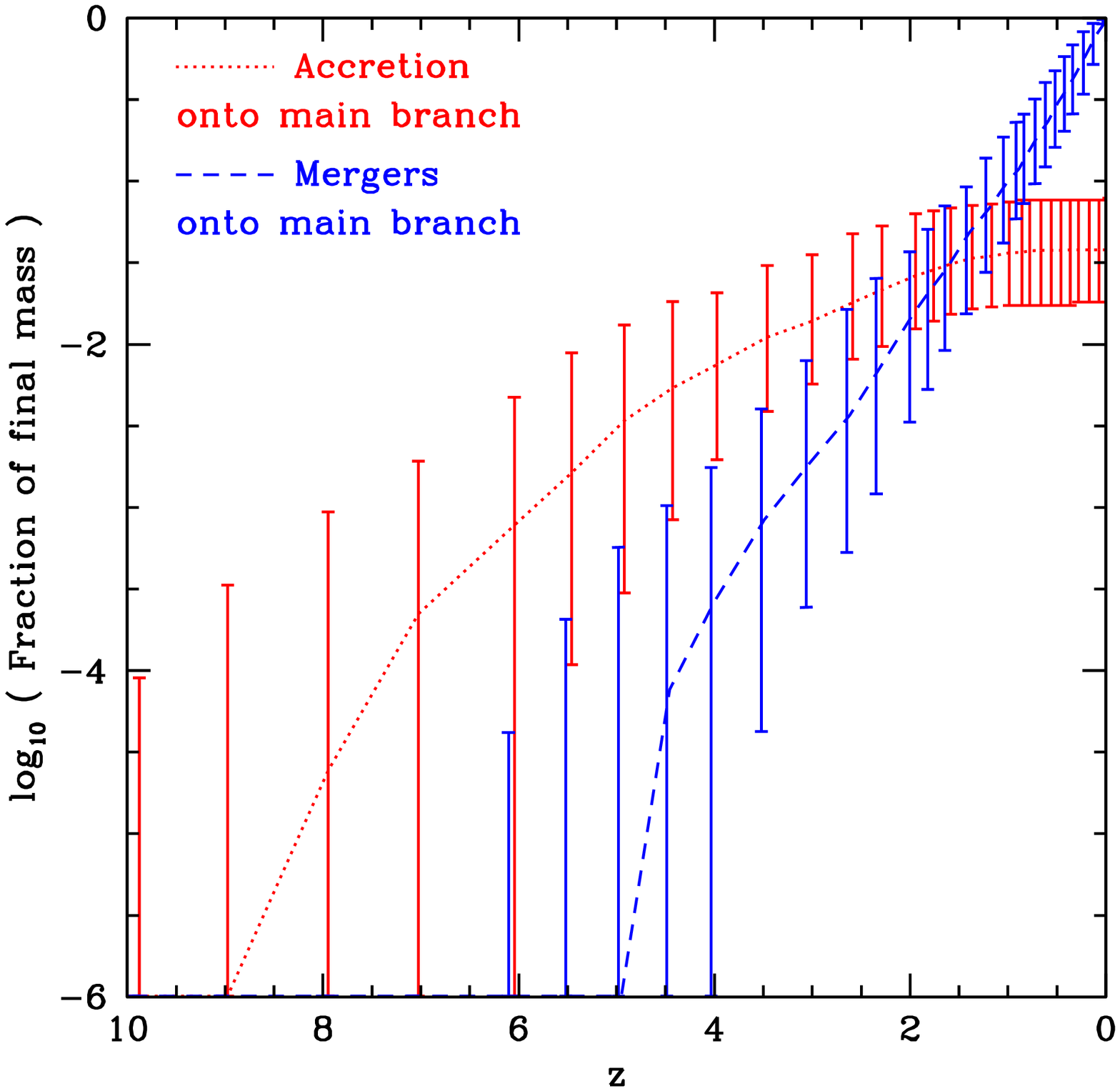}}
\end{picture}
\caption{The cumulative growth with redshift of the black hole mass in
  the `main branch' 
  divided into the contribution from mergers and accretion. We
  consider large samples of black holes with $z=0$ masses in the range 
 $10^{7}-10^{7.5}h^{-1}M_{\odot}$ (left) and
  $10^{9}-10^{9.5}h^{-1}M_{\odot}$ (right). The dotted and
  dashed lines connect the medians of the distribution, while the
  10--90 percentiles of the distribution are shown as errorbars.}
\label{fig:growz}
\end{figure*}

We come now to one of the principal results of our paper, the 
manner in which black holes acquire their mass. 
There are two distinct modes of mass assembly in our model: 
``accretion'', in which cold gas is turned into black hole 
mass during a starburst, and ``mergers'', in which existing black 
holes merge to build a more massive black hole. 
In the accretion mode, mass is being turned into black hole mass for 
the first time, whereas in the merger mode, pre-existing black hole 
mass is being rearranged or reassembled into a more massive 
black hole.   
In Fig.~\ref{fig:growz}, we plot the fraction of the mass in the `main branch' 
which is assembled by mergers or gas accretion as a function of redshift. 
We show results for black holes in two mass ranges at redshift zero: 
$10^{7} - 10^{7.5} h^{-1}  M_{\odot}$ (left) and
$10^{9} - 10^{9.5} h^{-1}  M_{\odot}$ (right). Fig.~\ref{fig:growz} shows 
that at high redshifts, growth by accretion dominates over growth through 
mergers. 
Mergers become increasingly important as 
redshift decreases, 
but for the $10^7 - 10^{7.5} h^{-1} M_{\odot}$ black holes, 
the cumulative growth by mergers never exceeds the cumulative growth 
by accretion, even at redshift zero. However, for black holes of mass 
$10^9 - 10^{9.5} h^{-1} M_{\odot}$ at $z=0$, the cumulative growth of 
their main progenitors by mergers exceeds that by accretion around a 
redshift of 1.7, and growth by accretion almost halts after this.\footnote{This
redshift can vary from black hole to black 
hole -- $z = 1.7$ is the redshift where the median growth by mergers exceeds
the median growth by accretion.} By redshift zero, the cumulative mass
assembled by mergers greatly exceeds that assembled by
accretion. The declining importance of growth by accretion for black holes of
mass $>10^{7.5} h^{-1}  M_{\odot}$ reflects the decline in the amount of gas
available in mergers as more and more of the gas in collapsed haloes is consumed
into stars. 

\begin{figure*}
\begin{picture}(0,270)
\put(-250,30){\epsfxsize=8.5 truecm \epsfbox{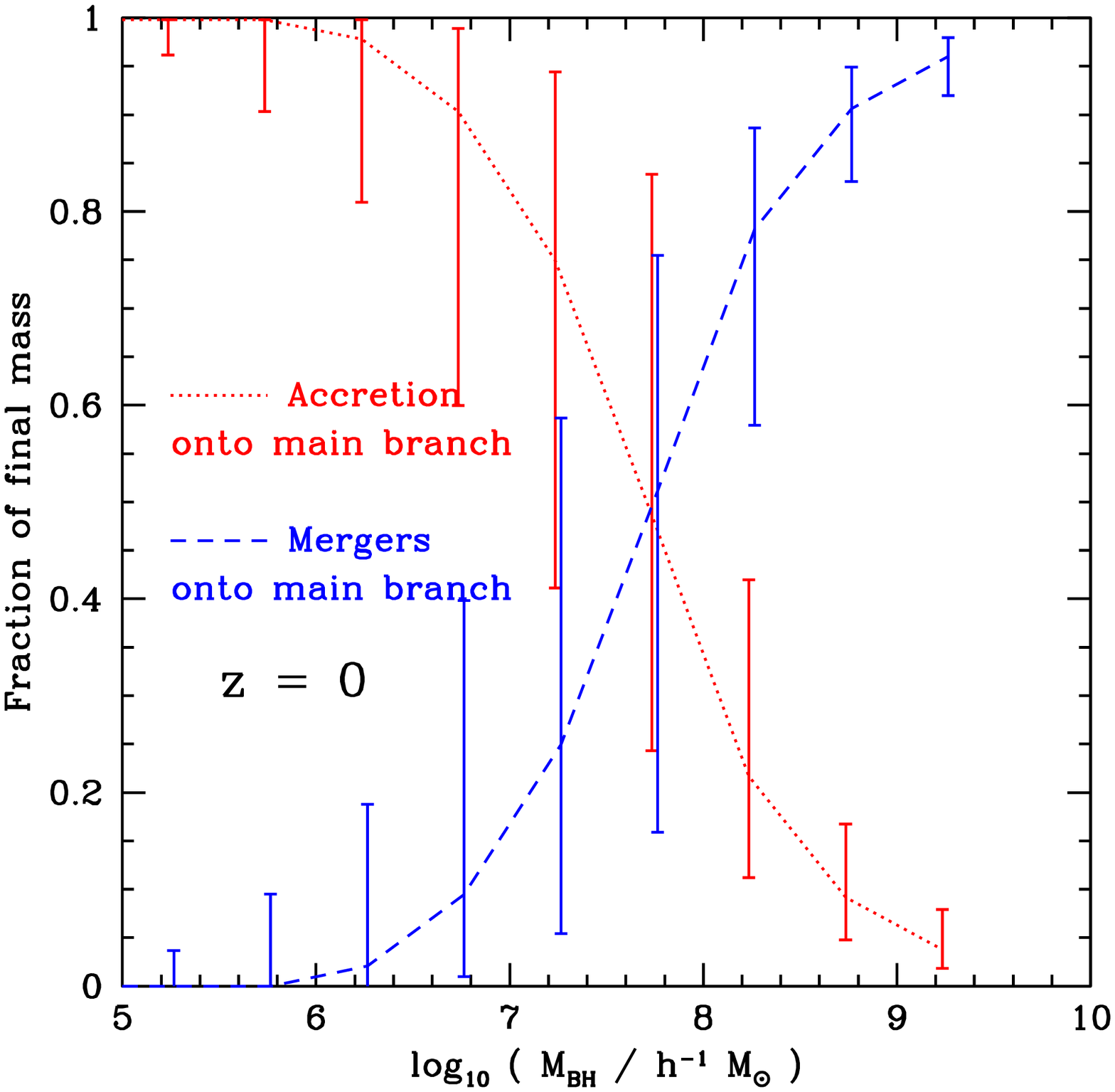}}
\put(10,30){\epsfxsize=8.5 truecm \epsfbox{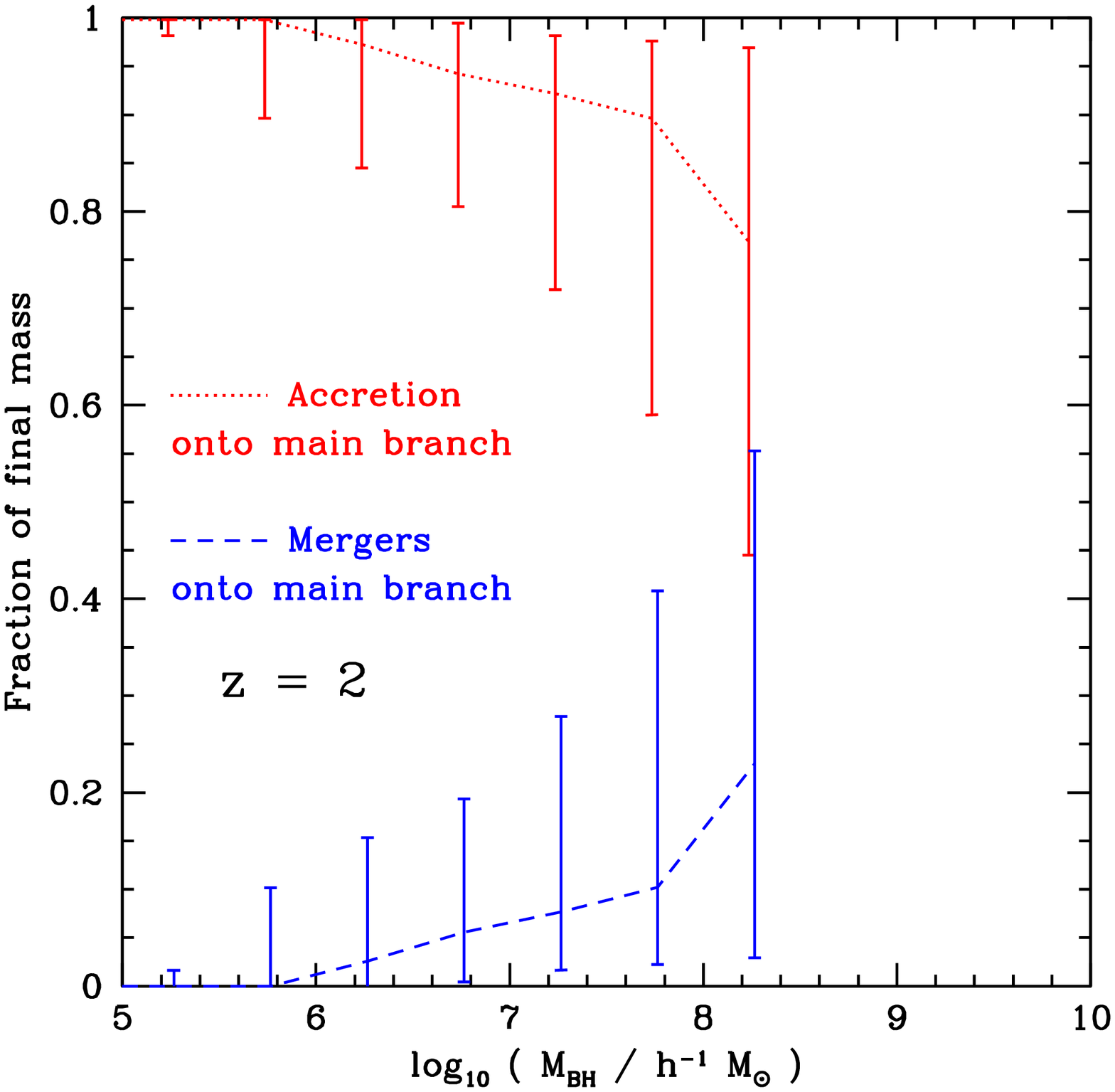}}
\end{picture}
\caption{The cumulative fraction of the mass assembled by mergers
and accretion, as a function of final black hole mass at $z = 0$ (left) and $z
= 2$ (right). The medians are connected by lines, and the 10--90 percentile
spread of the distribution is shown as an errorbar for each black hole mass.}
\label{fig:growint}
\end{figure*}

In Fig.~\ref{fig:growint}, we plot the fraction of the mass of black 
holes which, by $z = 0$ (left) and $z=2$ (right),
has been accumulated by mergers or accretion onto the `main branch'. 
This is shown as a function of black hole mass. Fig.~\ref{fig:growint}
(left) shows that, by redshift zero, low mass black
holes have accumulated nearly all of their mass by direct accretion
onto a single `main branch', while the most massive black holes
accumulate 80--90\% of their mass by mergers of less massive black
holes onto the `main branch'. The transition from 
accretion-dominated growth to merger-dominated growth occurs at a
$z=0$ mass of just over $10^8 h^{-1}
M_{\odot}$. Fig.~\ref{fig:growint} (right) shows that at $z=2$, all
black holes, even those more massive than $10^8 h^{-1}
M_{\odot}$, grow predominantly by accretion, although there is a contribution
from mergers which increases with black hole mass. Comparison of the
results in Fig.~\ref{fig:growint} for $z=0$ and $z=2$ shows that for
any given black hole mass, growth by accretion is more significant for
a black hole at $z=2$ than at $z=0$, and that this difference is
greater for the more massive black holes. This is consistent with
the idea that the luminous growth (i.e. growth by direct accretion of gas) of higher mass black holes switches
off towards lower redshifts (see \S 6).

\subsection{The redshift of black hole formation}
\label{sec:zform}

\begin{figure*}
\begin{picture}(0,650)
\put(-250,440){\epsfxsize=8.0 truecm \epsfbox{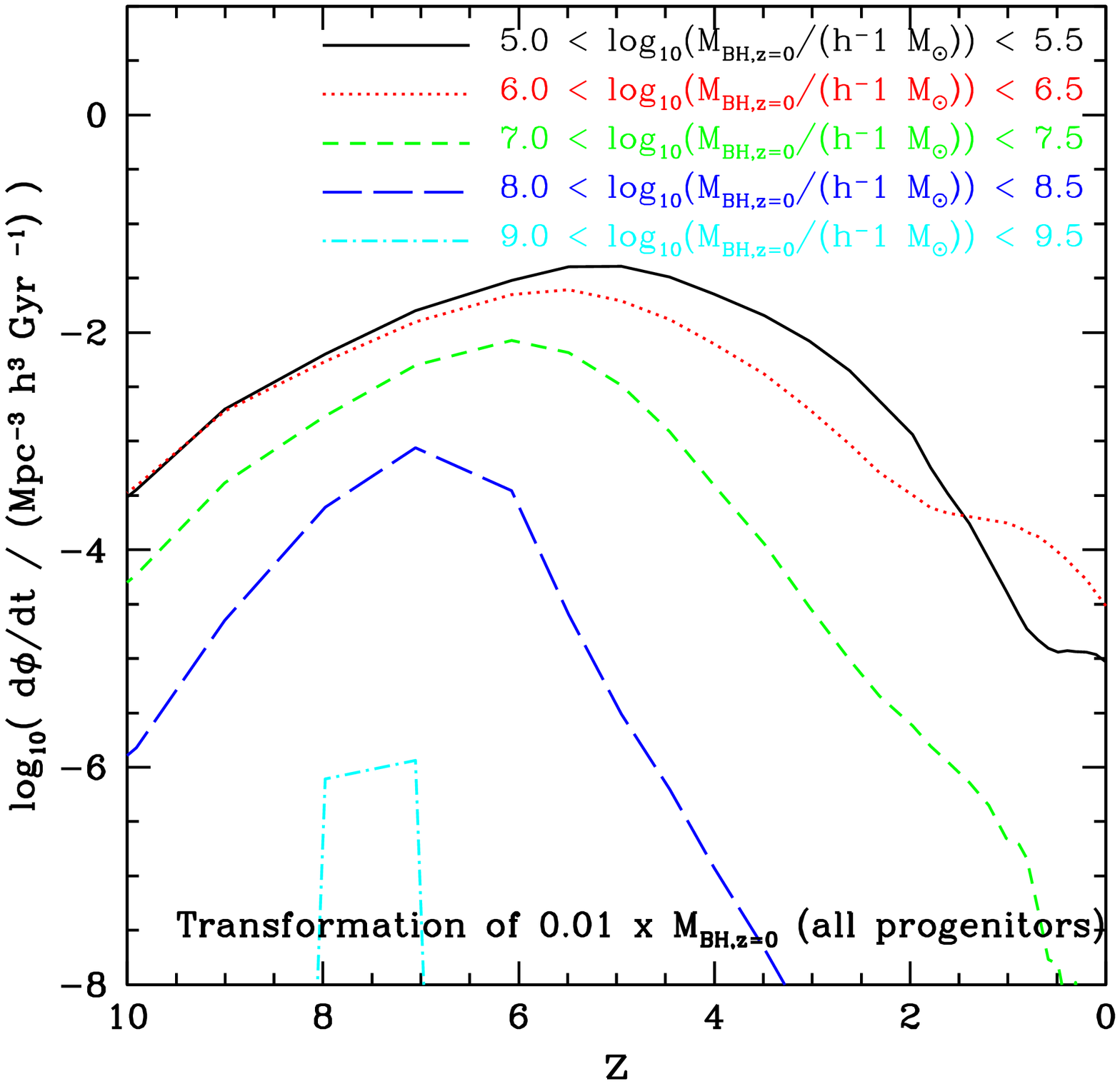}}
\put(0,440){\epsfxsize=8.0 truecm \epsfbox{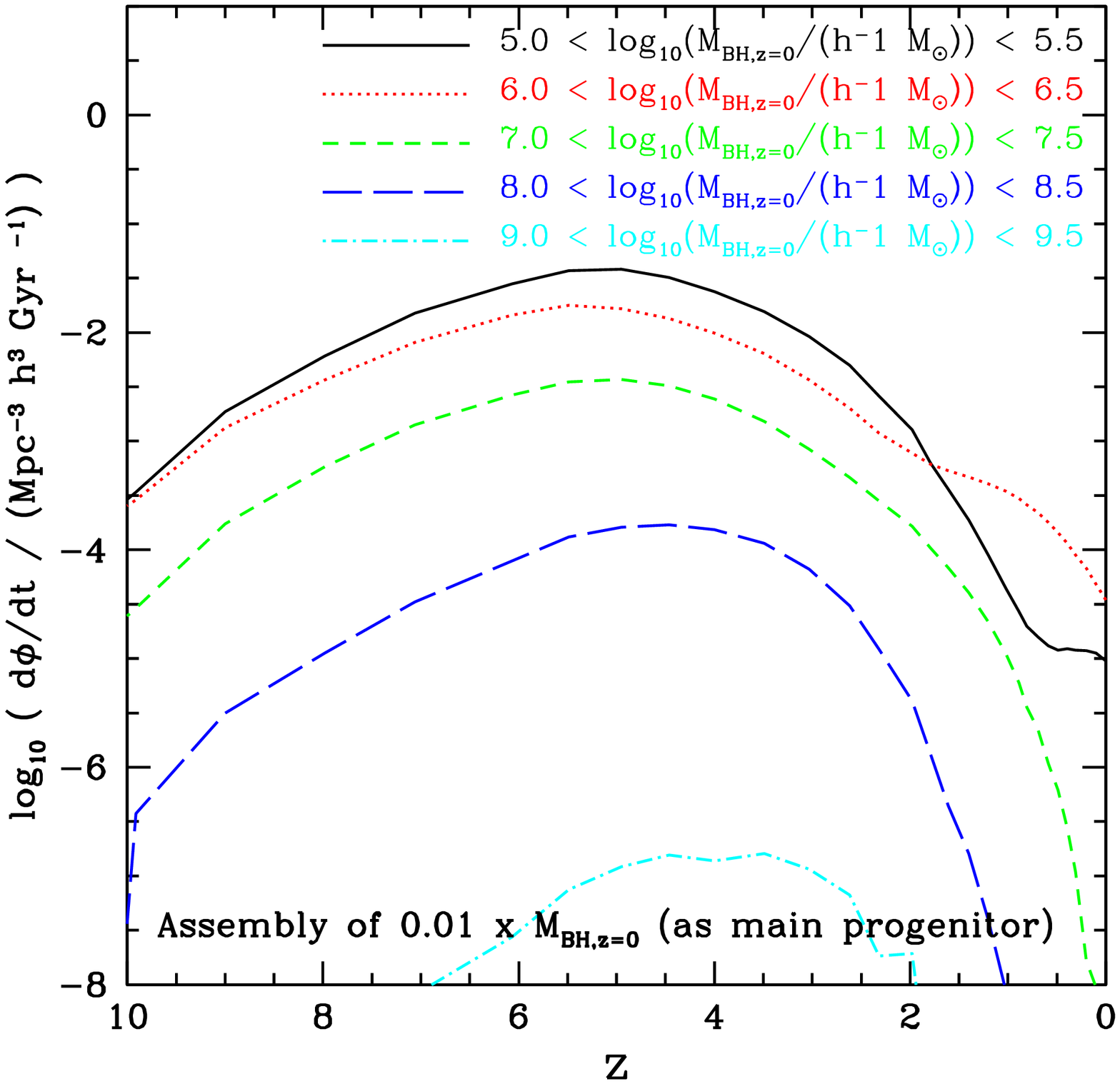}}
\put(-250,220){\epsfxsize=8.0 truecm \epsfbox{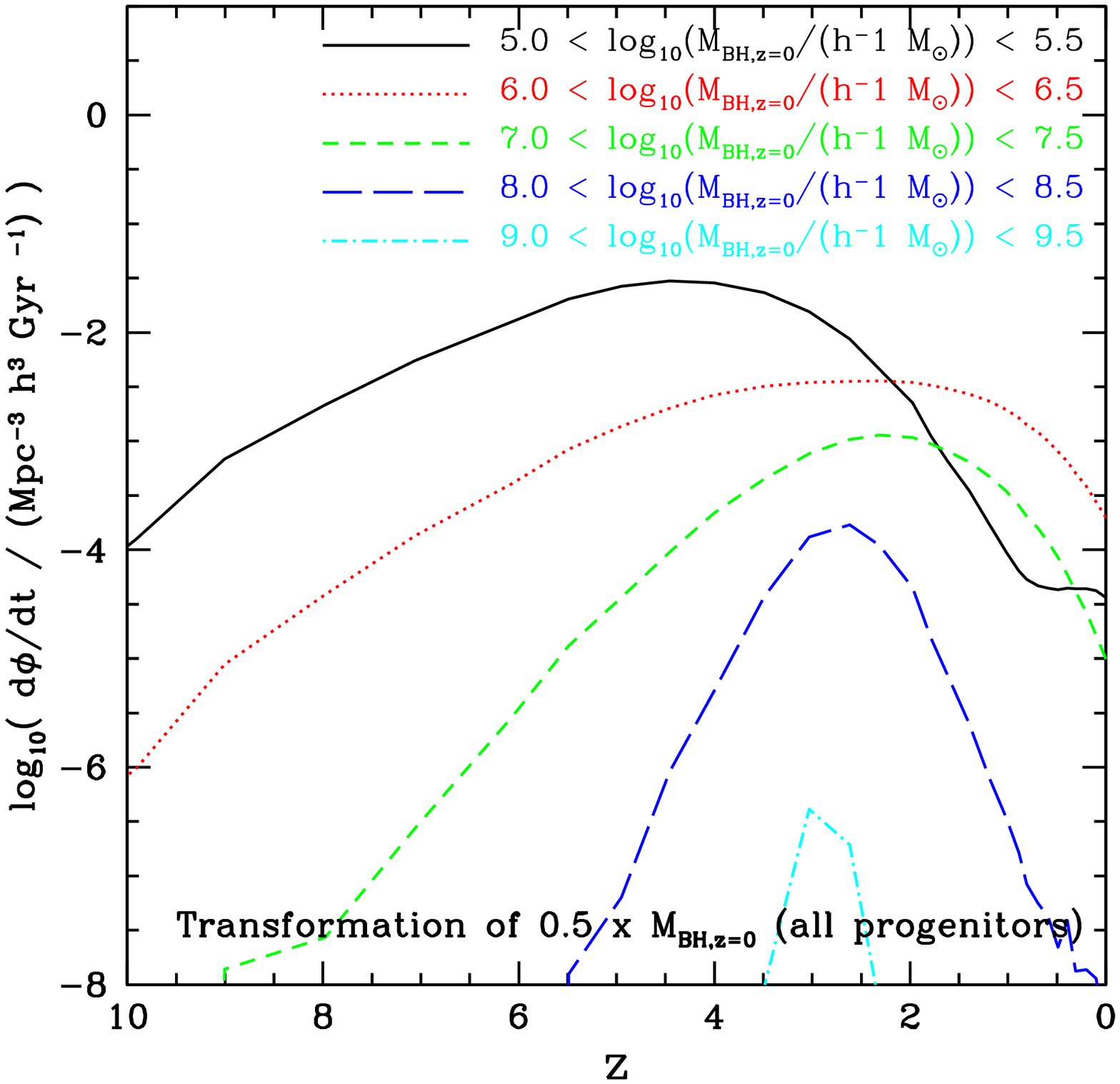}}
\put(0,220){\epsfxsize=8.0 truecm \epsfbox{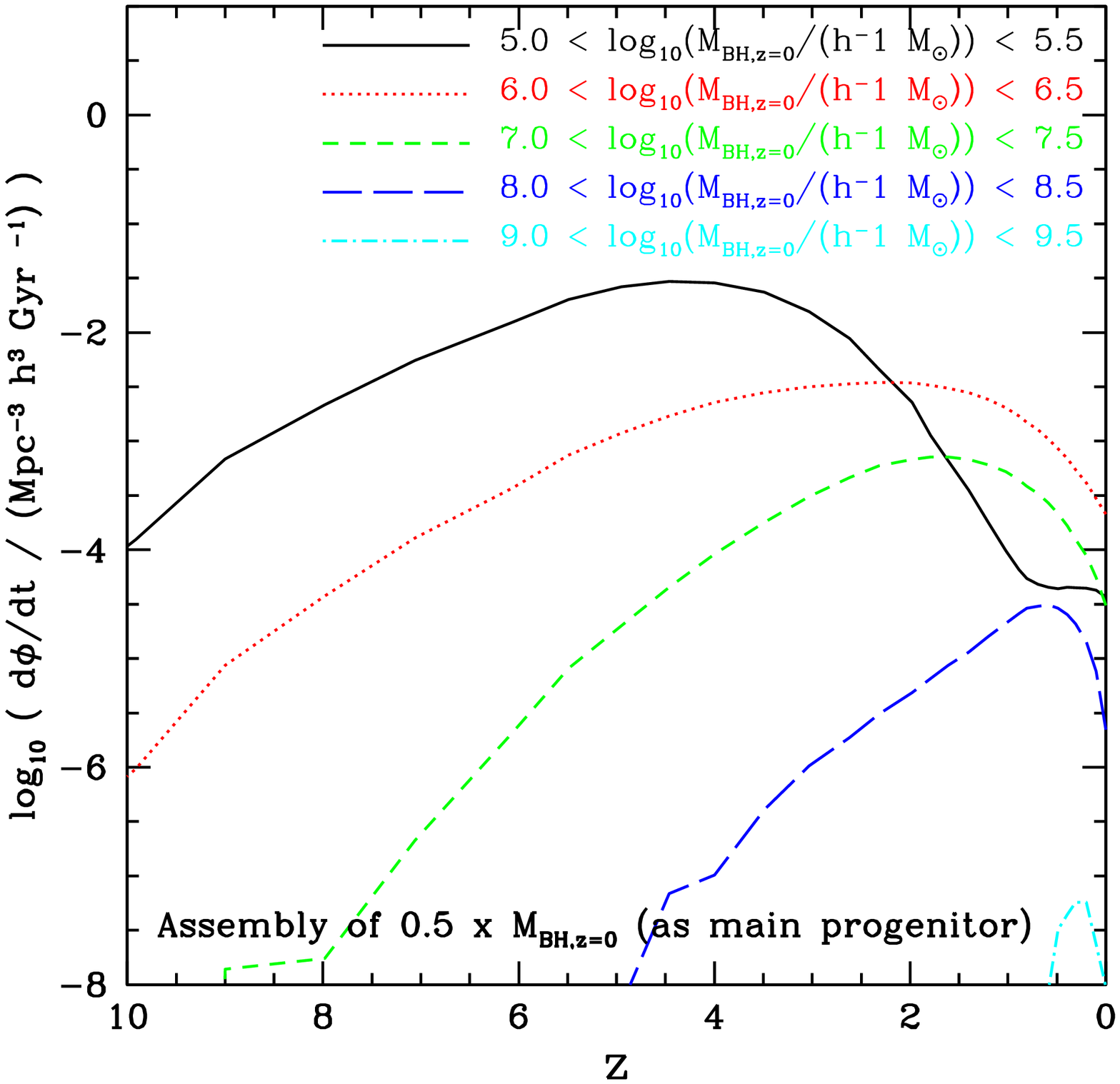}}
\put(-250,0){\epsfxsize=8.0 truecm \epsfbox{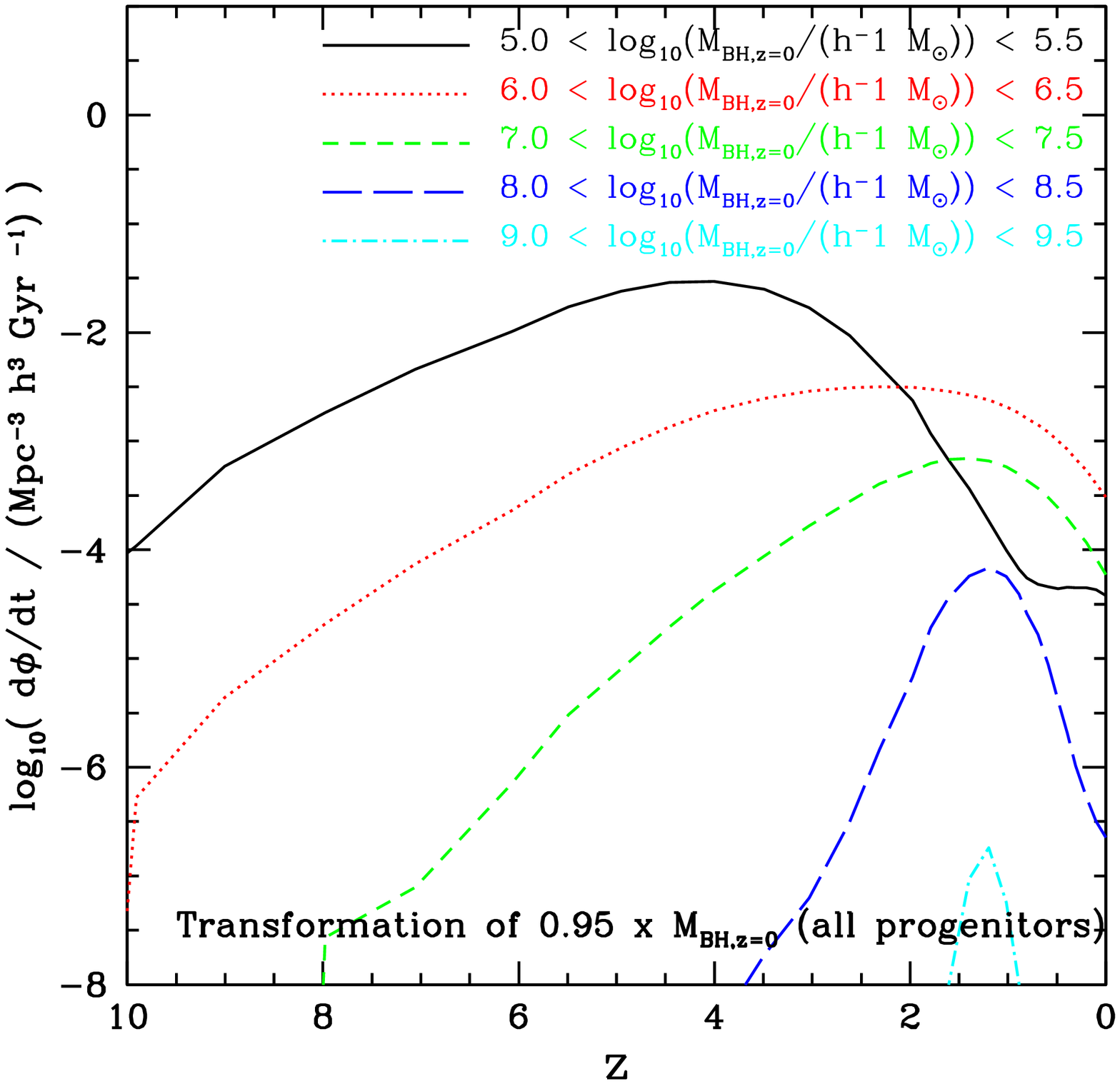}}
\put(0,0){\epsfxsize=8.0 truecm \epsfbox{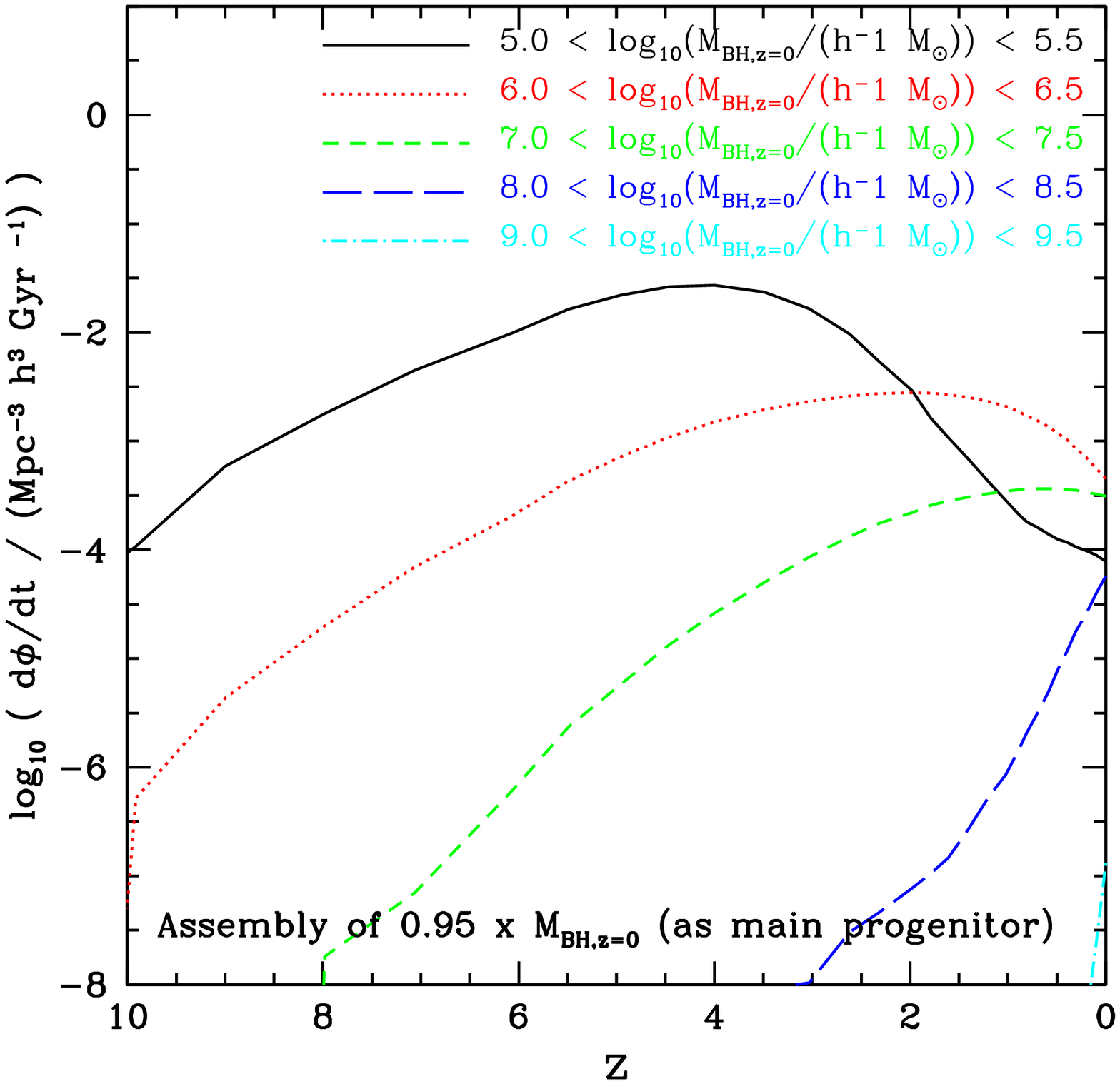}}
\end{picture}
\caption{The distribution of formation redshifts of black holes
in 5 different bins of $z=0$ mass, as indicated by the key. The differing
definitions of formation redshift used in each of the six plots are noted
briefly on each plot and explained more fully in \S\ref{sec:zform} of the
text.}
\label{fig:zformfacc}
\end{figure*}

In Fig. \ref{fig:zformfacc} we show the formation redshifts of black holes,
binned by $z = 0$ mass. Each of the six panels corresponds to a different
definition of formation redshift. `Formation' is defined as the time when either
the main progenitor (right three panels) or the sum of all existing progenitor
black holes (left three panels) first exceeds a given fraction of the final
black hole mass. Where the formation redshift is defined as the time when the
\emph{main progenitor} first exceeds a given fraction of the final mass, we refer to
this as the \emph{mass assembly} redshift, since this is the redshift where the
stated fraction of the mass has been assembled into a single object. Where the
formation redshift is defined as the time when the \emph{sum of all progenitors}
first exceeds a given fraction of the final mass, we refer to this as the
\emph{mass transformation} redshift.  This distinction between the \emph{mass transformation} time and
\emph{mass assembly} time for black holes is analagous to that 
between the star formation time and stellar mass assembly time for the stars in
a galaxy.  We consider three different mass fraction thresholds to define
formation times: 0.01 (top), 0.5 (middle) and 0.95 (bottom).
 
When we consider the assembly of 50\% or 95\% of the final black hole mass
(Fig. \ref{fig:zformfacc} -- middle-right \& bottom-right), we see clear
hierarchical behaviour; the more massive black holes at redshift zero peak in
their formation times at lower redshift than the less massive ones. This is
evidence for the hierarchical \emph{assembly} of black hole mass into a single
final object. However, when we consider the redshift at which 50\% of the final
black hole mass has \emph{accreted} onto \emph{any} black hole in the merger
tree (Fig. \ref{fig:zformfacc} -- middle-left), we see the opposite trend; in
the mass range $M_{\rm BH} = 10^7 - 10^{9.5} h^{-1} M_{\odot}$, the more massive
black holes display a distribution of formation redshifts which peaks at higher
redshift. When we consider the redshift at which 95\% of the final black hole
mass is \emph{accreted} onto \emph{any} black hole in the merger tree
(Fig. \ref{fig:zformfacc} -- bottom-left), we see further evidence of
downsizing. Although the formation rate peaks at a similar redshift ($z \sim 1.2$) for
all black holes in the mass range $M_{\rm BH} = 10^7 - 10^{9.5} h^{-1}
M_{\odot}$, the decline in the fraction forming per unit time as redshift approaches zero is far steeper for
more massive black holes within this mass range. The observational evidence for
``downsizing'' refers to the accretion of mass onto a particular progenitor, which
is accompanied by the release of energy. Hence, it is the latter trend which is
relevant -- as black hole mass increases, the redshift when mass is
\emph{accreted} onto \emph{any} progenitor increases. We return to this point in
\S\ref{sec:downsizing}.

We consider now the early growth of the black holes. Fig.~\ref{fig:zformfacc}
(top-left) shows the redshift when the first 1 per cent of the final black hole
mass has collapsed into any of the branches of the merger tree. There is a clear
trend for larger black holes to be seeded earlier. This is also a form of downsizing. All of the black holes in our
largest mass bin ($M_{\rm BH} = 10^9 - 10^{9.5} h^{-1} M_{\odot}$) and many of
those in the next mass bin, $M_{\rm BH} = 10^8 - 10^{8.5} h^{-1} M_{\odot}$,
are seeded before reionization occurs in the model at $z = 6$. Another
interesting feature of Fig.~\ref{fig:zformfacc} is that, for almost any
definition of formation time, less massive black holes have a much wider spread
in formation times than more massive black holes.

There is little difference in the distribution of formation times of black holes
of mass $M_{\rm BH} = 10^5 - 10^{6.5} h^{-1} M_{\odot}$ regardless of whether we
use a definition which relates to the `main branch' or to `all
progenitors'. This follows from our earlier result that black holes in this mass
range formed almost exclusively by accretion onto a single object, with little
contribution from mergers between black holes (\S\ref{sec:growth},
Fig. \ref{fig:growint}). The differentiation between the different definitions
of formation time begins to become apparent for black holes of mass $M_{\rm BH} =
10^7 - 10^{7.5} h^{-1} M_{\odot}$ and is increasingly more significant as black
hole mass increases further. This relates to our earlier result that the
contribution to the final black hole mass from mergers of pre-formed black holes
compared to the contribution from direct cold gas accretion onto the main branch
increases strongly with increasing black hole mass (\S\ref{sec:growth},
Fig. \ref{fig:growint}).

\subsection{Black hole merger rates}
\label{sec:merrate}

\begin{figure}
\epsfxsize=8.5 truecm \epsfbox{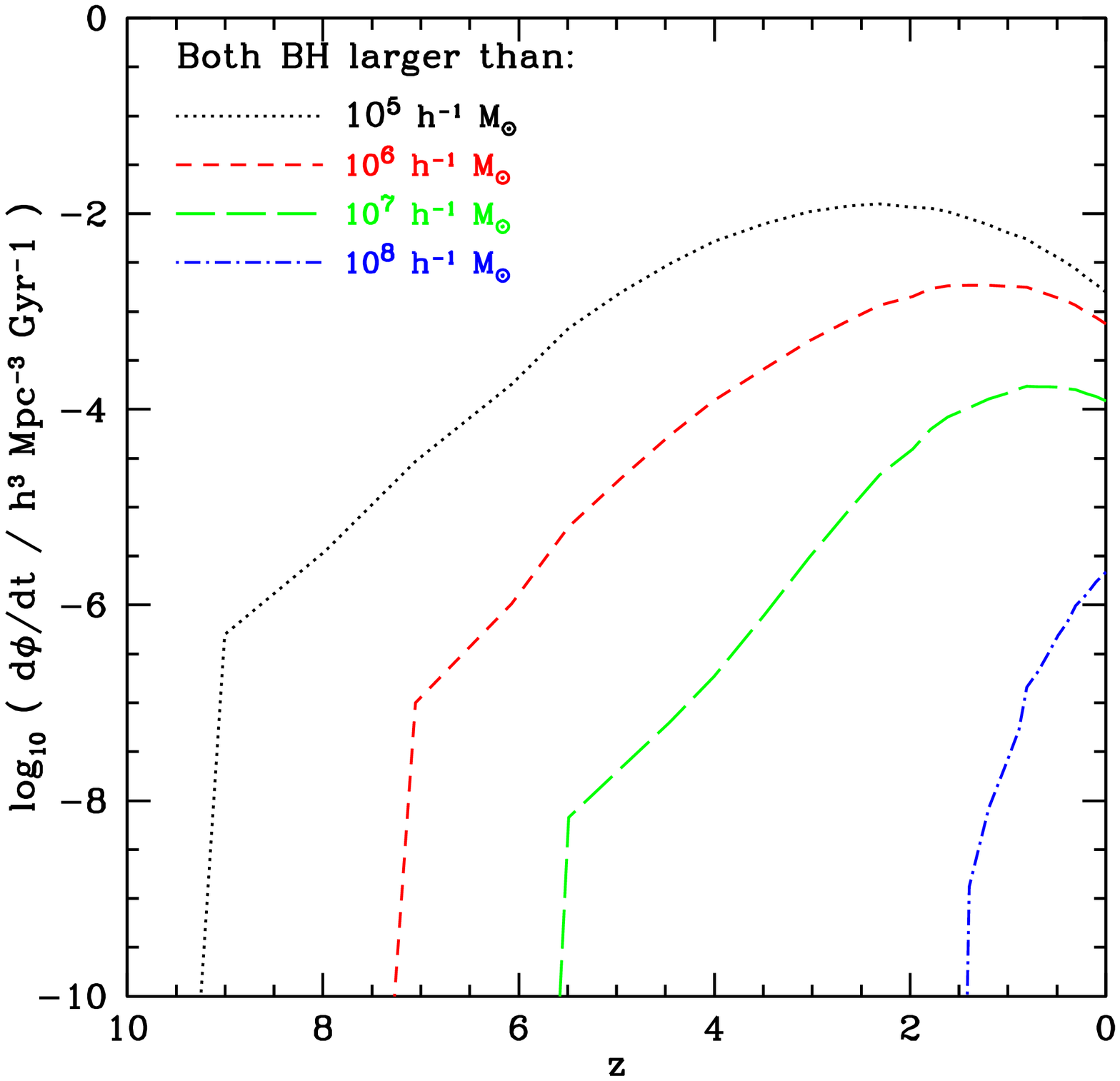}
\caption{Black hole merger rate per unit time as a function of
  redshift. The merger rate is plotted for 5 different mass
  thresholds, as shown by the key, which the (pre-starburst)
  masses of both black holes must exceed.}
\label{fig:mert}
\end{figure}

We show the merger rate per unit time of black holes as a function of redshift
in Fig.~\ref{fig:mert}. We show this for a number of mass thresholds which must
be exceeded by both of the black holes that take place in the merger. The merger
rates peak at lower redshift for more massive black holes, with the merger rate
for the most massive bin still rising at $z=0$.  This is consistent with the
trend seen at $z=0$ that larger mass black holes grow primarily by
mergers, while less massive black holes grow primarily by accretion
(\S\ref{sec:growth}).

This behaviour in the growth and merging of black holes of varying mass is
largely a reflection of the general hierarchical growth of structure, moderated
in the case of galaxies and black holes by baryonic processes. The results
presented in this subsection concern only mergers of black holes, not necesarily
their total growth which can also involve accretion. There is no evidence for
`anti-hierarchical' behaviour in the evolution of black hole mergers. However,
as we show in \S6, this is perfectly compatible with quasar downsizing -- black
hole merging can be a `dark' process in which no gas is present, whereas the
observational evidence for downsizing refers to processes involving star formation or gas
accretion.

\subsection{The fraction of baryons in black holes}
\label{sec:univ}
\begin{figure}
\epsfxsize=8.5 truecm \epsfbox{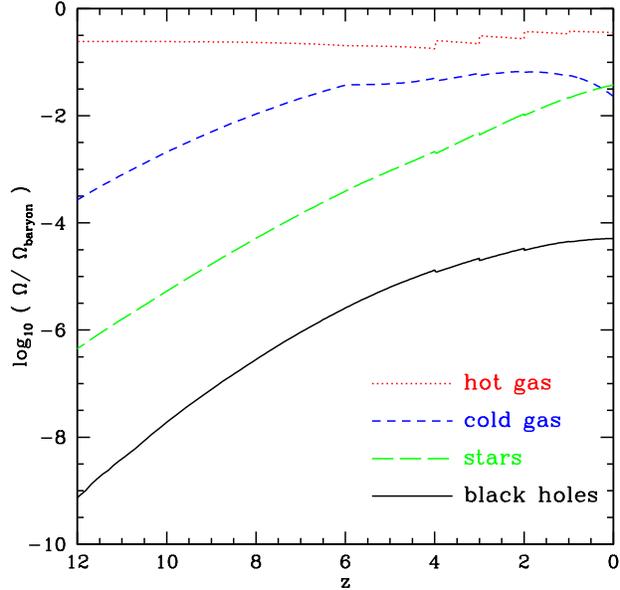}
\caption{The evolution of the fraction of the baryons in the universe
in hot gas, cold gas, stars (disc plus bulge) and black holes. The small
abrupt changes in some of the lines are due to simulation runs that finish at different times.}
\label{fig:globalBH}
\end{figure}

\begin{figure}
\epsfxsize=8.5 truecm \epsfbox{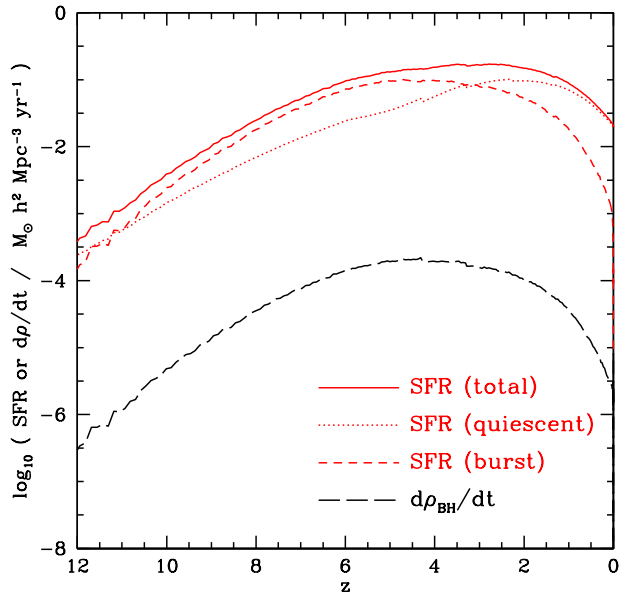}
\caption{
The variation with redshift of the global star formation rate (starbursts,
quiescent and total) and the global rate of black hole growth.}
\label{fig:globalBHdot}
\end{figure}

Having considered the formation of individual black
holes, we now look at the global picture. In Fig.~\ref{fig:globalBH}, 
we show the integrated cosmic density of all the baryonic components 
of the universe; hot gas, cold disc gas, stars and black holes. After $z \sim 4$, 
the growth of black hole mass in the universe slows down in comparison to 
that of stars, as quiescent star formation begins to dominate over star 
formation in bursts. The decline in cold gas from redshift 2 to 0 goes 
a small way towards explaining the decline in quasar activity 
over this redshift interval. The decline in the galaxy merger rate and 
the transition from burst-dominated star formation to quiescent 
star formation also play a role. In Fig.~\ref{fig:globalBHdot}, we show 
the star formation rate, divided into burst and quiescent modes, and the
rate of black hole growth. By construction in our model, black hole growth is more
strongly correlated with the star formation rate in bursts than with star 
formation in general. 
Very broadly, although perhaps less so at low redshifts, black hole accretion tracks the 
overall star formation over cosmic time, as observed (\cite{bt98}). 

The cosmological mass density of black holes at $z=0$ is a quantity of
interest. In our model, we find that $\rho_{\rm BH} = 2.83 
\times 10^{5} M_{\odot} \rm Mpc^{-3}$. Observationally, $\rho_{\rm BH}$ 
is determined by integrating the black hole mass function which, in turn, is
inferred from a combination of the velocity dispersion distribution of galaxies, the K-band
luminosity function or the bulge stellar mass function, and the appropriate
$M_{\rm BH}-\rm bulge$ relation. Observed values of $\rho_{\rm BH}/(10^{5}
M_{\odot} \rm Mpc^{-3})$, converted to $H_{0}=70 {\rm kms}^{-1}{\rm Mpc}^{-1}$,
are : $2.9 \pm 0.5$ (\cite{yt02}), $2.4 \pm 0.8$ (\cite{allerrichstone02}), $2.8
\pm 0.4$ (\cite{md04}), $4.2 \pm 1.1$ (\cite{shankar04}) and $4.6
^{+1.9}_{-1.4}$ (\cite{marconi04}). Our estimate is towards the lower end of
the broad range spanned by the observational estimates. We do not include any
measurement errors in our estimate. A detailed comparison would need to take
into account galaxy type (some estimates are based only on ellipticals), the
flux limits of the observational samples and the treatment of the dispersion in
the $M_{\rm BH}-\rm bulge$ relations when converting from bulge properties to
black hole masses. Many observational estimates assume that the scatter in $\rm
log(M_{\rm BH})$, at a given value of the bulge property under consideration, is
symmetrical. This assumption then leads to larger values of $\rho_{\rm BH}$ for
larger assumed values of the scatter (\cite{md04}). However, it is not at all
clear that the scatter in these relations is symmetrical.

\section{The evolution of the relation between black hole mass and bulge properties.}
\label{sec:magz}

\begin{figure*}
\begin{picture}(0,520)
\put(-250,280){\epsfxsize=8.5 truecm \epsfbox{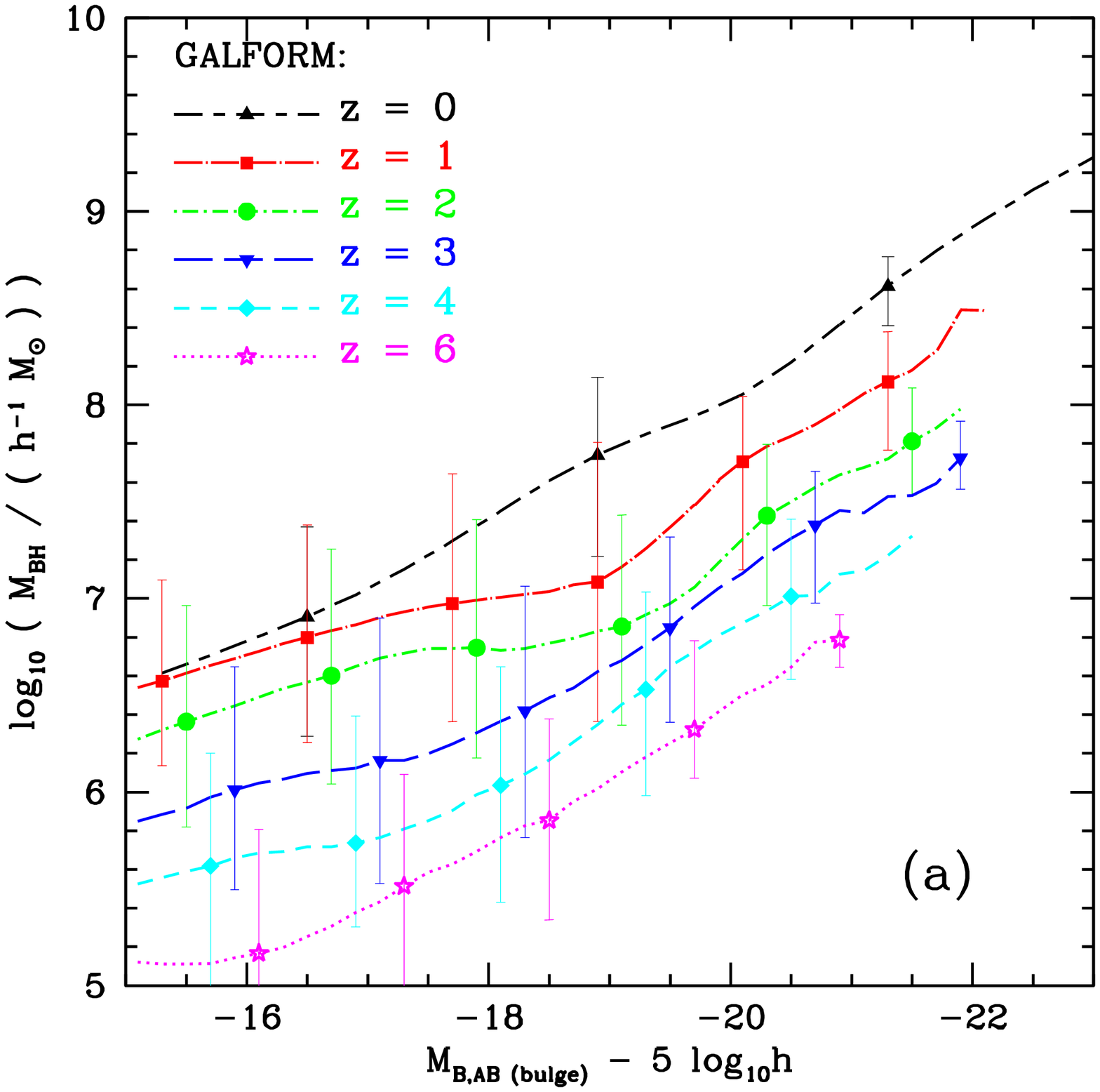}}
\put(0,280){\epsfxsize=8.5 truecm \epsfbox{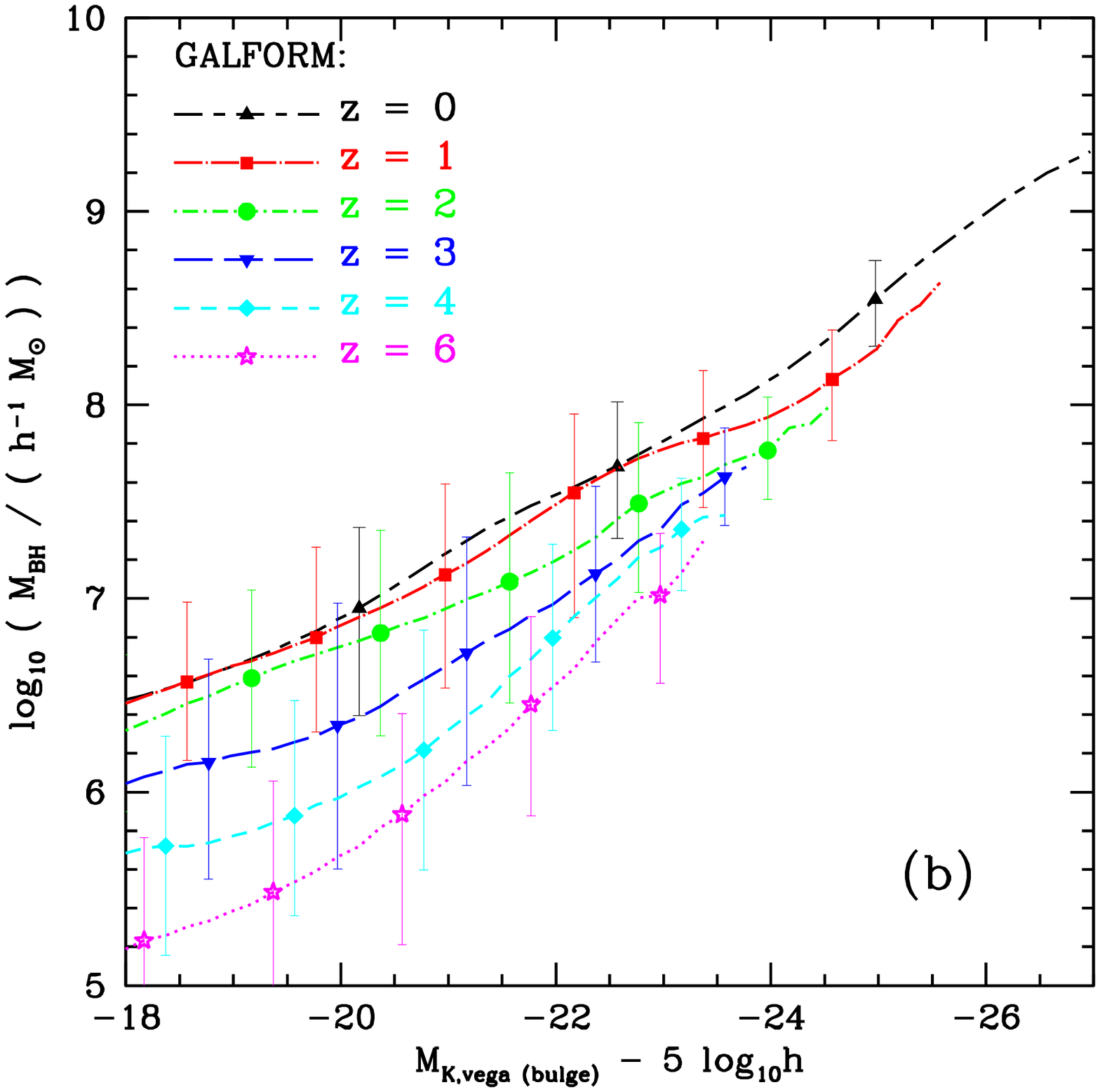}}
\put(-250,20){\epsfxsize=8.5 truecm \epsfbox{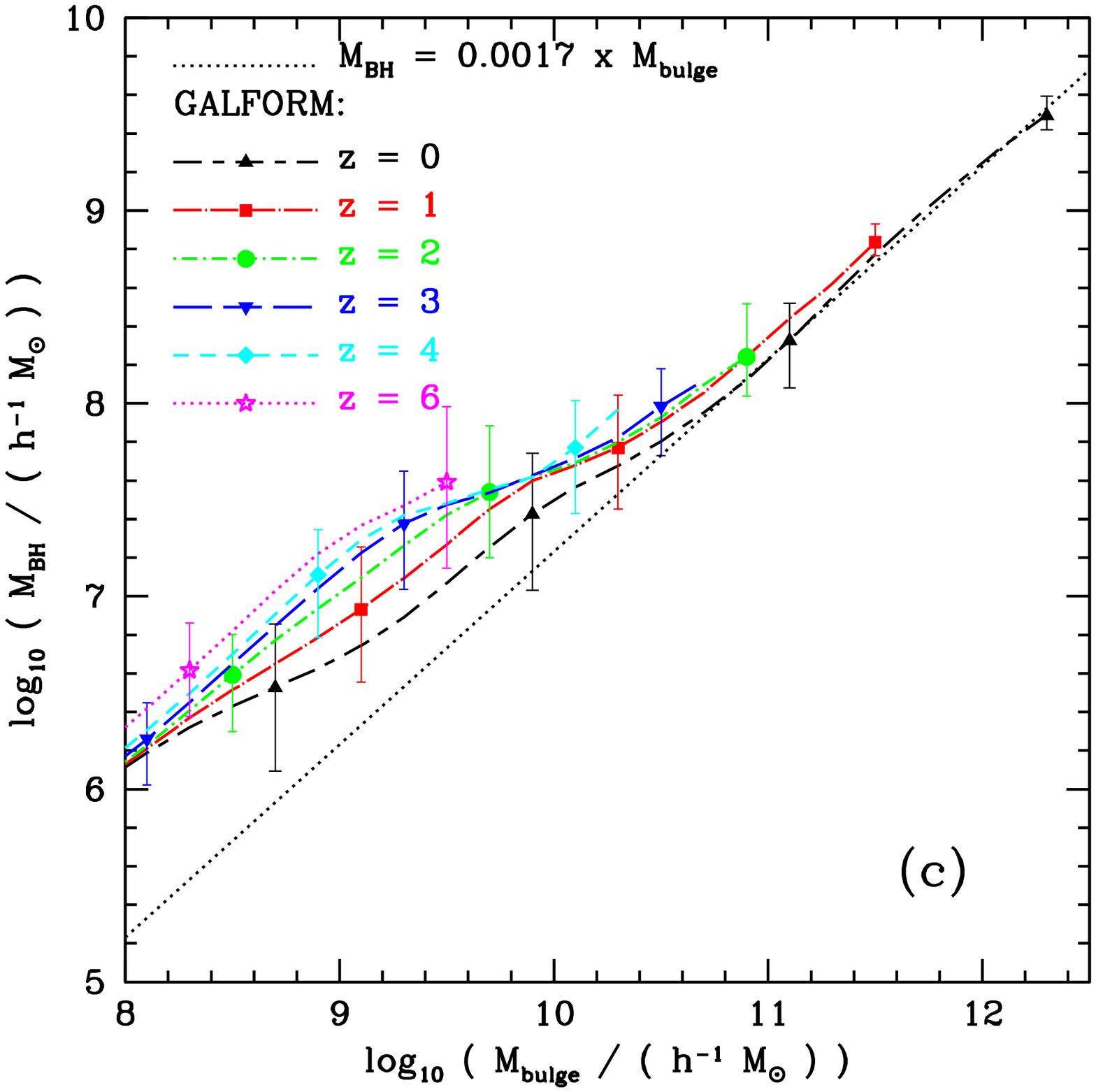}}
\put(0,20){\epsfxsize=8.5 truecm \epsfbox{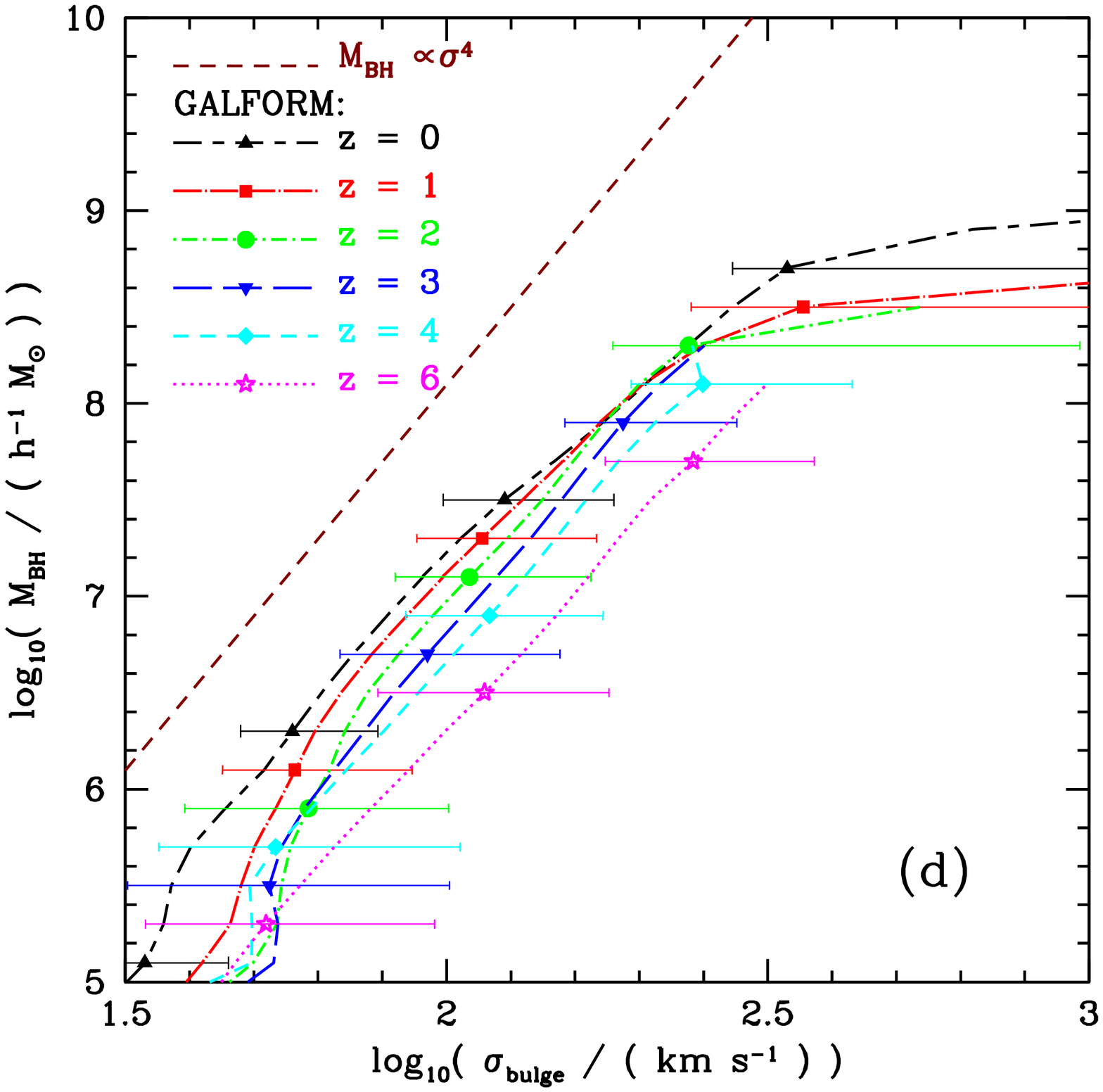}}
\end{picture}
\caption{The redshift evolution of the relations between central black hole
mass, $M_{\rm BH}$, and bulge properties. Each panel shows the relationship
between $M_{\rm BH}$ and a different property of the host spheroid : (a) the
bulge rest-frame B-band magnitude; (b) the bulge rest-frame K-band magnitude;
(c) the stellar mass of the bulge; (d) the velocity dispersion of the bulge.
The model predictions are shown by the symbols with errorbars; the lines show the
median relations and the errorbars the 10--90 percentile spread of the
distributions. Redshifts 0, 1, 2, 3, 4 and~6 are shown in different line types,
as indicated by the key.}
\label{fig:magz}
\end{figure*}

In this section, we discuss the evolution of the
relationship between black hole mass and various galaxy bulge
properties: K-band and B-band bulge magnitude, bulge stellar mass
and bulge velocity dispersion. We show these relationships in
Fig.~\ref{fig:magz}, in each case plotting the model predictions 
for the $M_{\rm BH}-\rm bulge$ relationships at $z=0, 1, 2, 3, 4$ and~
6. We discuss each of these in turn, briefly referring to any relevant
observational data. 
However, it is difficult to make rigorous comparisons to
current data. While relationships between black hole mass and bulge 
properties are fairly well determined at $z = 0$, this is currently not 
the case for $z > 0$, where observational samples are small and subject to 
selection effects. In particular, different surveys sample the population 
of galaxies and, where relevant, the AGN subpopulation, in ways that are not always straightforward to replicate in the models.

We show our model predictions for the $M_{\rm BH}-M_{\rm B,bulge}$ relation in Fig.~\ref{fig:magz}(a) 
and the $M_{\rm BH}-M_{\rm K,bulge}$ relation in Fig.~\ref{fig:magz}(b).
In the model, these relationships shift towards brighter 
magnitudes at higher redshifts. This is a reflection of the evolving 
stellar populations. The stellar populations of bulges at low redshift 
are older and thus less luminous than their high redshift counterparts. 
This effect more than compensates for any evolution in the opposite 
direction in the $M_{\rm BH}-M_{\rm bulge}$ relation, which we discuss below. 
The redshift evolution in the $M_{\rm BH}-M_{\rm B,bulge}$ relation is 
greater than that in the $M_{\rm BH}-M_{\rm K,bulge}$ relation because  
stellar populations dim more strongly with time in the B-band than 
in the K-band. Observationally, however, Peng \etal (\shortcite{peng06}), 
selecting high redshift quasars, find little trend in the 
$M_{\rm BH}-M_{\rm R,bulge,rest}$ relation with redshift, which
conflicts somewhat with our prediction of an evolution towards brighter magnitudes as redshift increases.

We show the $M_{\rm BH}-M_{\rm bulge}$ relation in Fig.~\ref{fig:magz}(c). There
is no significant evolution in either the slope or scatter at large bulge
masses. For $M_{\rm bulge} < 10^{10} h^{-1} M_{\odot}$, the black hole mass to
bulge mass ratio increases with increasing redshift. Observationally, Peng \etal
(\shortcite{peng06}) find that the ratio of $M_{\rm BH}$/$M_{\rm bulge}$ was
3--6 times larger at $z \gtrsim 2$ for AGNs than for quiescent galaxies at $z =
0$. McLure \etal (\shortcite{mclure06}), selecting radio galaxies at
$z>0$ from
the 3CRR catalogue,
argue that $M_{\rm BH}$/$M_{\rm bulge}$ increases with redshift, and is $\sim 4$
times greater for radio galaxies at $z=2$ than for quiescent galaxies at $z=0$. We find that
$M_{\rm BH}$/$M_{\rm bulge}$ was $\sim$ 2 times greater at $z=2$ than at $z=0$
for $M_{\rm bulge} < 10^{10} h^{-1} M_{\odot}$. This evolution is in the same
sense as and of comparable size to the observational trend, although the effect in
the model is perhaps not as strong. As discussed in \S\ref{sec:bhdata}, the
predicted variation in the $M_{\rm BH}$/$M_{\rm bulge}$ ratio reflects the
variation in the fraction of bulge stars which formed quiescently in
discs. Mergers at higher redshift, when discs are more gas-rich and have fewer
stars, deposit a lower fraction of (quiescently-formed) disc stars in the bulge
(e.g. \cite{croton0512}).

Close inspection of Fig.~\ref{fig:magz}(c) shows that a few objects at the
highest redshifts have black hole masses that exceed $F_{\rm BH} \times M_{\rm
bulge}$. This would appear to be impossible given our definition of $F_{\rm BH}$
in \S\ref{sec:BHmodel}. This apparent anomaly is due to our assumption that the
mass of the black hole increases instantaneously at the time of the
starburst. In our model, star formation in the bursts extends over $\sim 50$
dynamical times, while quasars shine only over $\sim 0.3$ dynamical times, so that the stellar mass builds up much more slowly than the black hole mass. It
seems likely, however, that a black hole will still be growing towards its final
mass towards the end of the starburst
(\cite{archibald02,alexander05nature,borys05}). We defer a study of the
co-evolution of the stars and the the black hole mass to future work.

Finally, we show the $M_{\rm BH}-\sigma_{\rm bulge}$ relation in 
Fig.~\ref{fig:magz}(d). There is no evolution in the slope of the relation, 
but the zeropoint does evolve and the scatter increases significantly towards 
higher redshift. For a given mass of black hole, the velocity dispersion 
of the bulge is greater at higher redshift. To some extent, this evolution 
reflects the expected variation in the properties of dark matter haloes: 
at a given mass, the halo velocity dispersions scales as 
$\sigma \propto (z_{\rm form}+1)^{1/2}$. Alternatively, the evolution 
could be viewed as a reduction in the black hole mass with increasing 
redshift, for a fixed bulge velocity dispersion. 

Shields \etal (\shortcite{shields03}) have compared the relative amounts of
black hole mass in distant quasars and in galaxies in the local Universe. They find 
a large scatter and an increase of $0-0.5~\rm dex$ in 
$M_{\rm BH}/\sigma_{\rm bulge}$ between $z = 0$ and $z \sim 3$. Similarly, Woo 
\etal (\shortcite{woo06}) have compared Seyferts at $z = 0.36$ with 
galaxies at $z = 0$. They too find an increase, of $0.62 \pm 0.10~\rm dex$, in
black hole mass at fixed $ \sigma_{\rm bulge}$ at $z = 0.36$ compared to $z =
0$. Thus, the observed trend in $M_{\rm BH}/\sigma_{\rm bulge}$, if any, is in
the opposite direction to the trend we find in our simulations. It is possible
that our model neglects effects that would cause black holes to be a larger
fraction of the galactic bulge mass at higher redshifts. However, it must be
remembered that $\sigma_{\rm bulge}$ is one of the more uncertain properties of
the galaxies in our model and that dynamical effects which are not included
could play a role in determining the properties of merger remnants
(\cite{dekelcox06,robertson06}).

\section{Downsizing in a hierarchical universe}
\label{sec:downsizing}

In cosmology, ``downsizing'' is an ill-defined term which has been applied to
describe the phenomenon whereby luminous activity 
(e.g. star formation or accretion onto black holes) appears to be occuring 
predominantly in progressively lower mass objects (galaxies or BHs) as the
redshift decreases. Claims of downsizing were first made in connection with the 
population of star-forming galaxies (\cite{cowie96}). More recently, 
the same trend has been inferred from the evolution of the X-ray luminosity 
function of quasars (\cite{cowie03,steffen03,ueda03,barger05,hasinger05}): 
the number of bright X-ray sources peaks at a higher redshift than the 
number of faint X-ray sources. The optical quasar luminosity function 
shows similar evolution, with more bright objects seen at increasing 
redshifts (e.g. \cite{croom04}).

The apparent downsizing in the quasar X-ray luminosity function has been
interpreted by some authors as implying that black holes acquire mass in an
`\emph{anti-hierarchical}' manner
(\cite{marconi04,merloni04,shankar04,hasinger05}). In this Section, we
demonstrate that the ``downsizing'' of the {\it luminous} growth of black holes
is actually a natural feature of our model, despite the fact that the overall
assembly of mass into black holes \emph{is hierarchical}. Downsizing in the
galaxy population in hierarchical models is promoted by the earlier collapse and
more active merging of objects in regions of high overdensity (\cite{kauffmann95,mouritaniguchi06,neistein06}). In recent models of galaxy formation, this natural
trend is accentuated by the feedback processes associated with AGN
activity in massive haloes (Bower \etal 2006; Croton \etal 2006).
However, we wish to emphasize that AGN activity in low redshift
cooling flows is very far from being the only ingredient required for
downsizing, and that we still find downsizing in our model. We now 
review some of the indirect evidence already presented in support of this
conclusion (\S\ref{sec:downsizeagree}), and go on to present explicit
predictions which reveal which black holes in our model are accreting mass most rapidly
(\S\ref{sec:downsizeplot}).

\subsection{Indirect evidence for downsizing in the model: the evolution of 
the optical luminosity function}
\label{sec:downsizeagree}

The optical quasar luminosity function, as we have already remarked, 
reveals a dramatic increase in the space density of bright quasars with 
increasing redshift (\cite{croom04}). In \S\ref{sec:qlf}, we presented 
the model predictions for the optical luminosity function, which are in 
good agreement with this trend in the observations. Two features of our 
model are responsible for this success: the increase in the halo merger 
rate (and hence the galaxy merger rate) with increasing redshift and the 
increase in the gas content of discs with increasing redshift (see also 
\cite{kh00}). In combination, these phenomena lead to an increase in the
frequency and strength of starbursts with redshift. In our model, a starburst
results in a {\it luminous} phase of growth of the supermassive black hole; a
fraction $F_{\rm BH}$ of the cold gas which is turned into stars during the
burst is accreted onto the black hole.

Galaxy mergers are still an important way of building black hole mass 
at low redshift. Our model predicts that BH-BH mergers are the most important 
channel for building black hole mass for the most massive black holes at the 
present day. This {\it dark} growth process represents the assembly of mass 
which is already locked up in black holes into larger units. Galaxy mergers 
at low redshift tend to be gas poor in our model simply because more time 
has elapsed to allow galactic discs to turn cold gas into stars quiescently. 
This effect is accentuated in the case of the most massive black holes which 
tend to reside in the more massive dark haloes. The process of galaxy 
formation starts earlier in the progenitors of massive haloes, since these 
objects collapse into bound structures earlier than is the case in less 
extreme environments. 

\subsection{Direct evidence for downsizing in the model: which black holes are 
accreting mass?}
\label{sec:downsizeplot}

\begin{figure*}
\begin{picture}(0,640)
\put(-240,430){\epsfxsize=8.0 truecm \epsfbox{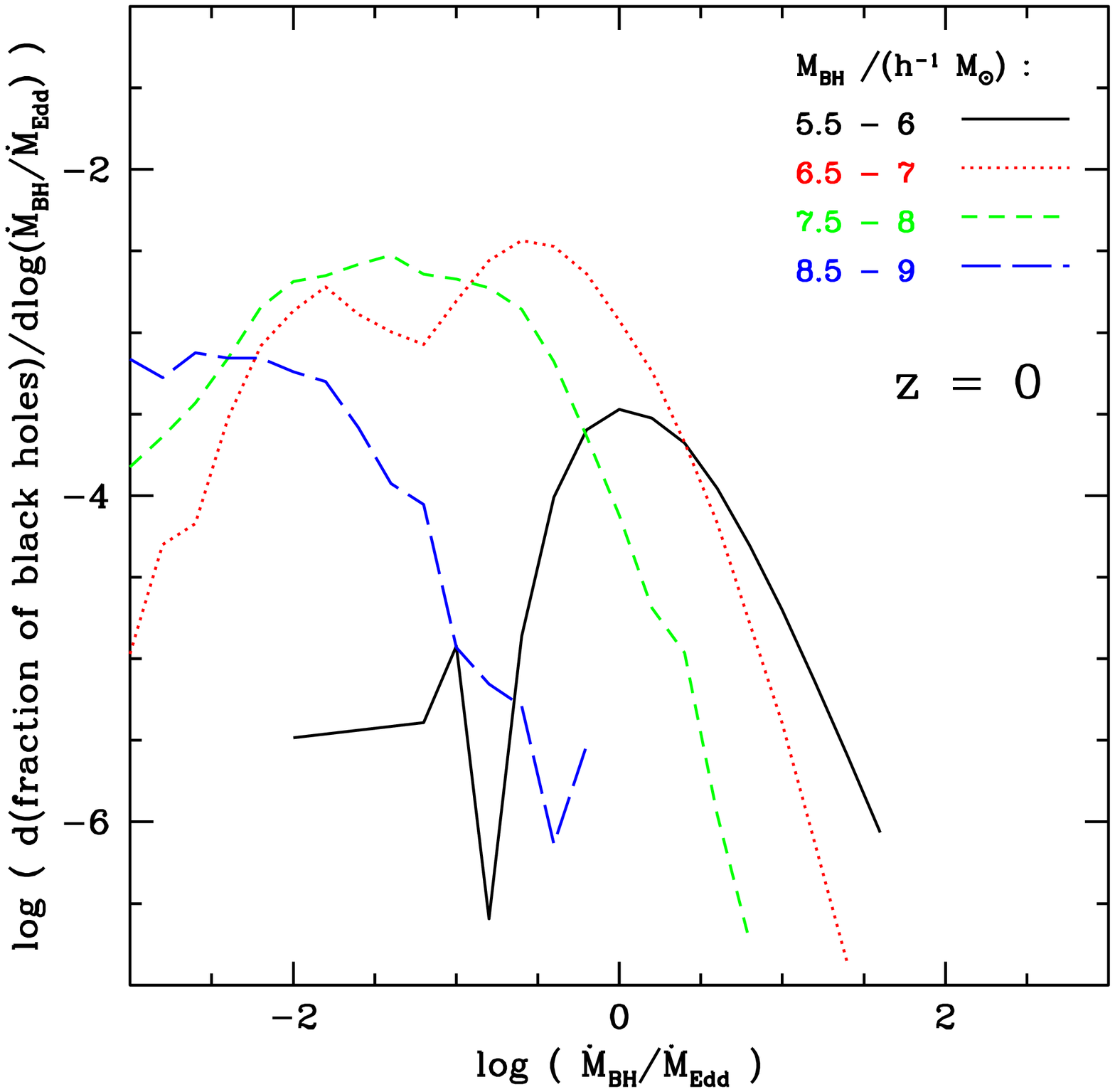}}
\put(10,430){\epsfxsize=8.0 truecm \epsfbox{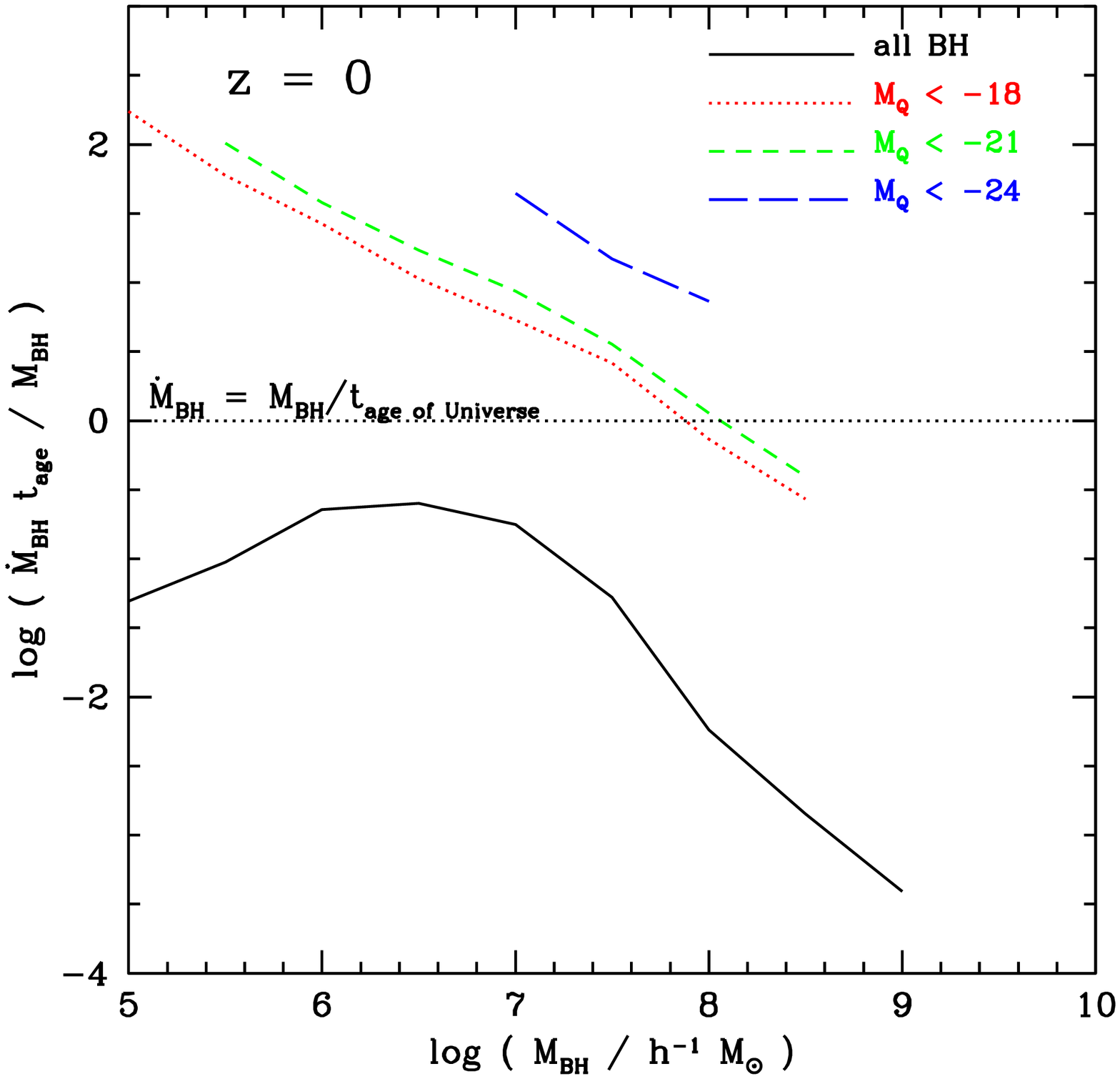}}
\put(-240,215){\epsfxsize=8.0 truecm \epsfbox{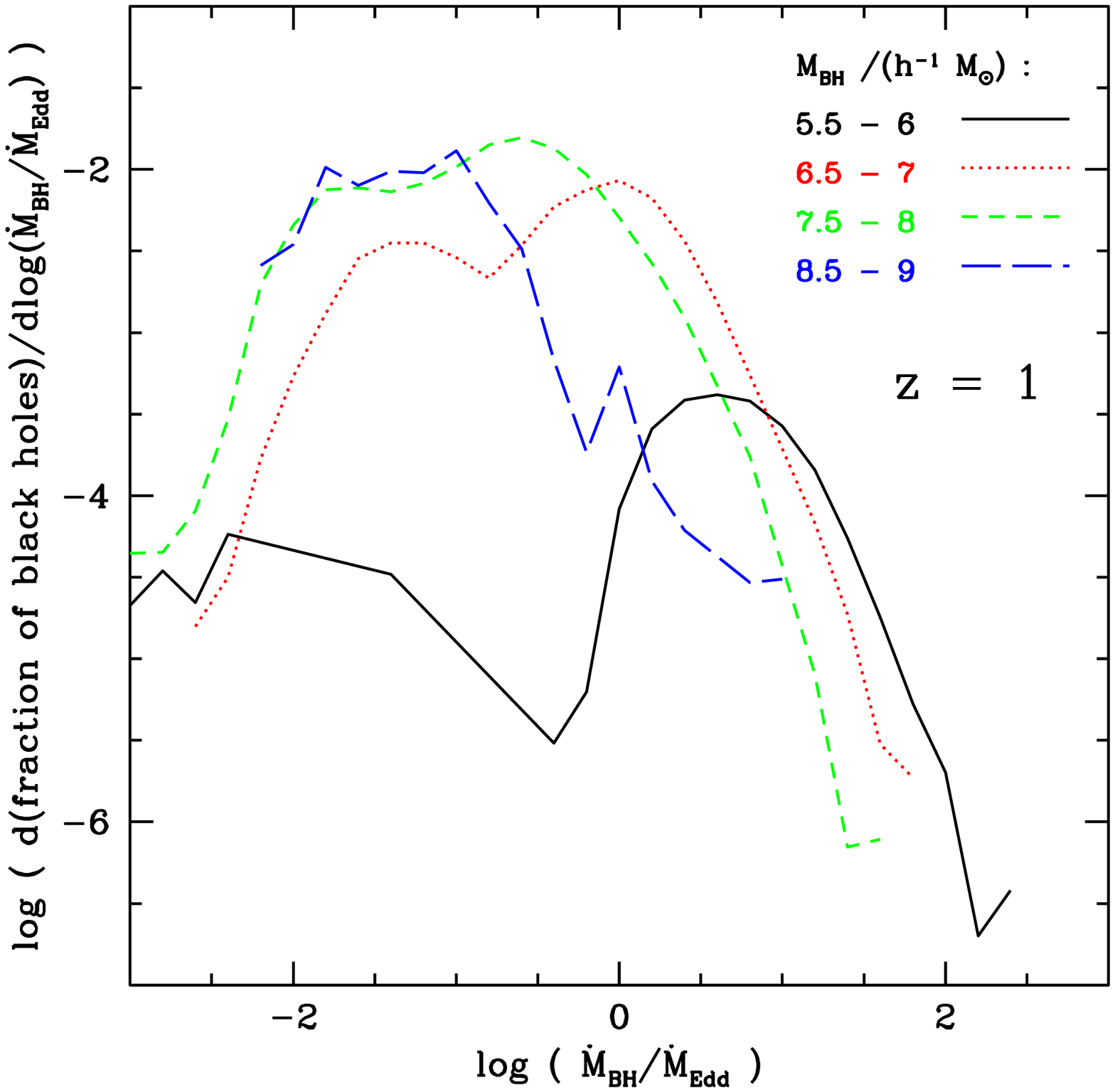}}
\put(10,215){\epsfxsize=8.0 truecm \epsfbox{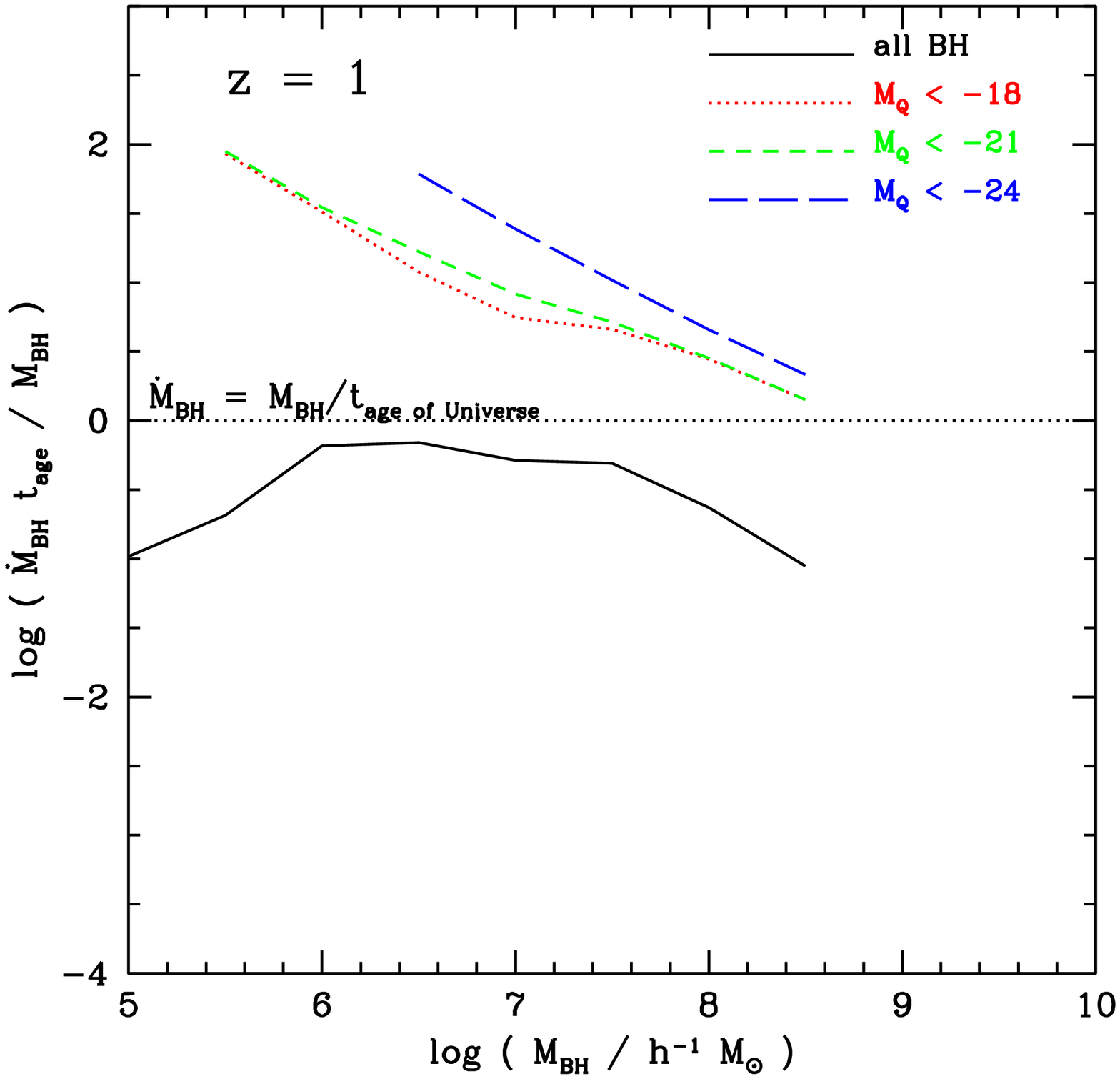}}
\put(-240,0){\epsfxsize=8.0 truecm \epsfbox{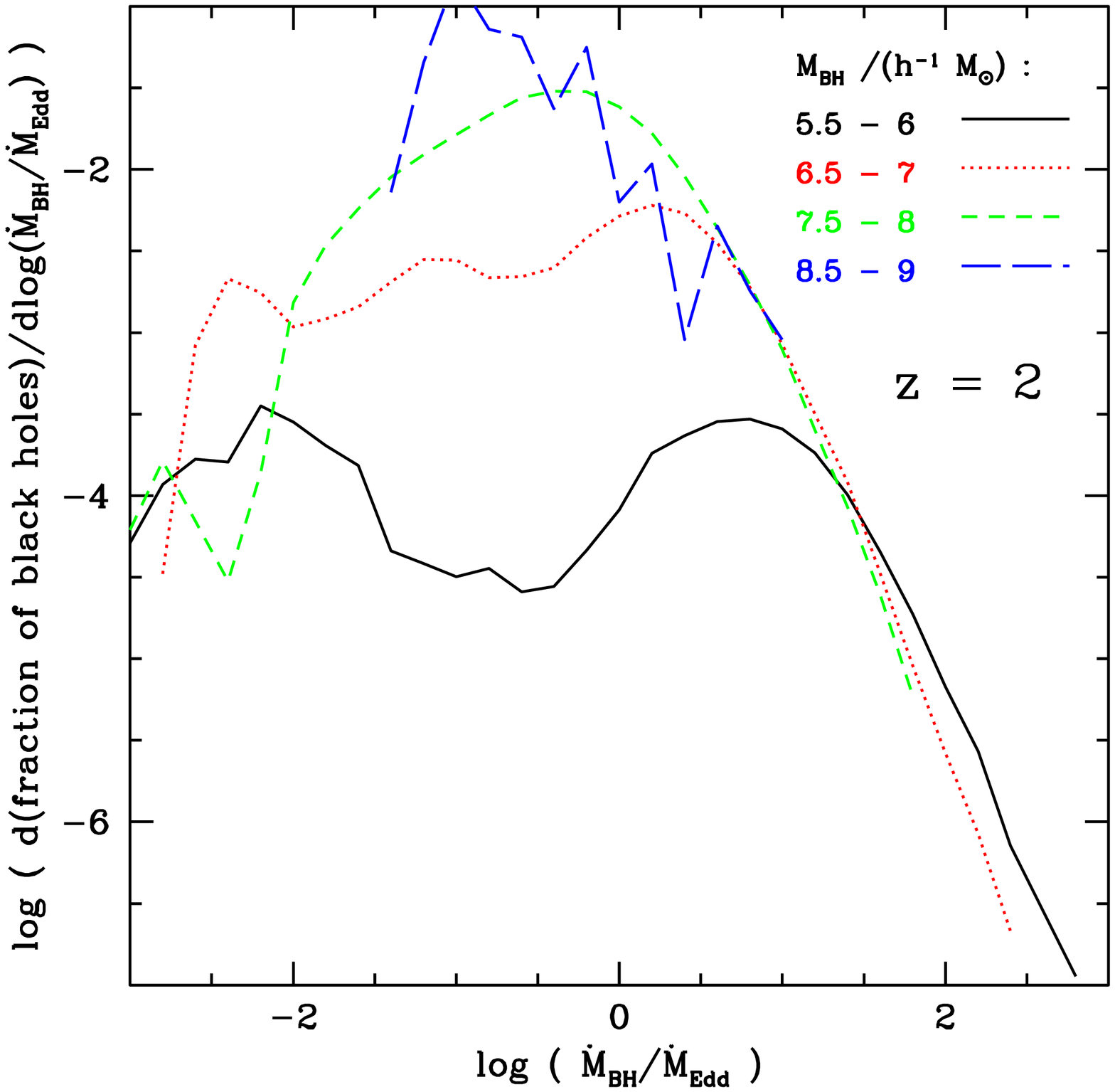}}
\put(10,0){\epsfxsize=8.0 truecm \epsfbox{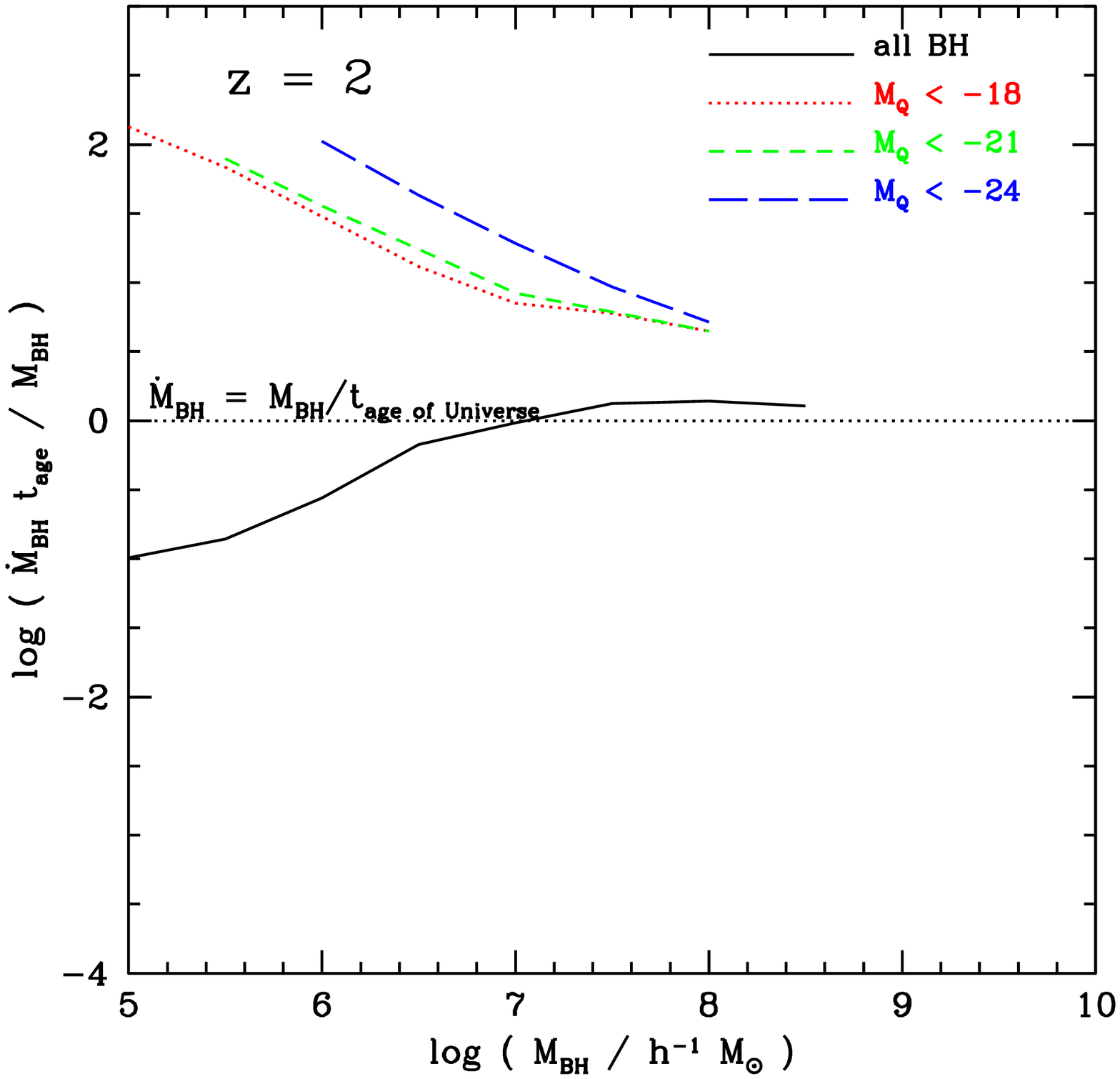}}
\end{picture}
\caption{Left panels: The distribution of accretion rate normalized 
by the Eddington mass accretion rate. Right panels: the current 
mean mass accretion rate normalized by the past average mass growth rate, 
plotted against black hole mass. In both cases, each row corresponds 
to a different redshift: $z=0$ (top), $z=1$ (middle) and $z=2$ (bottom).
In the left hand panels, each line shows the distribution of 
accretion rates for black in the mass interval shown by the key. 
In the right hand panels, the lines correspond to different 
cuts on quasar luminosity, again as shown by the key. 
}
\label{fig:downsize}
\end{figure*}

Our model allows us to separate the mass assembly of black holes into two 
contributions: accretion, in which cold gas is turned into black hole 
mass in a starburst and mergers, in which existing black holes merge 
to build a more massive black hole. Here we focus on the process of gas 
accretion. Fig.~\ref{fig:downsize} presents two views showing which mass  
of black holes are accreting material the most vigorously. The 
left-hand panels of Fig. \ref{fig:downsize} show the distribution of 
accretion rates, expressed in units of the Eddington mass accretion
rate. Since the Eddington accretion rate scales with mass, this is
easily scaled to give the distribution of fractional accretion rates. The right-hand panels compare 
the present accretion rate to the past average accretion rate
(calculated as $\langle \dot M \rangle \times
~t_{\rm age \thinspace of \thinspace Universe}/\langle M_{\rm BH} \rangle$, as a function of black 
hole mass. Each row corresponds to a different redshift (top: $z=0$, middle: 
$z=1$, bottom $z=2$). In these plots, we have \emph{not} limited the mass 
accretion rate to be less than or equal to the Eddington limit. 

The left-hand side of Fig.~\ref{fig:downsize} shows that, at all redshifts,
there is a large spread in the Eddington ratios at which black holes are
accreting. There is variation amongst mergers in gas supply, accretion timescale
and initial black hole mass. Furthermore, the Eddington ratio evolves during any
single accretion event. As expected, the mass accretion shifts towards higher
fractions of the Eddington limit at higher redshift since there is more gas
available in mergers. At $z=0$, we see that as black hole mass increases,
accretion shifts to lower fractions of the Eddington ratio. This trend is less
pronounced at $z=1$ and practically disappears by $z=2$. Thus, more massive
black holes were accreting mass more rapidly at $z=2$ than they are today. The
predicted distribution of mass accretion rates at $z=0$ agree reasonably well
with the observational results of Heckman \etal (\shortcite{heckman04}). The distribution is normalized to unit area, but most black holes, particularly at lower redshifts, are not accreting at all (i.e. they are in a $\delta$-function at $\dot M = 0$).

The right-hand column of Fig.~\ref{fig:downsize} shows the ratio of
the present accretion rate to the past average accretion rate
($\thinspace\langle \dot M \rangle \times
~t_{\rm age}/\langle M_{\rm BH} \rangle\thinspace$) as a function of black hole mass at $z=$ 0, 1 and 2. If
this ratio exceeds unity, then the current mass accretion rate exceeds the
average rate at which the black hole gained mass in the past (summed over all
progenitors). The predictions for this ratio are sensitive to the
black hole selection, for example, selection using a cut in quasar
luminosity. We show results for all black holes (solid lines) and also
for black holes selected as quasars brighter than a given luminosity. Note that the bulk of black
holes in the model are not accreting material at any given time. The solid lines
in each panel show that there is clear evidence for downsizing in the model. At
$z=0$, more massive black holes are growing less rapidly than less massive black
holes. By $z=1$, this trend is greatly diminished, and at $z=2$ it is reversed,
i.e. the most massive black holes have the highest fractional accretion
rate. When only quasars are selected, those with low mass black holes show
similar fractional accretion rates as redshift varies from 0 to 2,
while quasars with massive black holes
show a strong decline in fractional accretion rate towards the present day.

\section{Summary and Discussion}
\label{sec:discuss}

We have described an extension to the {\tt GALFORM} semi-analytical model of
galaxy formation in the $\Lambda$CDM cosmology to track the growth of black
holes (BH). Our model for black hole growth has one free parameter, $F_{\rm
BH}$, the mass accreted onto the black hole as a fraction of the stellar mass produced during a
starburst. We set the value of $F_{\rm BH}$ so as to reproduce the zeropoint of
the present day $M_{\rm BH}-\rm bulge$ relations. The slope, scatter
and evolution of the
$M_{\rm BH}-\rm bulge$ relations are model predictions.

In our model, black holes grow only during and following a galaxy
merger. They grow through two distinct channels: mergers of pre-existing black holes and accretion
of cold gas if a starburst is triggered by the merger. The importance of growth
through black hole mergers increases with the mass of the black hole; at $z=0$ the growth
of black holes less massive than $5 \times 10^7 h^{-1} M_{\odot}$ is
dominated by accretion, while the growth of more massive black holes is
dominated by mergers. In general, the growth of black hole mass by mergers
becomes more important at low redshifts as the supply of gas available for
accretion is consumed by star formation. Our model neglects black hole growth
from gas accreted directly in a cooling flow from a hot gas reservoir,
as may be expected in massive haloes at late times. This is the ``feedback mode'' of
black hole growth invoked by Bower \etal (\shortcite{bower06}) in their model
that explains why there is an exponential cutoff at the bright end of the galaxy
luminosity function. Apart from this new growth channel and an
explicit treatment of disc instabilities, the calculation of
black hole growth in the Bower \etal model is very similar to ours. However,
around 20\% of the global mass in black holes in the Bower \etal study is due to
the ``feedback mode'' of growth. It is as important as mass assembly due to galaxy mergers for the most massive black holes, which accumulate $\sim 50\%$ of their mass via this channel (Richard Bower, priv. comm.).

Essentially all current observational estimates of the accumulation of black hole mass are
sensitive to \emph{luminous} growth, i.e. mass accretion. However, we predict
that the importance of growth through BH-BH mergers grows with
decreasing redshift and with increasing black hole mass. BH-BH mergers represent
a \emph{dark} mode of growth that is more difficult to observe and confirm. The
most obvious way to detect BH-BH mergers is through the emission of
gravitational waves (e.g. \cite{haehnelt94}). This may be possible in ten years
with the planned LISA gravitational wave interferometer. When gas is present
during a black hole merger, a circumbinary accretion disc could form and the BH
merger may produce high velocity gas outflows (\cite{an02}) followed by an X-ray
afterglow which could be detected by the next generation of X-ray observatories
(\cite{mp05}). Winged or X-shaped radio sources (\cite{me02}) and cores in
elliptical galaxies (\cite{faber97,milos02}) may be indirect evidence of
gas-poor mergers.

Our model predicts that the most important growth mechanism for the most massive
black holes is the ``dark'' mode or mass assembly through BH mergers. A testable
prediction of our model is that a tail of black holes with masses
above a few times $10^{9} M_\odot$ should be found at $z=0$ once high quality
observations covering a large volume of the local Universe become available
(\S\ref{sec:bhmf}). Furthermore, we expect that these will be more
massive than any found in 
quasars at high redshift. To date, only black holes less massive than $\sim 3
\times 10^9 M_{\odot}$ have been unambiguously observed in galaxies at $z=0$
(\cite{tremaine02}) and in luminous, optically-selected quasars over the
redshift interval $0<z<2$ (\cite{md04}). However, this implied limit
on black hole mass is far from robust. The most massive black holes at $z=0$ tend to
reside in massive and hence rare elliptical galaxies, which could easily have
been missed in existing surveys. Larger volumes ($\gtrsim$ a few times
$10^6 h^{-3} \rm Mpc^3$) need to be surveyed to find such
objects, which are therefore likely to lie at large distances. This, coupled with their
expected low surface brightness (more massive ellipticals tend to have lower
surface brightness cores), could make it difficult to measure their
central mass using methods based on stellar dynamics (\cite{kg01}). The high
redshift quasar data do not give a complete census of the black hole
populations, as quasar observations are only able to probe accreting black
holes.

There is an important distinction to be made in our model between \emph{mass
transformation} and \emph{mass assembly}. Mass transformation refers to the
process of turning cold gas into black hole mass; at any one time in the
formation history of a black hole, this phenomenon could be occuring across a
number of progenitor black holes. Mass assembly refers to the
accumulation of mass in a black hole's main progenitor, and may occur via both direct
accretion of gas and merging of pre-existing black holes. Black hole mass \emph{assembles} hierarchically; more
massive black holes are assembled at lower redshifts than less massive black
holes. However, if we choose to define the formation time of a black hole in
terms of the mass \emph{transformation} redshift when some fraction of its mass has been \emph{accreted}
onto \emph{any} progenitor, we find that, for $M_{\rm BH} > 10^7 h^{-1}
M_{\odot}$, more massive black holes form earlier. This dichotomy mirrors the
growth of stellar mass in galactic spheroids in hierarchical models. In the
semi-analytical models, galaxy mergers produce spheroids. At high redshift, the
mergers tend to be gas rich and new stars are produced as a result of
the merger event. At low redshift, galactic discs tend to be gas
poor and consist mainly of stars, with the result that the merger simply rearranges the pre-existing stars (\cite{baugh96,kauffmann96,bell06,delucia06}). 

While we find that black hole mass is assembled hierarchically, our model
clearly exhibits a ``downsizing'' in the mass of black holes which are
undergoing luminous accretion.  At the present day, we find that low mass black
holes are accreting material at a higher proportion of their Eddington
luminosity than high mass black holes. This distinction is less apparent at
higher redshifts. Another way to demonstrate this downsizing is to examine the
rate at which black holes are accreting mass, expressed as a fraction of the
mass already in place: $\langle \dot M_{\rm BH}\rangle / \langle
M_{\rm BH} \rangle$. At $z=0$, $\dot M_{\rm
BH}/M_{\rm BH}$ is largest for low mass black holes and drops rapidly with
increasing mass. This trend is removed with increasing redshift; accretion
becomes an increasingly important mode of mass assembly for all masses of
black hole at earlier epochs in the universe. 

A number of authors have claimed that black holes grow in an ``anti-hierarchical''
fashion (\cite{marconi04,merloni04,shankar04}). This conclusion is reached by
comparing an inferred present day black hole mass function with the black hole
mass function expected from AGN relics under the assumption that black holes
grow exclusively by accretion. These calculations ignore any contribution to the
mass of black holes arsising from BH-BH mergers. Furthermore, the assumption
that all black holes accrete at a constant fraction of their Eddington ratio
(\cite{marconi04,shankar04}), or the use by Merloni (\shortcite{merloni04}) of
the `fundamental plane of black hole activity' (\cite{mhd03}), which has a very
large scatter, will introduce errors that may become cumulatively very large as
the black hole mass function is integrated backwards in time.

While the zeropoints of the present day $M_{\rm BH}-\rm bulge$
relations are set by
adjusting a single free parameter, the slope, scatter and evolution of
these relations are genuine predictions of the model. We find little evolution with
redshift in the slope of any of the $M_{\rm BH}-\rm bulge$
relations, although our model predicts differing evolution in their zeropoints, depending upon which particular bulge property is being considered. If we
focus attention on a fixed black hole mass, we find that with increasing
redshift, the typical host bulge is more luminous in the rest-frame B-band and
K-band, shows little change in stellar mass (except in the case of low mass
black holes, where the stellar mass is lower) and has a somewhat higher velocity dispersion.

Our model predicts the presence of massive black holes at high
redshift. However, our simulations are not large enough to check if a population
of sufficiently massive black holes has formed at high redshift to account for
observations of quasars that have been interpreted as implying the presence of
very massive black holes at early times (Fan \etal 2001, 2004).  Fan \etal have
discovered quasars of magnitude $M_{1450} \sim -27$ at $z\sim 6$, albeit at a
low space density, $\phi \approx 1.6 \pm 0.5 \times 10^{-10} h^3 \rm Mpc^{-3}$
(\cite{fan04}). Assuming that these objects are radiating at the Eddington
luminosity with an Elvis \etal (\shortcite{elvis94}) spectrum, and that beaming
and gravitational lensing are insignificant, Fan \etal (\shortcite{fan01})
inferred that these quasars host black holes masses of $\sim 1-3 \times 10^9
h^{-1} M_{\odot}$. In this paper, we probe the mass function of $z=6$ black
holes only down to a space density of $\phi \sim 1 \times 10^{-8} h^3 \rm
Mpc^{-3}$, which is not sparse enough to compare with these data. We plan to
address this problem by performing simulations of a larger volume than those analyzed here, which will probe
the high mass end of the halo mass function in detail.

In summary, we have presented a new model for the concurrent growth of galaxies
and black holes in the $\Lambda$CDM cosmology. We have previously shown that
this model can successfully account for many observed properties of the galaxy
population over a large range of wavelengths, from the local optical and
infrared galaxy luminosity function to the number counts of submillimeter
galaxies and the UV luminosity function of Lyman-break galaxies at redshift
$z \sim 3$ (Baugh \etal 2005). In this paper we have focussed on the
properties of the black hole population that grows in unison with the spheroidal
component of the galaxies. This model can account for a variety of observables
that involve black holes, such as the relationship between the mass of the
central black hole in galaxies and the properties of the bulge, the quasar
luminosity function, and the apparent ``anti-hierarchical'' growth of black
holes. The model may be tested by future observations of the evolution of the
$M_{\rm BH}-\rm bulge$ relations and, perhaps, by the detection of
gravitational waves associated with the mergers of massive black holes that play
a prominent role in our model.

\section{Acknowledgements}
We thank Andrew Benson, Richard Bower, Marta Volonteri, Darren Croton,
Gregory Novak and the anonymous referee for
useful comments and discussions and John Helly for providing tree-plotting
routines. We are indebted to Lydia Heck for providing extensive help with
computing and for maintaining the PC cluster on which the calculations in this
paper were performed. RKM was supported by a PPARC studentship. 
CMB is supported by the Royal Society. CSF is a Royal Society-Wolfson 
Research Merit Award holder. This research was supported by the PPARC 
rolling grant.

\end{document}